\documentclass[11pt]{article}
\pdfoutput=1

\usepackage{jheppub}
\usepackage{url,comment}
\usepackage{times}
\usepackage{latexsym}
\usepackage{graphicx, graphics, hyperref, amsmath, amssymb, slashed, xcolor, bbm,bm,amsthm, array}
 \usepackage{subfigure}
 \usepackage{listings} 
 \usepackage{trimclip}

\usepackage{rotating}
\usepackage{afterpage}
 
\newcommand{\nc}{\newcommand}

\nc{\beq}{\begin{equation}}
\nc{\eeq}{\end{equation}}
\nc{\barray}{\begin{eqnarray}}
\nc{\earray}{\end{eqnarray}}
\nc{\barrayn}{\begin{eqnarray*}}
\nc{\earrayn}{\end{eqnarray*}}
\nc{\bcenter}{\begin{center}}
\nc{\ecenter}{\end{center}}
\nc{\mc}{\mathcal}
\nc{\er}[1]{(\ref{eq:#1})}
\nc{\onehalf}{\frac{1}{2}} 
\nc{\partialbar}{\bar{\partial}}
\nc{\psit}{\widetilde{\psi}}
\nc{\Tr}{\mbox{Tr}}
\nc{\hc}{\mbox{H.c.}}
\nc{\ev}{\;\mathrm{eV}}
\nc{\mev}{\;\mathrm{MeV}}
\nc{\gev}{\;\mathrm{GeV}}
\nc{\kev}{\;\mathrm{keV}}
\nc{\tev}{\;\mathrm{TeV}}
\nc{\pev}{\;\mathrm{PeV}}
\nc{\eev}{\;\mathrm{EeV}}

\def\chii0{\chi_i^0}
\def\chij0{\chi_j^0}

\newcommand{\gsim}{\lower.7ex\hbox{$\;\stackrel{\textstyle>}{\sim}\;$}}
\newcommand{\lsim}{\lower.7ex\hbox{$\;\stackrel{\textstyle<}{\sim}\;$}}
\nc{\ttbar}{t\bar t}

\newcommand{\eref}[1]{Eq.~(\ref{e.#1})}

\newcommand{\cref}[1]{Chapter~\ref{c.#1}}

\graphicspath{{plots/}}

\title{A Letter of Intent for MATHUSLA: a dedicated displaced vertex detector above ATLAS or CMS}

\author[a]{Cristiano Alpigiani,}
\author[o]{Austin Ball,}
\author[c]{Liron Barak,}
\author[ah]{James Beacham,}
\author[c]{Yan Benhammo,}
\author[c]{Tingting Cao,}
\author[f,g]{Paolo Camarri,}
\author[f]{Roberto Cardarelli,}
\author[h]{Mario Rodr\'iguez-Cahuantzi,}
\author[d]{John Paul Chou,}
\author[b]{David Curtin,}
\author[e]{Miriam Diamond,}
\author[f]{Giuseppe Di Sciascio,}
\author[x]{Marco Drewes,}
\author[u]{Sarah C. Eno,}
\author[c]{Erez Etzion,}
\author[q]{Rouven Essig,}
\author[v]{Jared Evans,}
\author[w]{Oliver Fischer,}
\author[k]{Stefano Giagu,}
\author[d]{Brandon Gomes,}
\author[l]{Andy Haas,}
\author[z]{Yuekun Heng,}
\author[aa]{Giuseppe Iaselli,}
\author[m]{Ken Johns,}
\author[u]{Muge Karagoz,}
\author[d]{Luke Kasper,}
\author[a]{Audrey Kvam,}
\author[ae]{Dragoslav Lazic,}
\author[af]{Liang Li,}
\author[f]{Barbara Liberti,}
\author[y]{Zhen Liu,}
\author[a]{Henry Lubatti,}
\author[n]{Giovanni Marsella,}
\author[o]{Matthew McCullough,}
\author[p]{David McKeen,}
\author[q]{Patrick Meade,}
\author[c]{Gilad Mizrachi,}
\author[p]{David Morrissey,}
\author[c]{Meny Raviv Moshe,}
\author[j]{Karen Salom\'e Caballero-Mora,}
\author[ab]{Piter A. Paye Mamani,}
\author[k]{Antonio Policicchio,}
\author[a]{Mason Proffitt,}
\author[ad]{Marina Reggiani-Guzzo,}
\author[a]{Joe Rothberg,}
\author[f,g]{Rinaldo Santonico,}
\author[ag]{Marco Schioppa,}
\author[t]{Jessie Shelton,}
\author[s]{Brian Shuve,}
\author[ab]{Martin A. Subieta Vasquez,}
\author[r]{Daniel Stolarski,}
\author[o]{Albert de Roeck,}
\author[h]{Arturo Fern\'andez T\'ellez,}
\author[h]{Guillermo Tejeda Mu\~{n}oz,}
\author[h]{Mario Iv\'{a}n  Mart\'inez Hern\'andez,}
\author[c]{Yiftah Silver,}
\author[d]{Steffie Ann Thayil,}
\author[a]{Emma Torro,}
\author[u]{Yuhsin Tsai,}
\author[i]{Juan Carlos Arteaga-Vel\'azquez,}
\author[a]{Gordon Watts,}
\author[e]{Charles Young,}
\author[w,ac]{Jose Zurita.}

\affiliation[a]{University of Washington, Seattle}
\affiliation[b]{University of Toronto}
\affiliation[c]{Tel Aviv University}
\affiliation[d]{Rutgers University}
\affiliation[e]{SLAC National Accelerator Laboratory} 
\affiliation[f]{Istituto Nazionale di Fisica Nucleare, Sezione di Roma Tor Vergata, Roma, Italy}
\affiliation[g]{Universit\`{a} degli Studi di Roma Tor Vergata, Roma, Italy}
\affiliation[h]{Benem\'erita Universidad Aut\'onoma de Puebla, Mexico (BUAP)}
\affiliation[i]{Universidad Michoacana de San Nicol\'as de Hidalgo, Mexico (UMSNH)}
\affiliation[j]{Universidad Aut\'onoma de Chiapas, Mexico (UNACH)}
\affiliation[k]{Universit\`{a} degli Studi di Roma La Sapienza, Roma, Italy}
\affiliation[l]{New York University}
\affiliation[m]{University of Arizona}
\affiliation[n]{Universit\`{a} del Salento, Lecce, Italy}
\affiliation[o]{CERN}
\affiliation[p]{TRIUMF}
\affiliation[q]{SUNY, Stony Brook}
\affiliation[r]{Carleton University}
\affiliation[s]{Harvey Mudd College}
\affiliation[t]{University of Illinois Urbana-Champaign}
\affiliation[u]{University of Maryland}
\affiliation[v]{University of Cincinnati}
\affiliation[w]{Institute for Nuclear Physics (IKP), Karlsruhe Institute of Technology}
\affiliation[x]{Centre for Cosmology, Particle Physics and Phenomenology, Universit\'{e} catholique de Louvain, Louvain-la-Neuve B-1348, Belgium}
\affiliation[y]{Fermilab}
\affiliation[z]{Institute of High Energy Physics, Beijing}
\affiliation[aa]{Politecnico di Bari, Italy}
\affiliation[ab]{Instituto de Investigaciones F\'isicas (IIF), 
Observatorio de F\'isica C\'osmica de “Chacaltaya”,
Universidad Mayor de San Andr\'es (UMSA)}
\affiliation[ac]{Institute for Theoretical Particle Physics (TTP), Karlsruhe Institute of Technology}
\affiliation[ad]{University of Campinas}
\affiliation[ae]{Boston University}
\affiliation[af]{Shanghai Jiao Tong University}
\affiliation[ag]{INFN and University of Calabria}
\affiliation[ah]{Ohio State University}

\abstract{
In this Letter of Intent (LOI) we propose the construction of MATHUSLA (MAssive Timing Hodoscope for Ultra-Stable neutraL pArticles)~\cite{Chou:2016lxi}, a dedicated large-volume displaced vertex detector for the HL-LHC on the surface above ATLAS or CMS. Such a detector, which can be built using existing technologies with a reasonable 
budget in time for the HL-LHC upgrade, could search for neutral long-lived particles (LLPs) with up to several orders of magnitude better sensitivity than ATLAS or CMS, while also acting as a cutting-edge cosmic ray telescope at CERN to explore many open questions in cosmic ray and astro-particle physics.
We review the physics motivations for MATHUSLA and summarize its LLP reach for several different possible detector geometries, as well as outline the cosmic ray physics program. 
We present several updated background studies for MATHUSLA, which help inform a first detector-design concept utilizing modular construction with Resistive Plate Chambers (RPCs) as the primary tracking technology.
We present first efficiency and reconstruction studies to verify the viability of this design concept, and we explore some aspects of its total cost.
We end with a summary of recent progress made on the MATHUSLA test stand, a small-scale demonstrator experiment currently taking data at CERN Point 1, and finish with a short comment on future work.
}

\begin{document}

\begin{flushright}
\small{.}
\end{flushright}

\maketitle


\section{Introduction}
\label{s.introduction}


In this Letter of Intent we propose the construction of MATHUSLA (MAssive Timing Hodoscope for Ultra-Stable neutraL pArticles)~\cite{Chou:2016lxi}, a dedicated large-volume displaced vertex detector (DV) for the HL-LHC on the surface above ATLAS or CMS. Such a detector, which can be built using existing technologies with a reasonable 
budget in time for the HL-LHC upgrade, could search for neutral long-lived particles (LLPs) without trigger limitations and with very low or zero backgrounds, allowing it to probe LLP cross sections and lifetimes up to several orders of magnitude beyond the reach of ATLAS or CMS. 
MATHUSLA would also act as a cutting-edge cosmic ray telescope at CERN, exploring many open questions in cosmic ray and astro-particle physics.

The Large Hadron Collider (LHC) and its main detectors ATLAS and CMS were built primarily to elucidate the mechanism of electroweak symmetry breaking and search for signals of beyond-Standard-Model (BSM) physics. 
The discovery of the 125 GeV Higgs boson in 2012 marked a triumphant first step in realizing the first of these goals~\cite{Aad:2012tfa, Chatrchyan:2012xdj}. While measurements of the new state's couplings and decays are roughly consistent with the Standard Model (SM) expectation, the Higgs is uniquely sensitive in these respects to modifications from new physics (see e.g. \cite{Zeppenfeld:2000td, Giardino:2013bma, Contino:2013kra, Curtin:2013fra}), and its continued study forms the basis of a rich and ongoing physics program at the LHC. 
Remarkable progress also has been made towards the second goal of LHC searches for BSM scenarios covering a great variety of possible signals with impressive reach that directly access mass scales well beyond a TeV. Unfortunately, to date all BSM searches have produced null results. 

This lack of BSM signals is especially vexing in light of strong motivations for new physics at the LHC. These motivations are even more compelling post-Higgs-discovery than when the LHC was built. Solving the hierarchy problem (HP) to stabilize the light Higgs mass via mechanisms like supersymmetry (SUSY)~\cite{Martin:1997ns} or compositeness~\cite{Panico:2015jxa} requires the existence of new states at the TeV scale. 
On purely empirical grounds, the existence of dark matter (DM), the dynamical creation of a baryon asymmetry in the early universe (BAU), and non-zero neutrino masses have been established beyond any reasonable doubt, and all of them require BSM physics. Many theoretical frameworks that explain one or all of these fundamental mysteries are expected to include new particles that are  discoverable at the LHC. Where are these new states?

One simple possibility to explain this LHC conundrum is that our searches for BSM physics have been operating under incorrect assumptions about what those signals look like. Most searches focus on the production of heavy new states at or above the weak scale, which promptly decay and give rise to high-energy visible final states including jets, leptons, photons, or stable invisible particles that are detectable as missing energy. This reasonable expectation, 
informed and validated by the historical discovery program that established the SM in the last few decades, is implicit in the design choices made for the LHC main detectors. It is, however, far from the only possibility for BSM signals. 
A generic but much less explored alternative is that BSM signals first show up as new Long-Lived Particles (LLPs) that decay into SM final states some macroscopic distance away from the production point. Historically, such signals find their analogue in earlier explorations of the SM, such as the discovery and characterization of the muon as well as the zoo of hadron states. This makes clear that LLPs
are realized many times in the SM, and more broadly are a generic bottom-up possibility in BSM theories as well. 
In recent years it has become increasingly apparent that far from being a curiosity, LLPs are as broadly and fundamentally top-down motivated as prompt BSM signals. Theories with LLPs, such as many versions of hidden sectors or hidden valleys \cite{Strassler:2006im,Strassler:2006ri}, Neutral Naturalness \cite{Chacko:2005pe, Burdman:2006tz, Cai:2008au} and SUSY \cite{PhysRevD.40.2987, Dreiner:1997uz, Barbier:2004ez, Csaki:2013jza, Csaki:2015fea, Giudice:1998bp, Kim:1983dt,Chun:1991xm}, WIMP Baryogenesis \cite{Cui:2012jh}, many models of DM~\cite{Griest:1990kh,Kaplan:2009ag,Kim:2013ivd,Hall:2009bx,Hochberg:2014dra,Kuflik:2015isi,Dror:2016rxc,Dienes:2011ja,Dienes:2011sa} and neutrino extensions~\cite{Minkowski:1977sc,Mohapatra:1979ia,Yanagida:1979as,GellMann:1980vs,Glashow:1979nm,Batell:2016zod,Pati:1974yy,Mohapatra:1974gc,Helo:2013esa,Chun:2017spz, Asaka:2005an,Asaka:2005pn,Cai:2017mow},
can address some or all of the fundamental shortcomings of the SM, including the HP, DM, BAU and the origin of neutrino masses (see \cite{Curtin:2018mvb} for a more comprehensive discussion). 
Due to their connections to the weak or TeV scale, many of these LLP signals can only be produced at energy frontier machines like the LHC, HL-LHC or a possible HE-LHC.
However, even if LLPs are already being produced in LHC collisions as the first harbinger of new physics, their signals could easily have been overlooked until today.

LLPs could be colored, electrically charged, or neutral. In the former two cases, the passage of these particles shows up in the main detectors (if the lifetime is long enough). Neutral LLPs, on the other hand, can only be directly observed once they decay. In most cases, they are searched for by reconstructing a DV from decay products away from the LHC Interaction Point (IP). The LLP search program at the LHC is growing \cite{Aaboud:2018iil, Aaboud:2017mpt, Aaboud:2017iio, Aaboud:2016uth, Aad:2015uaa, Aad:2015rba, Aad:2015asa, Aad:2013txa, Sirunyan:2018ldc, Sirunyan:2017jdo, Khachatryan:2016sfv, CMS:2018pmg, CMS:2015sjc, CMS:2014wda, CMS:2014hka}, and while coordinated efforts are underway to systematically explore the large LLP signature space \cite{LHCLLPwhitepaper}, the existing coverage is still nowhere near as comprehensive as the prompt search program for evidence of theories like minimal supersymmetry. The geometrical nature of the DV signature makes LLP decays spectacular signals that, while sometimes challenging to reconstruct in the detectors, often have much lower backgrounds than most prompt searches. In some cases, the LHC searches are essentially background-free, but other searches for neutral LLPs have to contend with complicated backgrounds, and limitations of a trigger system that is optimized for high-energy prompt particle production. These limitations are exacerbated if the decay length $\lambda$ is much larger than the size $L$ of the detector, due to a resulting very low decay acceptance $\sim L/\lambda$. In most theories, LLP lifetime is essentially a free parameter, and probing the long-lifetime regime is vital for exploiting the full discovery potential of the LHC.

The importance of LLP searches and the inherent background, trigger and acceptance limitations of the main detectors in probing neutral LLPs with long lifetimes motivated the original MATHUSLA proposal~\cite{Chou:2016lxi} for a dedicated, large-volume, simply instrumented LLP detector for the HL-LHC. In the most basic terms, MATHUSLA is a large, robust tracking system monitoring an empty decay volume on the surface to reconstruct LLP decays as upwards-traveling particles originating from displaced vertices. By operating without trigger limitations in a very low background environment, shielded from the IP by more than 100 m of rock, MATHUSLA can enhance sensitivity to LLP production rates by up to three orders of magnitude compared with the main detectors, at a relatively incremental cost. MATHUSLA's purely parasitic nature also means that the detector, once constructed, could continue functioning without modification even as the HL-LHC is replaced by the HE-LHC.

The physics motivation for MATHUSLA was comprehensively examined in a dedicated theory white paper~\cite{Curtin:2018mvb}, as well as several studies of individual models~\cite{Co:2016fln, Dev:2016vle, Dev:2017dui, Caputo:2017pit, Curtin:2017izq, Evans:2017kti, Helo:2018qej, DAgnolo:2018wcn, Berlin:2018pwi, Deppisch:2018eth, Jana:2018rdf}, and we briefly review it in Section~\ref{s.motivation}. The basic detector principles, including possible geometries and locations, and its inherently scalable and modular characteristics, are outlined in Section~\ref{s.basicprinciples}. 
The search for LLPs is MATHUSLA's primary physics goal, and we present its reach for some particularly well-motivated signal benchmark models, as well as approximate model-independent reach estimates, in Section~\ref{s.LLPphysics}. 
Cosmic ray (CR) physics is MATHUSLA's secondary physics goal, and represents a guaranteed return on the investment of building the detector. 
We describe some of the possible CR measurements that would enhance the experimental astro-particle physics program at CERN in Section~\ref{s.cosmicray}. 

The original MATHUSLA proposal~\cite{Chou:2016lxi} examined the most important backgrounds to LLP searches, which are cosmic rays, muons from the LHC, and neutrino scattering, in order to establish the plausibility of searching for LLPs on the surface with near-zero backgrounds. In Section~\ref{s.backgrounds} we report on the progress of several detailed follow-up studies to firm up those first estimates.

The discussions of the above sections are, at the present stage of development, largely technology-independent, and in fact inform the technical requirements of the final MATHUSLA design. We present a first pass of a modular design using Resistive Plate Chamber (RPC) technology in Section~\ref{s.design}, including a very rough cost estimate as well as various trigger and efficiency studies to establish the viability of the design.  However, we emphasize that this represents just one possibility for implementing the MATHUSLA idea, and other detector technologies are also being explored. The information gleaned from operating the MATHUSLA test stand detector at CERN Point 1 will help guide our background studies. We summarize details of its construction, operation, and first physics results in Section~\ref{s.teststand}. We close with a projected time line of studies and other goals towards a proposal  that defines the construction of a full-scale MATHUSLA detector at CERN over the next few years in Section~\ref{s.timeline}.

\section{Motivation for a Dedicated Surface LLP Detector at CERN}
\label{s.motivation}

The motivation for a MATHUSLA-like external LLP detector at $pp$ colliders was first outlined in the original proposal~\cite{Chou:2016lxi} and expanded upon in several follow-up studies \cite{Co:2016fln, Dev:2016vle, Dev:2017dui, Caputo:2017pit, Curtin:2017izq, Evans:2017kti, Helo:2018qej, DAgnolo:2018wcn, Berlin:2018pwi, Deppisch:2018eth, Jana:2018rdf}. Recently, a comprehensive theory white paper~\cite{Curtin:2018mvb} explored the bottom-up and top-down theoretical and experimental motivations in great detail. Here we briefly summarize the main arguments.

A variety of mechanisms can suppress the decay rate of a fundamental or composite particle, leading to macroscopic decay lengths. This includes accidental or approximate symmetries, phase space suppressions, small dimensionless couplings, and small effective couplings due to a large suppression scale compared to the particle mass. 
In the SM, these mechanisms lead to macroscopic lifetimes for the muon, tau, proton, neutron, several pion and kaon states, $B$-hadrons and other particles in great excess of the naive dimensional analysis (NDA) estimate. 
LLPs are therefore a very familiar and common phenomenon. 
The same mechanisms which lead to long lifetimes in the SM can also apply in almost any BSM theory. 
This bottom-up motivation for LLP signals is readily demonstrated by examining minimal simplified models like dark photons \cite{Holdom:1985ag,Galison:1983pa,Alexander:2016aln}, scalar extensions \cite{Evans:2017lvd,Krnjaic:2015mbs, Bezrukov:2009yw, Clarke:2013aya}, generic hidden valleys \cite{Strassler:2006im,Strassler:2006ri}, axion-like particles (ALPs) \cite{Peccei:1977hh,Peccei:1977ur,Weinberg:1977ma,Wilczek:1977pj,Bauer:2016ydr, Bauer:2016zfj, Bauer:2017nlg, Bauer:2017ris}, and heavy neutral leptons coupling via the neutrino portal \cite{Minkowski:1977sc,Mohapatra:1979ia,Yanagida:1979as,GellMann:1980vs,Glashow:1979nm,Batell:2016zod,Pati:1974yy,Mohapatra:1974gc,Helo:2013esa,Chun:2017spz, Asaka:2005an,Asaka:2005pn,Cai:2017mow}. All of these give rise to LLP signals in parts of their parameter space. In most cases the lifetime is an almost free parameter, and exploring the long-lifetime regime at a MATHUSLA-like detector is therefore inherently motivated on generic BSM grounds.

\begin{figure}
\begin{center}
\includegraphics[width=0.9 \textwidth]{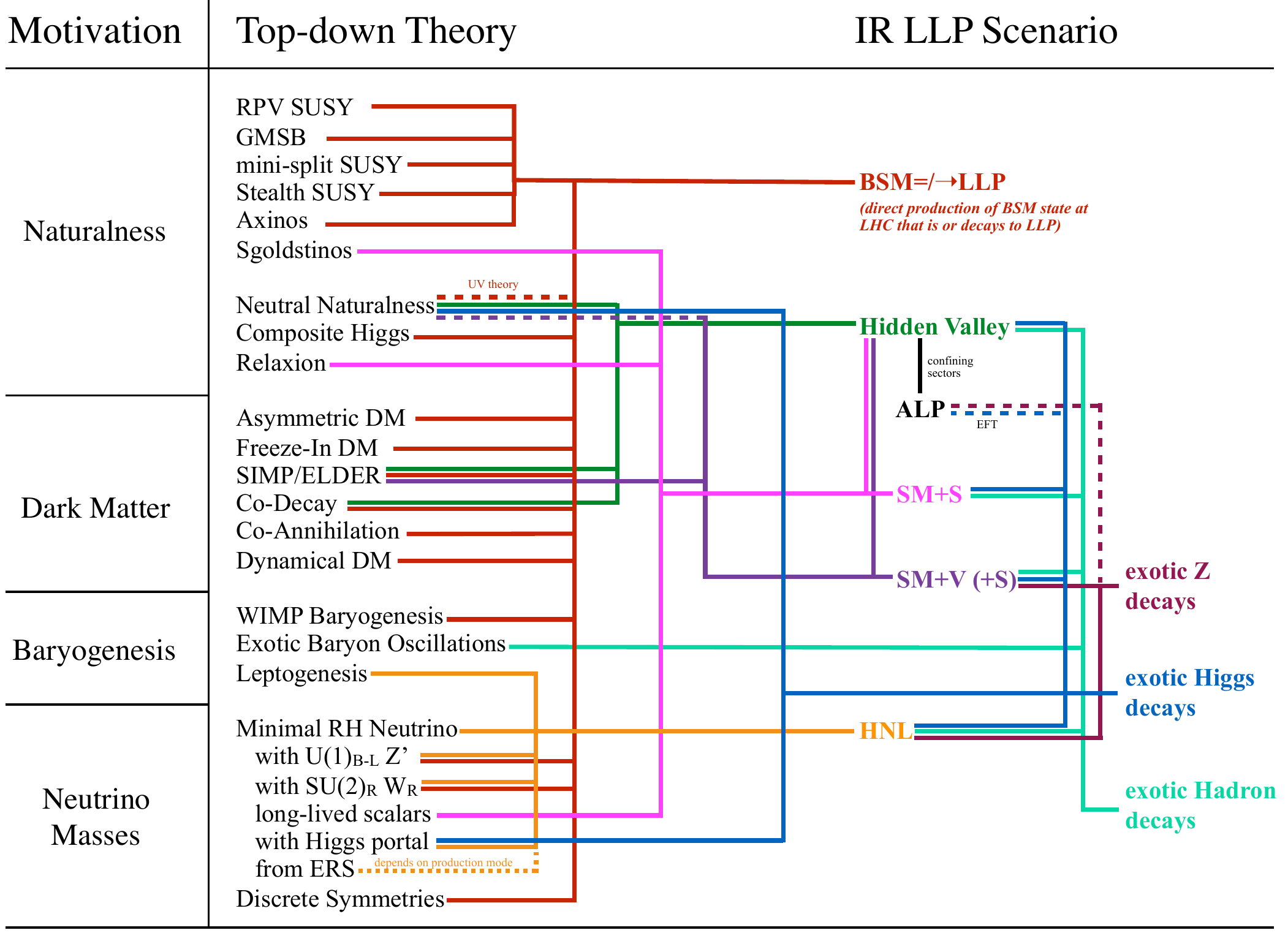}
\end{center}
\caption{
Summary of some top-down theoretical motivations for LLP signals at MATHUSLA. Figure taken from~\cite{Curtin:2018mvb}.
}
\label{f.theorysummary}
\end{figure}

The above arguments establish the \emph{plausibility} of LLP signals at the LHC main detectors and MATHUSLA. However, it is crucial to realize that these signals are also \emph{fundamentally motivated} on top-down theoretical grounds. The MATHUSLA physics case white paper~\cite{Curtin:2018mvb} examined a large variety of models that address the Hierarchy Problem, the nature of Dark Matter, the Baryon Asymmetry of the Universe, and the origin of neutrino masses. 
All of these fundamental mysteries of the SM can be addressed by theoretical frameworks that give rise to LLP production at the LHC. Furthermore, there are many scenarios where observation of LLP decays at MATHUSLA represents the first or only discovery opportunity. This is summarized schematically in Fig.~\ref{f.theorysummary}.

For example, the hierarchy problem could be addressed by Neutral Naturalness~\cite{Chacko:2005pe, Burdman:2006tz, Cai:2008au}, where the Higgs mass is stabilized at low scales by a hidden valley that is related to the SM by a discrete symmetry. Crucially, the top partners that cancel the top quark contributions to the Higgs mass are not charged under SM QCD, making these scenarios unconstrained by standard SUSY searches. However, the hidden valley gives rise to LLP signatures, such as mirror glueballs, which can be produced in exotic Higgs decays~\cite{Craig:2015pha, Curtin:2015fna}. In the long lifetime regime, MATHUSLA is the only way to discover these new states. 
Even standard SUSY scenarios like R-parity violation and gauge mediation give rise to LLP  signatures (a scenario made perhaps more likely by the discovery of a relatively light 125 GeV Higgs boson~\cite{Arvanitaki:2012ps,ArkaniHamed:2012gw,Draper:2011aa}). 
The same is true of many Dark Matter models, especially if the DM candidate is part of a dark sector with a variety of hidden states, some of which can be long-lived. In some theories, like Freeze-In DM \cite{Hall:2009bx, Bernal:2017kxu} 
or Dynamical DM~\cite{Dienes:2011ja,Dienes:2011sa} the LLP plays a crucial role in generating the DM abundance. 
%
Baryogenesis could proceed via the $B$-violating out-of-equillibrium decay of a WIMP-like LLP particle in the early universe, in a scenario called WIMP Baryogenesis \cite{Cui:2012jh}. 
Producing this parent particle and observing its decay would not only solve the mystery of the BAU, but would allow us to witness this cosmic act of creation first-hand in the laboratory. 
Finally, extensions of the SM to generate neutrino masses \cite{Minkowski:1977sc,Mohapatra:1979ia,Yanagida:1979as,GellMann:1980vs,Glashow:1979nm,Batell:2016zod,Pati:1974yy,Mohapatra:1974gc,Helo:2013esa,Chun:2017spz,Asaka:2005an,Asaka:2005pn,Cai:2017mow} are natural candidates for LLP searches, since the small couplings involved inherently give rise to a variety of possible LLP states and signatures~\cite{Bondarenko:2018ptm}.
In many of these scenarios, MATHUSLA is the first or only discovery opportunity, making its construction crucial to exploit the physics potential of the HL-LHC.

Finally, we comment on the purely experimental reasons for building MATHUSLA.
As summarized in~\cite{Curtin:2018mvb}, the most spectacular LLP searches are highly efficient and practically background-free at the LHC and HL-LHC. The combination of the geometrical DV signal requirement with another conspicuous feature of the final state, like well-separated leptons or high-energy jets that are produced either in association with the LLP or in its decay, effectively removes all backgrounds and allows conventional triggers to record the event.
However, any LLP signal that is less conspicuous is limited by both triggers and backgrounds. In particular, full displaced vertex reconstruction in the tracker is not available to the Level 1 trigger systems of ATLAS and CMS, and this capability would be very challenging to include in full for the HL-LHC. As a result, features of the signal event not related to the displaced nature of the LLP decay have to pass Level 1 triggers, severely limiting acceptance for many low-mass or even weak-scale  scenarios below a TeV. Furthermore, exploring the long-lifetime regime necessitates LLP searches that require just one LLP decay in the event, since requiring additional decays lead to further acceptance suppressions at long lifetime. This means that backgrounds can be significant, especially if the LLP decays dominantly hadronically and has a mass below a few hundred GeV, or if it decays to leptons but has a mass below $\sim 10 \gev$.
%
These ``less conspicuous'' LLP signals actually cover a large variety of theoretical scenarios, including the well-motivated class of LLP signatures arising from exotic Higgs decays~\cite{Curtin:2013fra}, and exhaustive exploration of their signature space is very difficult using the main detectors alone.

MATHUSLA has a similar acceptance for LLP decays in the long-lifetime limit as ATLAS or CMS, making up for its smaller solid angle coverage with its greater size. For the purpose of LLP searches, MATHUSLA can therefore be thought of as a heavily shielded (and much simpler) cousin of the main detectors that sacrifices sensitivity to shorter LLP lifetimes $\lesssim 10$m in order to operate in the zero-background-regime without trigger limitations. As a result, MATHUSLA is able to perfectly complement the main detectors, probing the well-motivated LLP scenarios that are most difficult to probe using the existing LHC experiments alone. This incremental upgrade to the LHC complex therefore allows the discovery potential of the 14 TeV $pp$ collisions, which could already be producing as-yet undiscovered new particles, to be fully exploited.

\section{Basic Detector Principles}
\label{s.basicprinciples}

\begin{figure}
\begin{center}
\includegraphics[width=0.8\textwidth]{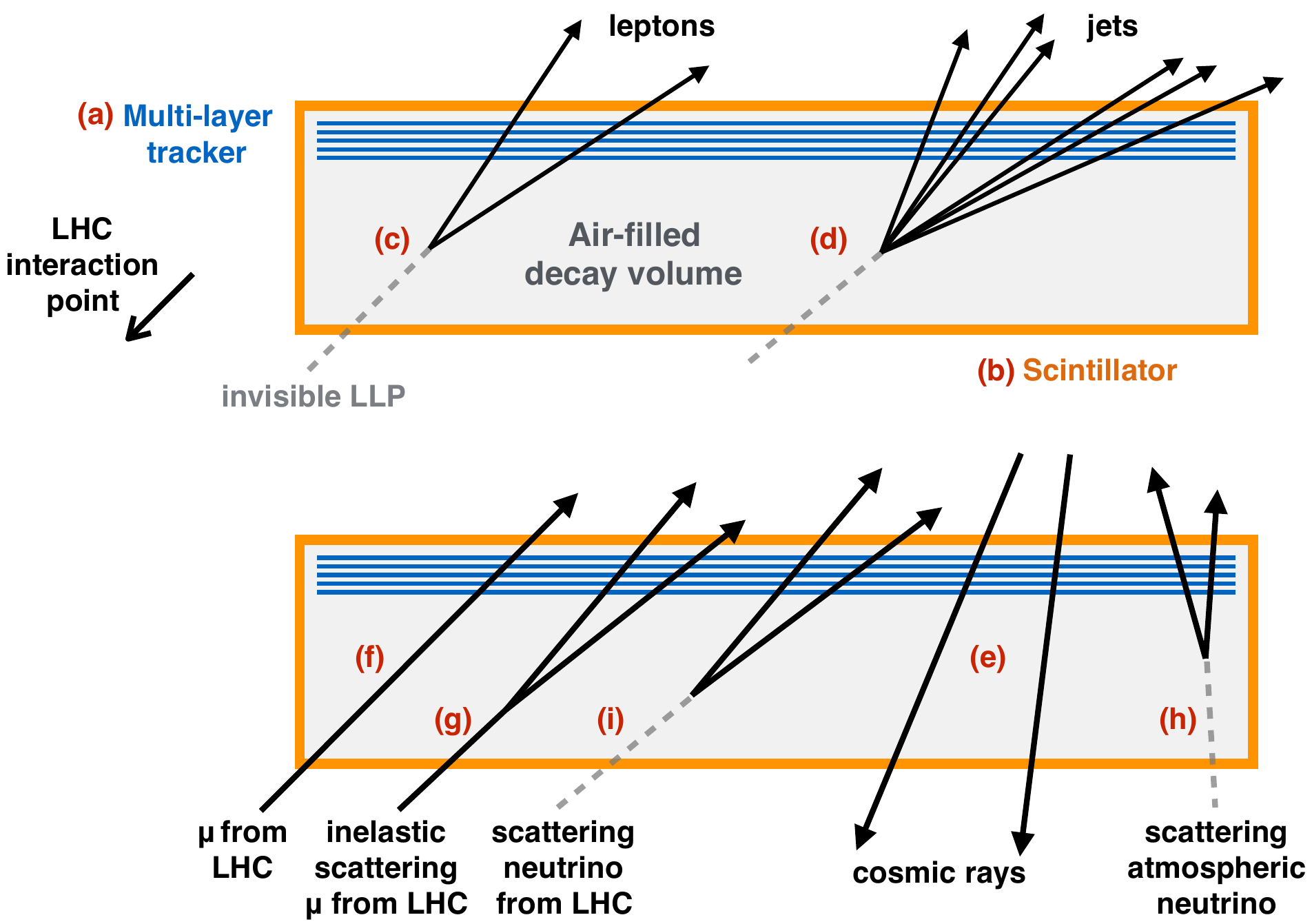}
\end{center}
\caption{
Schematic illustration of LLP decay signal (top) and main backgrounds (bottom) in a MATHUSLA-like detector consisting of a robust multi-layer tracker (a) above an air-filled decay volume. A veto of scintillator or tracker (b) may surround the decay volume to provide additional background rejection, but depending on the outcome of detailed background studies, this may not be required.
}
\label{f.bgsummary}
\end{figure}


The basic principle of the MATHUSLA detector concept is illustrated in Fig.~\ref{f.bgsummary}. Neutral LLPs produced at the LHC IP travel upwards and decay inside the air-filled\footnote{This is required to allow for the detection of LLPs that decay hadronically.} decay volume, giving rise to displaced vertices of upwards-traveling charged tracks that are reconstructed by a highly-robust multi-layer tracking system near the roof. 
The original simplified design proposed in~\cite{Chou:2016lxi} assumed 5 layers of Resistive Plate Chambers (RPCs), which are suitable for economical coverage of  very large areas with good position and timing resolution. This is also the basis for the more realistic preliminary detector design proposed in Section~\ref{s.design} of this letter, though other technologies are also being considered.

\begin{figure}
\begin{center}
\includegraphics[width= 1.0\textwidth]{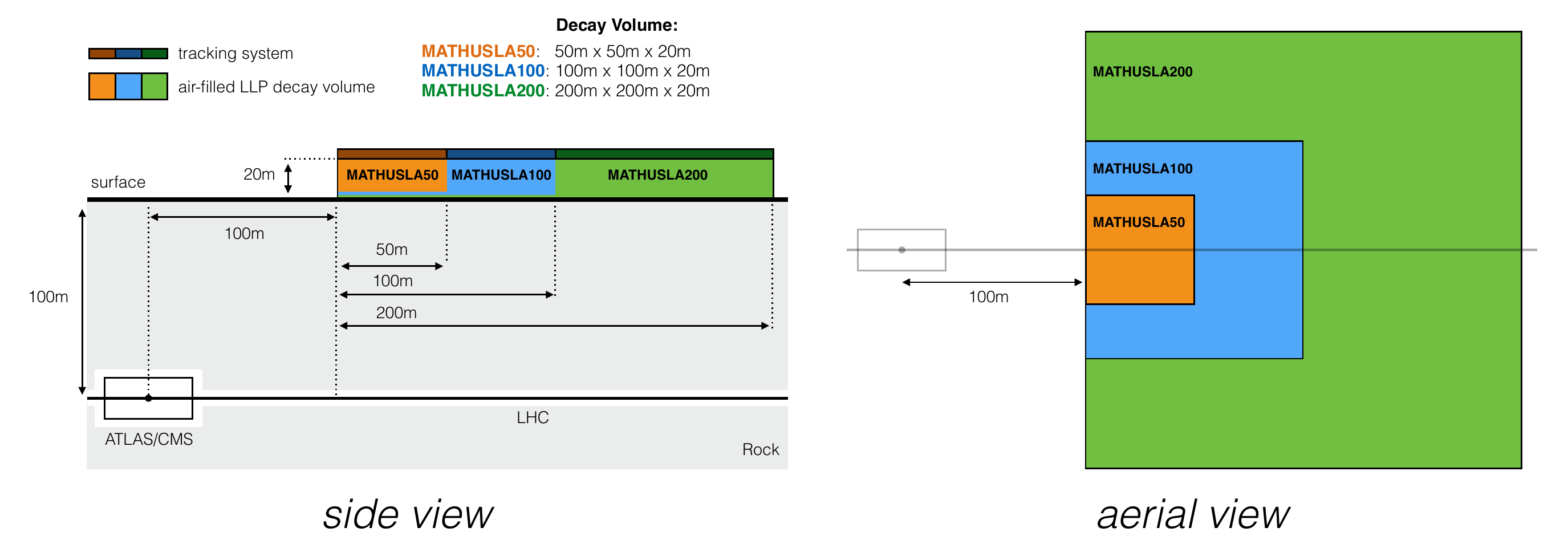}
\end{center}
\caption{
Possible simplified MATHUSLA geometries with different decay volumes. 
MATHUSLA100 is the default benchmark for this Letter of Intent, since it is approximately the right size for an available experimental site  near CMS.
MATHUSLA200 is the geometry proposed in \cite{Chou:2016lxi} and studied in~\cite{Curtin:2018mvb}. It can probe the Big Bang Nucleosynthesis (BBN) lifetime limit for LLPs produced in exotic Higgs decays, and could be implemented if sufficient area was made available.
The smaller MATHUSLA50 is considered alongside the other two geometries to demonstrate the scalability of the design.
}
\label{f.mathuslageometry}
\end{figure}


Fig.~\ref{f.mathuslageometry} shows the position of three simplified geometries for the MATHUSLA decay volume we will consider in this letter.
To a reasonable approximation, sensitivity to LLP production rate (long lifetime) scales inversely (linearly) with detector area, assuming fixed height. As long as the detector starts $\lesssim 100\mathrm{m}$ horizontally displaced from the IP on the surface, the sensitivity is not greatly dependent on the precise position of the detector.

\begin{figure}
\begin{center}
\includegraphics[width=0.7\textwidth]{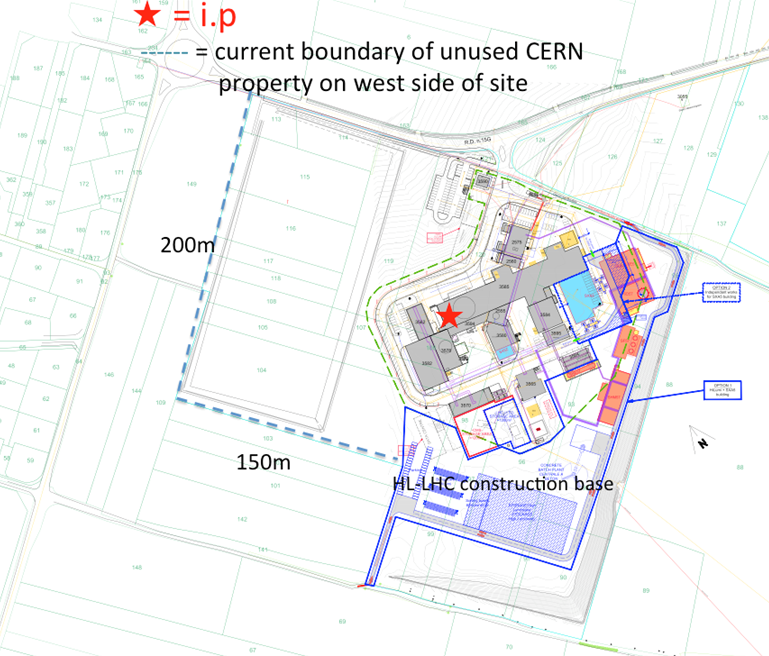}
\end{center}
\caption{Possible site for the MATHUSLA detector near CMS. The relevant area is to the west of the IP and enclosed by the blue dashed lines. The LHC beam runs roughly east-west. This site could accommodate a detector layout similar to our MATHUSLA100 benchmark.
}
\label{f.cmssite}
\end{figure}

MATHUSLA200 has a $200\mathrm{m} \times 200\mathrm{m} \times 20\mathrm{m}$ decay volume. This is the geometry proposed in~\cite{Chou:2016lxi} and studied in most previous works, including the physics case white paper~\cite{Curtin:2018mvb}. Significantly, this large size would allow MATHUSLA to probe LLP lifetimes close to the Big Bang Nucleosynthesis (BBN) limit $c \tau \lesssim 10^7 \mathrm{m}$ for LLPs produced in exotic Higgs decays (see Section~\ref{s.LLPphysics}). It could be implemented if sufficient land was made available, but the default benchmark we consider in this letter is MATHUSLA100, which is smaller by a factor of 4 with a $100\mathrm{m} \times 100\mathrm{m} \times 20\mathrm{m}$ decay volume. 
The reason for making this choice is that an available experimental site near CMS may be able to accommodate a detector layout similar to MATHUSLA100,  see Fig.~\ref{f.cmssite}. We will study explicitly site-specific detector geometries in the future.

Clearly, MATHUSLA is a very flexible detector concept that is inherently \emph{scalable} and can be adjusted to the desired sensitivity as well as available land area and budget. 
We further demonstrate this point by providing sensitivity estimates for a smaller MATHUSLA50 geometry with a $50\mathrm{m} \times 50\mathrm{m} \times 20\mathrm{m}$ decay volume in the sensitivity estimates of Section~\ref{s.LLPphysics}.
However, we strongly emphasize that given the relative simplicity of the detector, a size at least as large as MATHUSLA100 should be the realistic goal. It takes full advantage of readily available land while greatly enhancing the discovery potential of the LHC.

If the layers of the tracking system span a vertical distance of a few meters,  resolutions in the cm and ns range allow for full 4-dimensional displaced vertex reconstruction, requiring all associated tracks to intersect in both space and time. 
This is an extremely stringent signal requirement, especially for hadronic LLP decays with $\mathcal{O}(10)$ charged final states, which is crucial for rejecting backgrounds to effectively near-zero levels, most importantly the $\sim 10^{15}$ downwards-traveling charged particles from cosmic rays incident on the detector during the HL-LHC run. These and other backgrounds are discussed in more detail in Section~\ref{s.backgrounds}. 
It is important to note that despite this large integrated total particle count, the occupancy of the detector is very low by high energy physics experimental standards. This makes triggering and DAQ challenges manageable, as we discuss in Section~\ref{s.design}. 

As shown in Fig.~\ref{f.bgsummary} (b), the original proposal included a layer of scintillator (or tracker) that hermetically encloses the decay volume to provide additional vetoing and tracking information.
Since all charged tracks associated with a true LLP DV 
would have to satisfy such a fiducial veto, this significantly increases background rejection.
However, detector construction and operation would be greatly simplified if the decay volume did not have to be physically enclosed by any sensors or support structure. 
The latest results presented in Section~\ref{s.backgrounds} suggest that such a hermetic veto may not be required to reject muon backgrounds, but significant additional study is required to understand whether the same holds true for the cosmic ray-related backgrounds as well. 
At the present time, the minimal realistic detector design in Section~\ref{s.design} does not include the hermetic veto, though it can be extended to include this feature.


The scalable nature of the MATHUSLA concept also means the detector lends itself to a \emph{modular} implementation. Assembling the necessary  decay volume out of closely spaced submodules with an area in the $\mathcal{O}(10\mathrm{m} \times 10\mathrm{m})$ range has many practical advantages, allowing their production to be standardized and benefit from economies of scale, while easily adjusting their deployment to the particular characteristics of the available experimental site. A modular design also allows for incremental ramp-up of detector operations,  since even a small fraction of the final experiment allows for tunning trigger strategies and providing early results. Finally, individual modules can be upgraded much more easily and cheaply than the whole detector, which realistically allows for additional capabilities to be added, including higher spatial tracking resolution for low-mass LLP searches, or possibly additional material layers for limited particle ID and photon detection~\cite{Curtin:2017izq}, see Section~\ref{s.LLPphysics}. The more realistic design presented in Section~\ref{s.design} is based on such a modular construction.
%
%


Finally, we point out that both the simplified geometries presented in this section and the more realistic design presented in Section~\ref{s.design} have to be regarded as  preliminary. There are many aspects of the detector that are subject to detailed optimization and engineering studies, including the final shape of the decay volume footprint (e.g. a pie shape may be more optimal, depending on the experimental site shape and position), as well as its height. It is also likely that the precise coverage of the tracker can be optimized. For example, the results of Section~\ref{s.efficiencystudies} demonstrate that instrumenting the vertical decay volume boundary facing away from the IP with a ``wall'' of tracking layers could significantly enhance DV reconstruction efficiency for higher-mass leptonic LLP decays with a very modest increase in sensor cost. Even so, further work is needed to determine whether this is justified, given that these LLP scenarios will be well-covered by ATLAS/CMS LLP searches.

\section{Primary Physics Goal: LLP Searches}
\label{s.LLPphysics}

The sensitivity of MATHUSLA to LLP signals was first discussed in the original proposal~\cite{Chou:2016lxi} for the well-motivated benchmark model of LLP production in exotic Higgs decays. The comprehensive physics case white paper~\cite{Curtin:2018mvb} as well as several individual follow-up studies~\cite{Co:2016fln, Dev:2016vle, Dev:2017dui, Caputo:2017pit, Curtin:2017izq, Evans:2017kti, Helo:2018qej, DAgnolo:2018wcn, Berlin:2018pwi, Deppisch:2018eth, Jana:2018rdf} discuss this in great detail, but the most important lessons can be stated very succinctly. In this section we repeat some of the model-independent arguments of~\cite{Curtin:2018mvb} regarding MATHUSLA's LLP physics reach, slightly adapting the discussion to the different benchmark geometries of Fig.~\ref{f.mathuslageometry} that we study in this letter.

For simplicity, and also to make these results as independent of the particular detector technology as possible, we assume perfect LLP detection efficiency and rejection of all backgrounds within the MATHUSLA decay volume. 
The sensitivity projections in this section should therefore be regarded as an ideal that can be approached with a detector of sufficient design. 
That being said, even the relatively modest design of Section~\ref{s.design} closely approaches these sensitivities for hadronic LLP decays to many charged particles, since the design driver of the tracking system is cosmic ray rejection by directionality, which in turn leads to excellent reconstruction efficiency for high-multiplicity final states, see Section~\ref{s.efficiencystudies}.

\subsection{LLP Sensitivity Benchmarks}


We now examine MATHUSLA's LLP sensitivity for a few representative benchmark models that serve to illustrate the broad range of both TeV- or weak-scale and low-scale new physics scenarios such a detector could probe. See~\cite{Curtin:2018mvb} for a much more detailed discussion.

Exotic Higgs decays~\cite{Curtin:2013fra} are a highly motivated possibility for BSM particle production with great discovery potential, owing to the suppressed Higgs width in the SM, large production cross section at the LHC and lack of symmetry protection for the $|H|^2$ operator, making Higgs portal couplings to any new physics very generic. 
Exotic Higgs decays to LLPs could easily have branching ratios up to $\sim \mathcal{O}(0.1)$ and not be in conflict with current Higgs coupling measurements \cite{Khachatryan:2016vau}. The LLP lifetime and decay modes can be regarded as free parameters, but even from this model-independent perspective, the Higgs portal coupling makes it very plausible that the LLP decays to Yukawa-ordered and hence dominantly hadronic SM final states (see e.g. mirror glueballs in Neutral Naturalness \cite{Craig:2015pha, Curtin:2015fna}). 
Ref.~\cite{Coccaro:2016lnz} studied the reach of the HL-LHC main detectors to such an LLP scenario in the long lifetime regime and obtained sensitivity projections for a single-DV search in the ATLAS Muon System. 
Since LLPs that decay to jets in the tracker and are produced in exotic Higgs decays are too light to efficiently pass Level 1 triggers 
without e.g. being limited to associated Higgs production, the ATLAS Muon System is uniquely well suited for an inclusive main detector search. It is shielded from QCD activity by the calorimeters, and can trigger at Level 1 on a single LLP decay.  However, the search is still limited by $\mathcal{O}(100\mathrm{fb})$ of backgrounds at the HL-LHC.

\begin{figure}
\begin{center}
\includegraphics[width=0.8\textwidth]{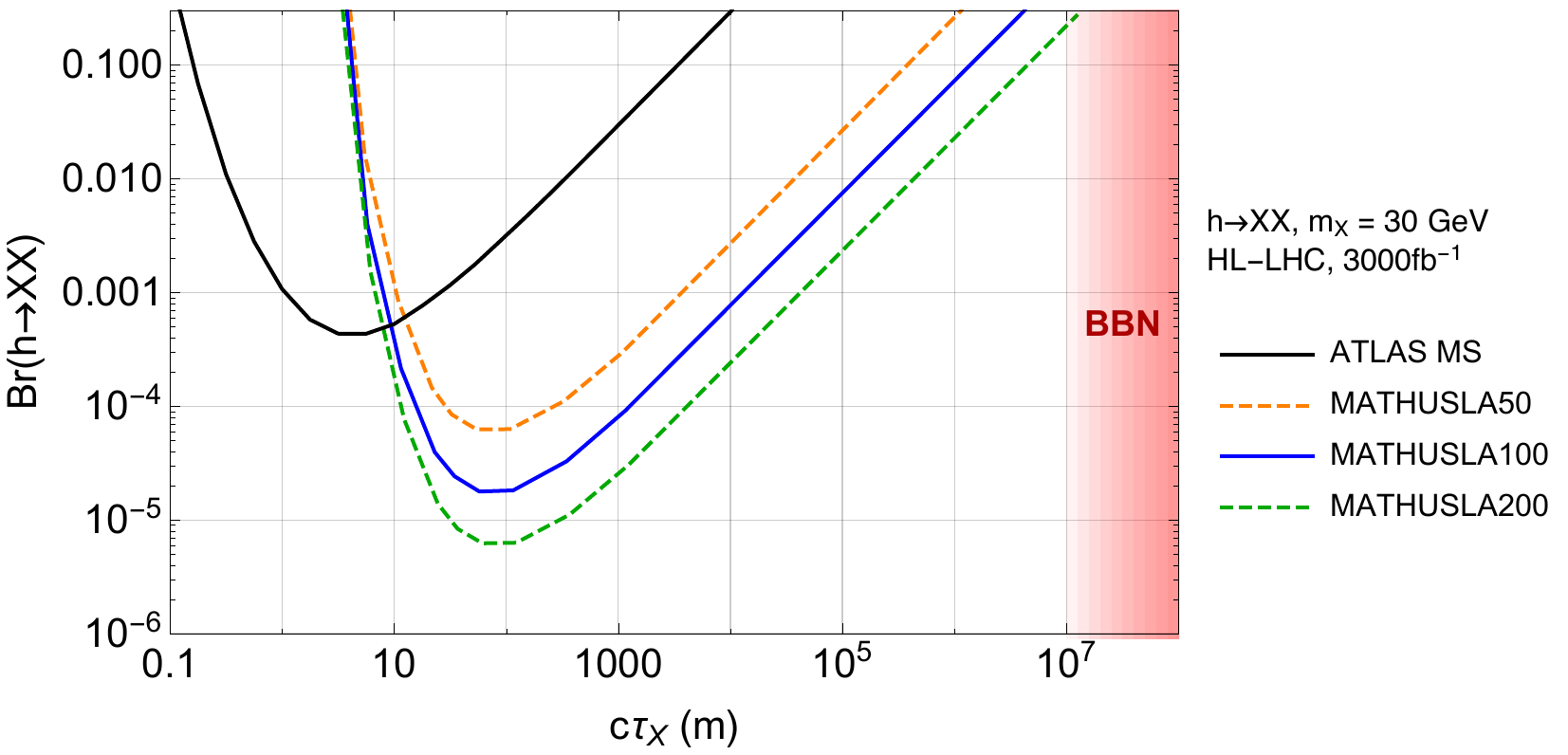}
\end{center}
\caption{
Comparison of reach for 30 GeV LLPs produced in exotic Higgs decays and decaying hadronically, for various MATHUSLA geometries (curves correspond to 4 LLPs decaying in the detector volume) and the LHC main detector exclusion projection using a single-DV search in the ATLAS Muon System~\cite{Coccaro:2016lnz}. 
}
\label{f.MATHUSLAhiggssensitivity}
\end{figure}

By comparison, MATHUSLA is able to search for these LLPs without background (or trigger) limitations. 
We compute the number of LLPs decaying in MATHUSLA by convolving the geometries from Fig.~\ref{f.mathuslageometry} with the kinematic distribution of LLPs produced in exotic Higgs decays following~\cite{Chou:2016lxi}. Sensitivity projections assuming 4 events decaying in MATHUSLA are shown in Fig.~\ref{f.MATHUSLAhiggssensitivity}.
MATHUSLA200 has \emph{three orders of magnitude} better sensitivity to LLP production cross section or long lifetime than ATLAS. 
MATHUSLA100 and MATHUSLA50, despite being $1/4$ and $1/16$ the size of MATHUSLA200, have about $0.4$ and $0.1$ times the sensitivity of MATHUSLA200, due to the slightly greater importance of the common decay volume closest to the IP.
This is perhaps the simplest demonstration of the enormous physics potential MATHUSLA contributes to the LHC.

MATHUSLA100 and in particular MATHUSLA200 are able to probe LLP lifetimes approaching the $c\tau \lesssim 10^7 \mathrm{m}$ upper limit from BBN~\cite{Fradette:2017sdd}. 
Apart from being the ceiling of the majority of LLP parameter space, this reach is significant in another way. If the HL-LHC detects a nonzero invisible Higgs boson decay at the 10\% level, then regardless of detection at MATHUSLA, its LLP searches will help resolve the nature of the newly discovered signal. If MATHUSLA sees no LLP decays, then due to the near-universal nature of the BBN bound, it can be regarded as highly likely that the Higgs decays to a stable invisible particle, hence confirming the production of a dark matter candidate at the HL-LHC. On the other hand, if the invisible Higgs boson decay produces an unstable constituent of a hidden sector, MATHUSLA is very likely to observe its decay, allowing data from both detectors to diagnose the newly discovered physics. 

\begin{figure}
\begin{center}
\hspace*{-16mm}
\begin{tabular}{m{12cm}m{5cm}}
\includegraphics[width=12cm]{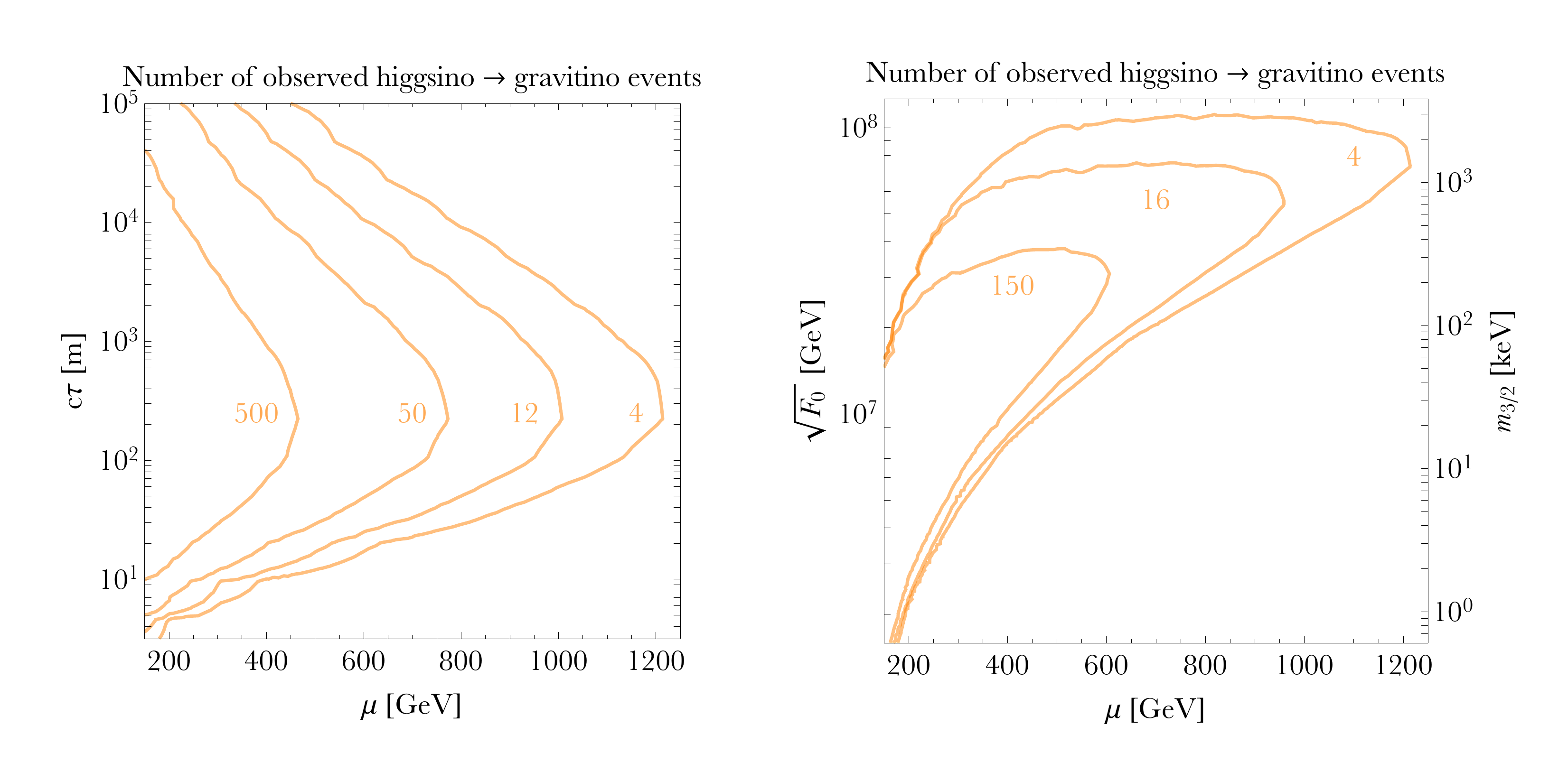}
&
\vspace{3mm}
\includegraphics[width=5cm]{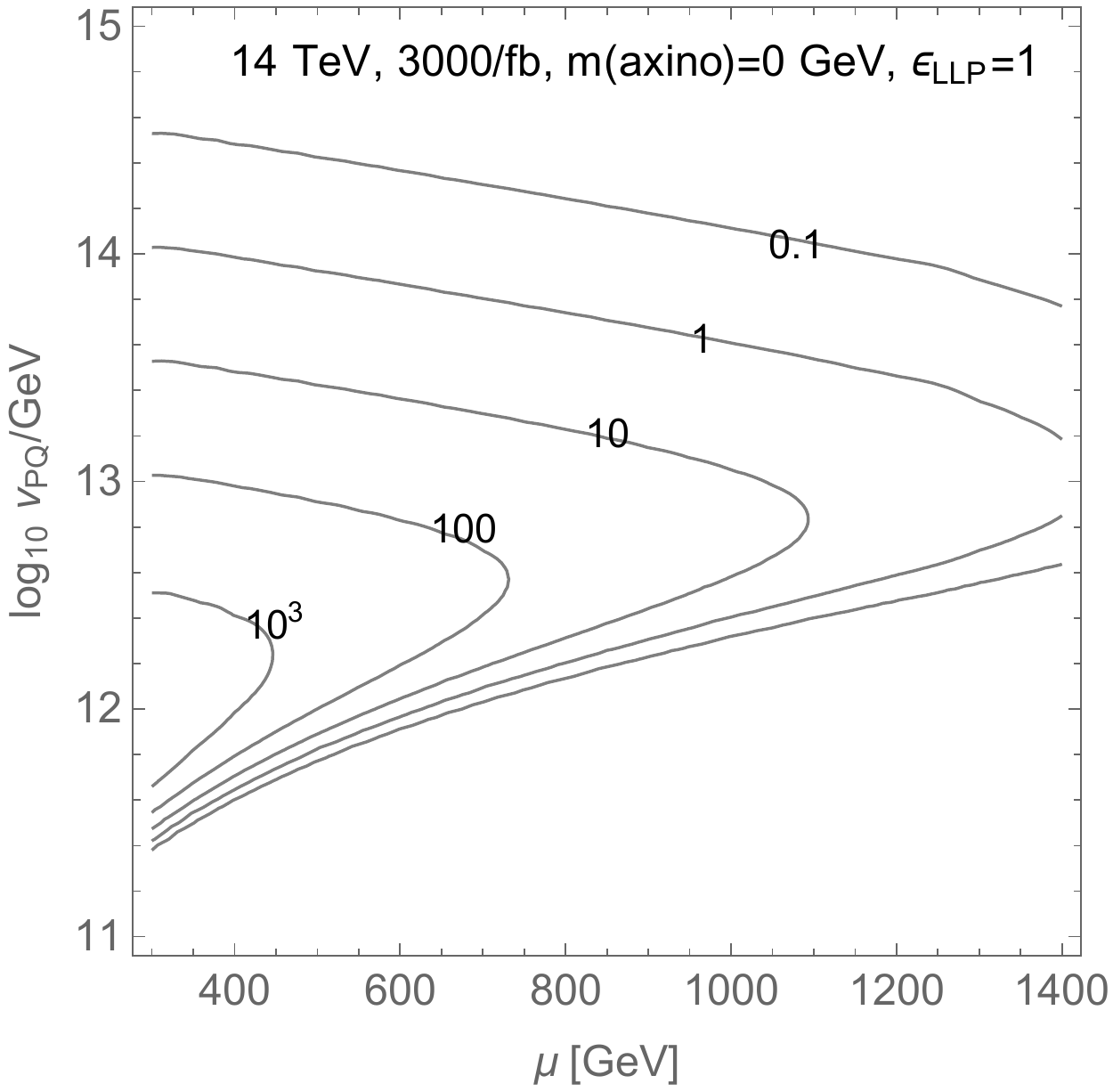}
\end{tabular}
\end{center}
\caption{
MATHUSLA200 reach for Higgsino LLPs. (The MATHUSLA100 and MATHUSLA50 reach can be obtained by rescaling the shown event yield by about $1/3$ and $1/10$ respectively.) 
\emph{Left:} contours show the number of Higgsino decays in MATHUSLA as a function of Higgsino mass $\mu$ and decay length $c\tau$ at the HL-LHC. 
\emph{Middle:} The same event yield as a function of the gauge-mediated SUSY breaking scale $F_0$, if the Higgsino is long-lived due to a suppressed decay to a light gravitino and $Z/h$. 
\emph{Right:} Event yield as a function of the Peccei-Quinn symmetry breaking scale $v_{PQ}$ in supersymmetric axion models where the Higgsino is long-lived due to a suppressed decay to an axino and $Z/h$.
Plots reproduced from~\cite{Curtin:2018mvb}.
}
\label{f.MATHUSLAHiggsinosensitivity}
\end{figure}

\begin{figure}
\begin{center}
\includegraphics[height=10cm]{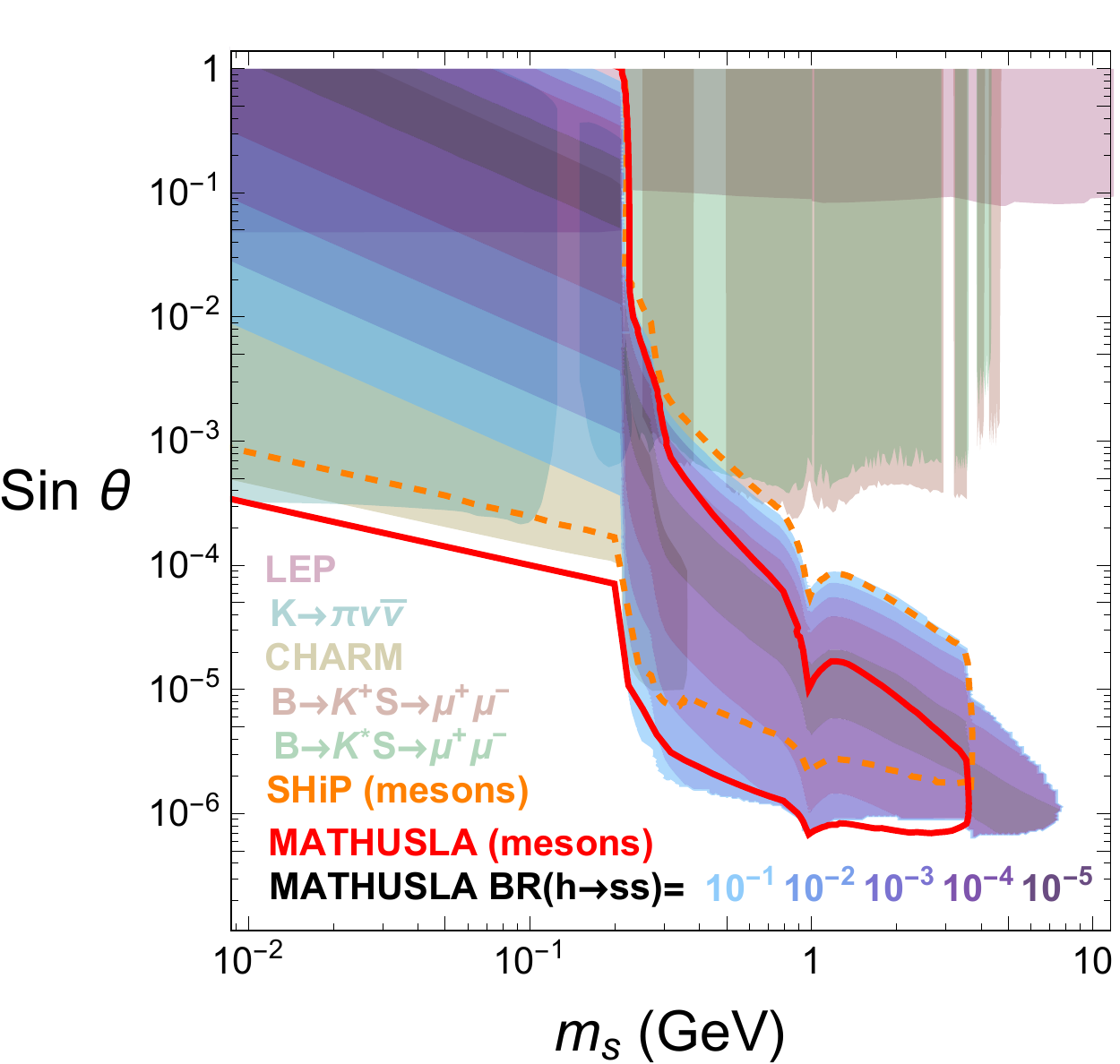}
\end{center}
\caption{
Projected MATHUSLA200 sensitivity to scalar LLPs in the minimal SM+S extension at the HL-LHC assuming 4 LLP decays in the detector volume, as a function of scalar mass $m_S$ and mixing angle $\sin \theta$ with the 125 GeV Higgs boson.
The red contour is the sensitivity to $B$-meson decays. The different blue-purple contours illustrate the minimum BR$(h\to ss)$ value to which MATHUSLA200 would be sensitive.
The projected constraint contour for the SHiP experiment \cite{Lanfranchi:2243034,Anelli:2015pba,Alekhin:2015byh} is shown by the dashed orange contour. 
Plot from~\cite{Evans:2017lvd}, see original reference or \cite{Curtin:2018mvb} for additional details.
}
\label{f.MATHUSLASMSsensitivity}
\end{figure}


Another important benchmark  are Higgsinos in supersymmetry, which can be long-lived in a variety of scenarios, including gauge-mediation~\cite{Giudice:1998bp}, R-parity violation~\cite{PhysRevD.40.2987, Dreiner:1997uz, Barbier:2004ez} or a supersymmetric axion models~\cite{Kim:1983dt,Chun:1991xm}.
MATHUSLA's reach is shown in Fig.~\ref{f.MATHUSLAHiggsinosensitivity}, extending to TeV-scale masses depending on the lifetime (or, equivalently, SUSY-breaking scale in gauge mediation or the scale of Peccei-Quinn (PQ) symmetry~\cite{Peccei:1977ur, Wilczek:1977pj, Weinberg:1977ma} breaking).

Finally, many low-mass LLPs, including sterile neutrinos~\cite{Minkowski:1977sc,Mohapatra:1979ia,Yanagida:1979as,GellMann:1980vs,Glashow:1979nm,Batell:2016zod,Pati:1974yy,Mohapatra:1974gc,Helo:2013esa,Chun:2017spz,Asaka:2005an,Asaka:2005pn,Cai:2017mow} and dark scalars~\cite{Evans:2017lvd,Krnjaic:2015mbs, Bezrukov:2009yw, Clarke:2013aya} can be produced in exotic $B$-meson decays. 
To illustrate reach for this LLP production mode, we show in Fig.~\ref{f.MATHUSLASMSsensitivity} the sensitivity of MATHUSLA200 to scalar LLPs produced either in exotic $B$-meson decays or exotic Higgs decays within the minimal SM+S simplified model, which features a single real scalar mixing with the Higgs, owing both its production and decay to this Higgs portal coupling~\cite{Evans:2017lvd}.
MATHUSLA200 probes longer lifetimes and hence smaller mixing angles than the SHiP experiment~\cite{Lanfranchi:2243034,Anelli:2015pba,Alekhin:2015byh}, allowing both experiments to complement each other significantly for this particular benchmark model. 
While MATHUSLA100 will have slightly less reach than MATHUSLA200, broadly the same conclusions apply for this geometry as well.

\subsection{Model-independent Reach and Comparison to Other Experiments}


\begin{figure}
\begin{center}
\includegraphics[width=0.8\textwidth]{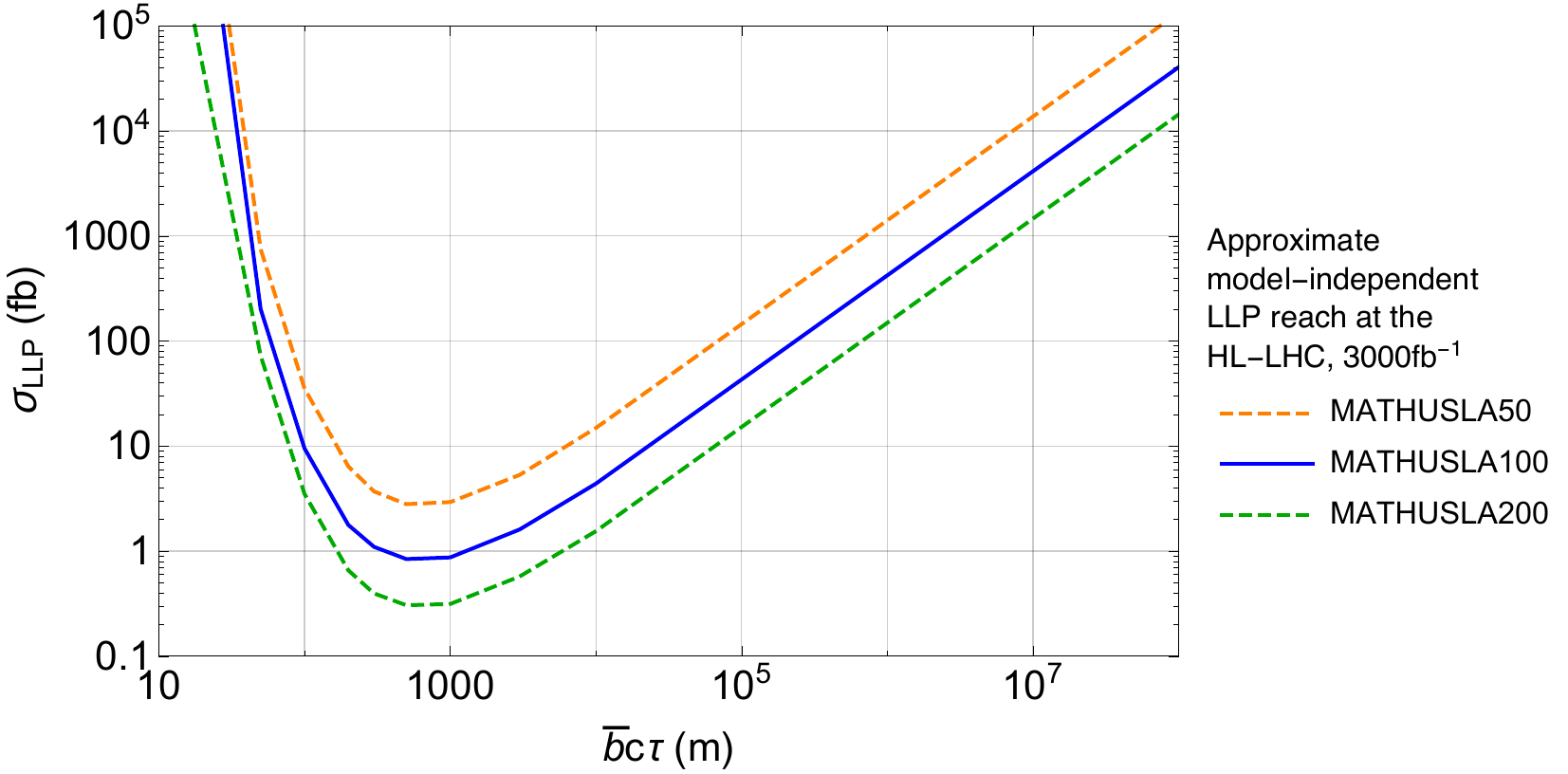}
\end{center}
\caption{
Approximate model-independent LLP sensitivity of different MATHUSLA detector geometries, assuming $\mathcal{O}(1)$ produced LLPs per production event at the HL-LHC. $\bar b$ is the mean boost of the produced LLPs traveling through MATHUSLA. 
The shape of the exclusion/discovery region at short lifetimes depends on the detailed boost distribution, but for long lifetimes $\bar b c \tau \gg 200 m$ depends only on the mean boost and is quite insensitive to the production mode, up to an $\mathcal{O}(1)$ factor. 
Note that LLPs near the BBN lifetime limit of $c \tau \sim 10^7$m can be probed if they are produced with cross-sections in the pb range at the HL-LHC.
}
\label{f.modelindependentreach}
\end{figure}

\begin{figure}
\begin{center}
\includegraphics[width=13cm]{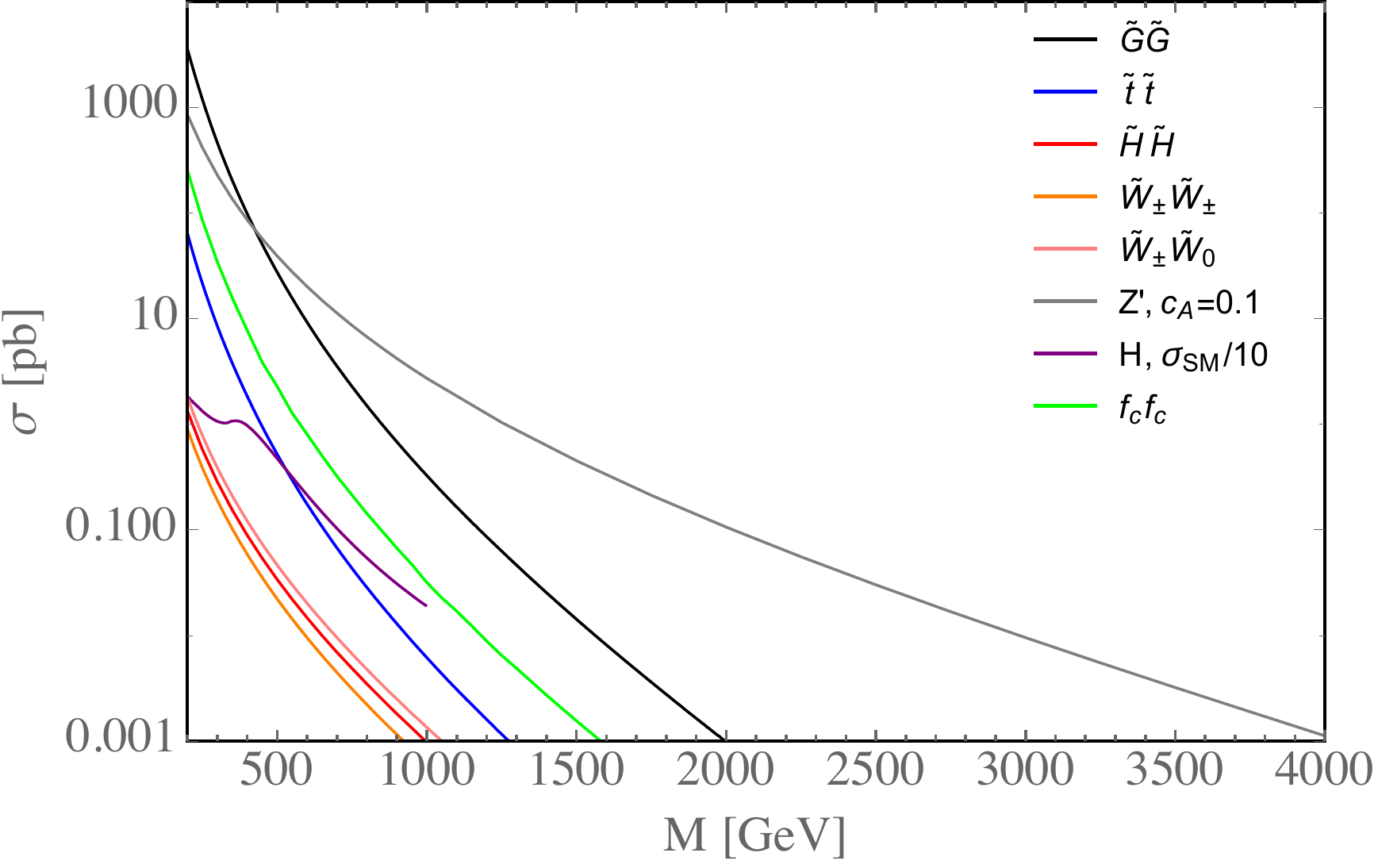}
\end{center}
\caption{
Benchmark LLP production cross-sections at the 14 TeV HL-LHC, as a function of either parent particle or LLP mass. Plot from~\cite{Curtin:2018mvb}.
}
\label{f.benchmarkxsecs}
\end{figure}

The background-free nature of LLP searches in MATHUSLA, and the purely geometrical nature of the signal, make it very straightforward to understand the sensitivity of different MATHUSLA geometries to LLP production rate in a relatively model-independent fashion~\cite{Curtin:2018mvb}.
To $\mathcal{O}(1)$ precision, the reach of MATHUSLA in the long-lifetime regime depends only on the production rate and average boost of the LLPs. We can therefore obtain an approximately model-independent sensitivity estimate by showing the exotic Higgs decay bounds from Fig.~\ref{f.MATHUSLAhiggssensitivity} as a function of production rate and $\overline b c \tau$ in Fig.~\ref{f.modelindependentreach}.\footnote{For example, in these units, the bounds for different LLP masses coincide very closely.}
For lifetimes in the few hundred meter range, MATHUSLA can probe 0.1 - 1 fb production cross sections at the LHC. For 10 pb cross sections, it is sensitive to BBN lifetimes. 

This LLP cross section reach can be translated to mass reach for either the LLP or the parent particle producing the LLP in its decay~\cite{Curtin:2018mvb}. The benchmark cross sections shown in Fig.~\ref{f.benchmarkxsecs} for a variety of possible LLP production modes demonstrate that MATHUSLA probes BSM physics well into the TeV scale. 

For low-scale models, it is of particular interest to understand the minimum LLP mass that can be probed at MATHUSLA. The opening angle $\theta$ of LLP decay products scales with the LLP boost $\theta \sim 1/b$.
Crucial to background rejection is the reconstruction of each LLP decay as a well-resolved multi-pronged DV.
The limiting factor is the spatial resolution $\Delta x$ of the tracking system. The requirement of resolving LLP decay products translates to an approximate maximum LLP boost:
\begin{equation}
\label{e.maxboost}
b_\mathrm{LLP}^\mathrm{max} \sim 1000 \ \left(\frac{1 \mathrm{cm}}{\Delta x}\right) \ .
\end{equation}
LLPs produced in the decay of a significantly heavier parent particle with mass $m_\mathrm{parent}$ and boost $b_\mathrm{parent}$ have boost
\begin{equation}
b_\mathrm{LLP} \sim \frac{b_\mathrm{parent}  m_\mathrm{parent}}{2m_\mathrm{LLP}} \ .
\end{equation}
Eqn.~(\ref{e.maxboost}) can therefore be translated into a \emph{minimum LLP mass} that MATHUSLA can probe in the low-background regime: 
\begin{equation}
\label{e.minmass}
m_{LLP}^{min} \sim \frac{b_\mathrm{parent} m_{parent}}{2 b_{LLP}^{max}} \sim \left( \frac{b_\mathrm{parent}  m_{parent}}{2000} \right) \left( \frac{\Delta x}{1 \mathrm{cm}} \right)
\end{equation}
Assuming a 1~cm spatial resolution, this corresponds to a minimum LLP mass of $\sim 0.1 - 1 \gev$ for LLPs produced in the decay of a 0.1 - 1 TeV parent. For light LLPs produced in $B$-hadron decays, the minimum mass is about 10 MeV.
MATHUSLA could therefore probe relatively low LLP masses very effectively.


How does the reach of MATHUSLA compare to the reach of main detector LLP searches in general? This is a very difficult question to answer in detail, due to the relatively nascent state of the LHC LLP search program and uncertainties about main detector backgrounds and HL-LHC running conditions. However, some general arguments guided by existing LLP searches were put forth in~\cite{Curtin:2018mvb}. These can provide some qualitative insight. 

Broadly speaking, if a given search for a single LLP decay in the main detector is background free, \emph{and} has no limitations from the trigger or requirements on the final state that significantly reduce sensitivity (e.g. relying on a sub-dominant production or decay mode to have a lepton in the event, or a high-$p_T$ jet in a lower-scale LLP production process), then the ATLAS/CMS search will have similar sensitivity to the corresponding search at MATHUSLA200. For all other scenarios, MATHUSLA200 has better LLP cross section sensitivity. In particular, for LLPs giving rise to hadronic final states with less than a few 100 GeV of energy, MATHUSLA200 can be up to three orders of magnitude more sensitive (see Fig.~\ref{f.MATHUSLAhiggssensitivity}). 
For LLPs with masses below 10 GeV decaying to displaced lepton jets, MATHUSLA200 can be 1-2 orders of magnitude more sensitive. 
For MATHUSLA100, the same arguments apply, with the sensitivity gain reduced by a factor of $\sim 3$. 
Clearly, MATHUSLA greatly extends the physics reach of the HL-LHC main detectors in a very general way for many different kinds of LLP signals.


LLPs with very long decay lengths can also be searched for with missing energy searches at ATLAS or CMS. 
However, unless the amount of missing energy in the event is very large (order TeV) and/or the production rate is very sizable, the presence of backgrounds in MET searches limit their sensitivity to LLPs.
The reach of missing energy searches was examined in~\cite{Curtin:2018mvb} by considering DM simplified models where the invisible final-state particles are substituted with LLPs. 
In all cases, MATHUSLA greatly extended mass reach compared to the missing energy search alone, depending on the LLP lifetime.


Finally, we can ask how MATHUSLA's capabilities compare to other dedicated LLP experiments. 
The SHiP experiment~\cite{Lanfranchi:2243034,Anelli:2015pba,Alekhin:2015byh} has been proposed to look for hidden sectors and LLP signals at the few-GeV scale or below. 
Fig.~\ref{f.MATHUSLASMSsensitivity} demonstrates that MATHUSLA can be very competitive to SHiP's reach. This can be understood due to the large $B$-hadron production rate at the HL-LHC and MATHUSLA's relatively good geometrical acceptance. Ref.~\cite{Evans:2017lvd} showed that for LLPs with decay lengths $b c \tau \gg 100\mathrm{m}$ produced in $B$-hadron decays, MATHUSLA200 sees about two orders of magnitude more events than SHiP. Together, the two experiments cover shorter and longer lifetimes in a highly complementary fashion. 

In the last year, two promising new ideas for external LLP detectors at the LHC have appeared in the literature. 
CODEX-b~\cite{Gligorov:2017nwh} was proposed to take advantage of a $\sim (10 \mathrm{m})^3$ sized free space that will become available near the LHCb experiment. Similar to MATHUSLA in its basic principles, its smaller size and lower luminosity means it has a few orders of magnitude less reach to LLP production rates than MATHUSLA. However, its smaller size may also allow it to be equipped with more sophisticated instrumentation than MATHUSLA's simple tracker, while its close proximity to LHCb would allow it to act as a fully integrated sub-detector. This adds significant new capability to LHCb, and its reach for low-mass or shorter-lifetime LLPs is likely to be complementary to MATHUSLA. 
FASER~\cite{Feng:2017uoz, Feng:2017vli} is a very small proposed detector, cylindrical in shape with a radius of ten centimeters to one meter, and a length of several to ten meters. It would be placed a few hundred meters down the beamline in the LHC tunnel in an available access tunnel with minimal required excavation. Its downstream position on the beam axis of the collision point allows it to take advantage of large forward enhancements for light LLPs, such as dark photons or dark scalars. It would significantly complement MATHUSLA's sensitivity to these low-scale LLP scenarios, probing shorter lifetimes beyond the reach of a surface detector. 
Obviously, the best scenario for taking advantage of the existing LHC collisions would be the construction of FASER, CODEX-b as well as MATHUSLA. 

\subsection{Characterizing the LLP signal}
\label{s.LLPdiagnosis}

\begin{figure}
\begin{center}
\begin{tabular}{cc}
\includegraphics[height=4.5cm]{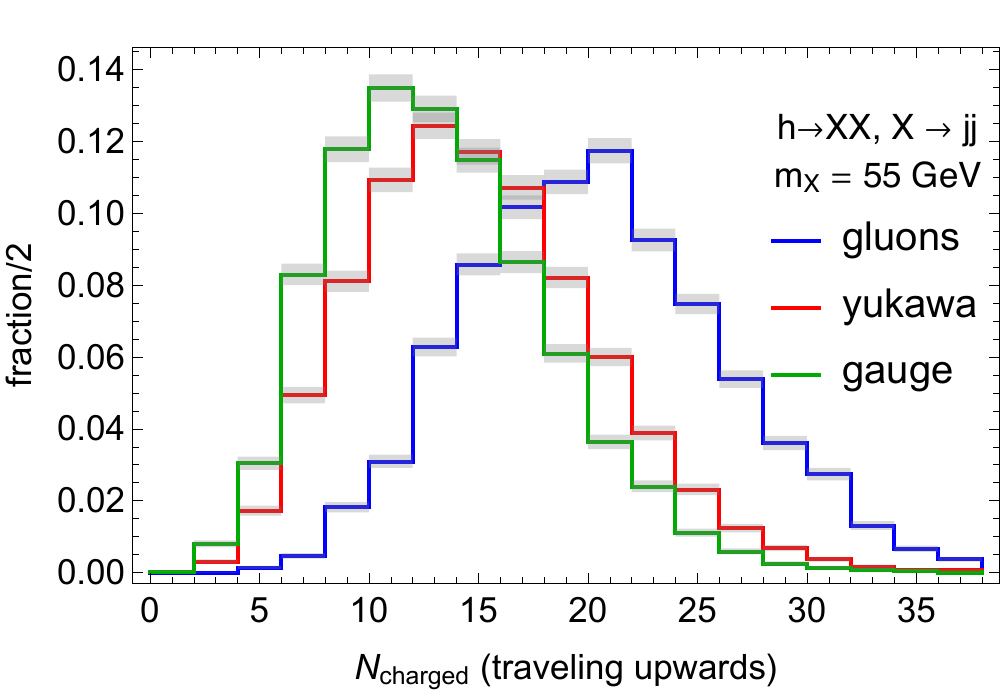}
&
\includegraphics[height=4.5cm]{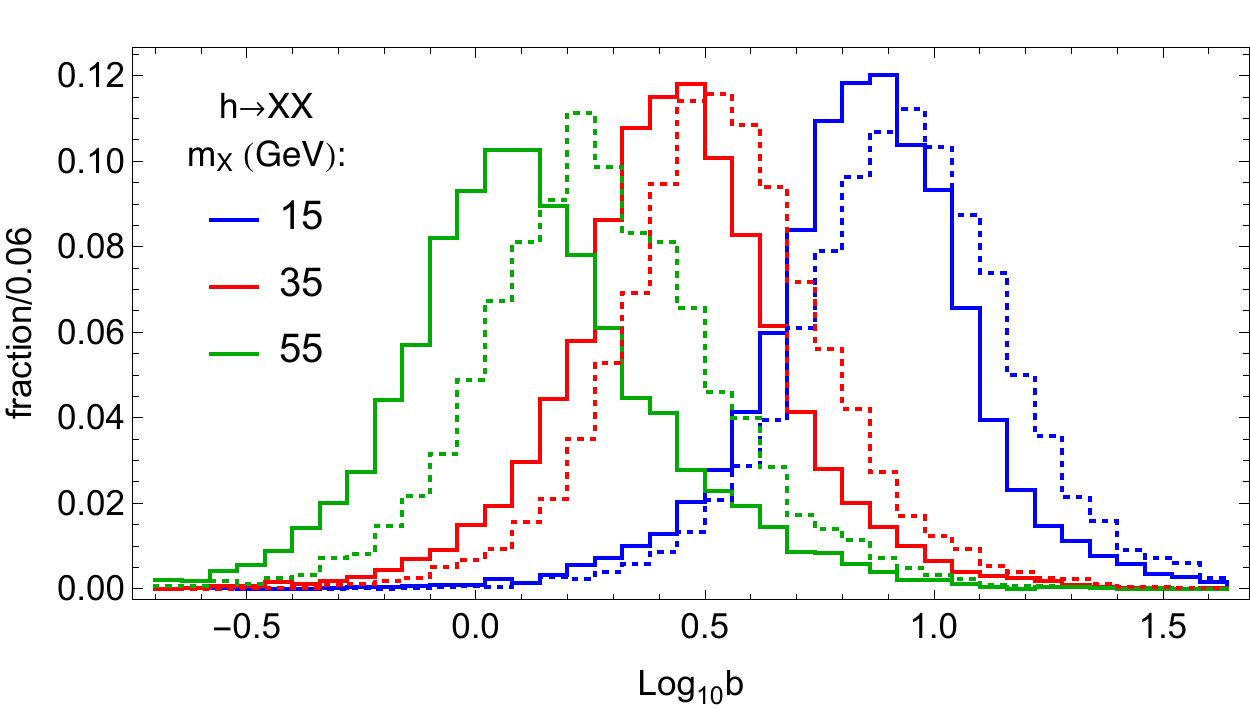}
\end{tabular}
\end{center}
\caption{
Left: multiplicity of charged final states traveling upwards for a 55 GeV LLP decaying to different flavors of jets. Right: for LLPs with 15, 35 and 55 GeV mass produced in exotic Higgs decays, the true boost distribution in MATHUSLA200 (solid) is compared to the reconstructed boost distribution using vertex-track opening angles (dotted - as described in text). 
Figures from~\cite{Curtin:2017izq}.
}
\label{f.LLPdiagnosis}
\end{figure}

%

The benchmark version of MATHUSLA consists of only a tracker above the LLP decay volume. Despite this minimal instrumentation, MATHUSLA has surprising capabilities to determine the properties of any detected LLP signal. This was first systematically explored in~\cite{Curtin:2017izq}, and we summarize the main conclusions here. 


The most obvious observable to diagnose the properties of an observed LLP decay is final state multiplicity. A decay to two well-separated charged tracks strongly suggests a decay to two leptons. A few tightly collimated tracks in each leg of the DV may indicate tau leptons, while decay to jets leads to large $\mathcal{O}(10)$ multiplicities. 
Detailed analysis may even reveal the \emph{flavor} of the final state jet. Fig.~\ref{f.LLPdiagnosis} (left) shows the distribution of final state multiplicities for a 55 GeV LLP that decays either into gluons, into jets with Yukawa-ordered branching ratios (i.e. $b$-dominated), and into jets with gauge-ordered branching ratios (i.e. light-flavor dominated). With $\sim 10-100$ observed LLP decays, these possibilities can be effectively distinguished.


Additional information can be discerned from the geometry of the displaced vertex. For decays to two charged final states, there is a direct relationship between the boost of the LLP and the opening angle, which can be used to determine the LLP boost with high fidelity, event-by-event, as long as the LLP mass is much greater than the final state masses. 
For hadronic decay, a slightly more sophisticated technique produces almost equivalently good results. Since MATHUSLA lacks energy measurement, its signal will be dominated by large multiplicities of soft pions produced with near-thermal energy spectra. Assuming the final states to be massless, one can Lorentz-boost the displaced vertex backwards towards the IP until the final state distribution is approximately spherically symmetric, and obtain an estimate of the LLP boost for this specific event. 
As Fig.~\ref{f.LLPdiagnosis} (right) demonstrates, this simple method agrees surprisingly well with the true boost.


Fig.~\ref{f.LLPdiagnosis} (right) also makes clear that for a given LLP production mode, the boost distribution can reveal the LLP mass. In fact, for the studied exotic Higgs decay scenario, only about 20 observed LLP decays, either leptonic or hadronic, are required to measure the mass with 10\% precision.
This suggests that once LLP decays are observed, the mass measurement could be obtained for a variety of different hypothetical production modes. 
The event-by-event boost measurement allows for the identification of the individual HL-LHC bunch crossing where the LLP was created. Off-line correlation of MATHUSLA and main detector data could then narrow down the choices of LLP production modes, or even determine it completely with enough data. 

Further study is needed to fully understand the diagnostic capabilities of MATHUSLA. For example, it would be important to understand whether an invisible component in the LLP decay can be reliably detected. At any rate, the tracks from LLP decays clearly contain a significant amount of information, allowing MATHUSLA to not only detect but diagnose many new physics scenarios.

\subsection{Possible Extensions}


One of the most important features of the MATHUSLA proposal is its entirely parasitic nature. The detector operates independently of the LHC and the main detectors, and neither its construction nor operation interferes with running the accelerator or other CERN experiments. This also ensures that MATHUSLA remains operational and useful if the HL-LHC is succeeded by an HE-LHC. 

It is, however, possible to imagine linking MATHUSLA more closely with its neighboring main detector. Triggering on a possible LLP decay could be done very rapidly at MATHUSLA, relying on only an upwards going charged track. This may allow MATHUSLA to contribute a trigger signal to the main detector. It is even imaginable to use MATHUSLA as a Level 1 trigger, given that the light-speed travel time between the main detector and MATHUSLA is $\sim 0.5 - 1.2$$\mu$s, but the viability of this idea has to be studied further, it may add significant cost to the electronics of the MATHUSLA detector itself, and will require the approval of the LHC main detector collaboration.
It might therefore be considered as an upgrade once an LLP signal is detected, since triggering with MATHUSLA would greatly enhance the main detector's ability to help diagnose the LLP production mode as discussed in Section~\ref{s.LLPdiagnosis}.


Finally, we point out that various extensions of the basic MATHUSLA design outlined in this section could be considered. For example, adding a few cm of material like iron between the tracking layers~\cite{Curtin:2017izq} could allow for photons to be detected, as well as aid in electron/muon discrimination, which could enhance both LLP diagnostic capabilities and enhance the cosmic ray physics program discussed in Section~\ref{s.cosmicray}. However, it is not yet clear whether the added cost, weight and complication makes this idea viable.  
One could also equip one or several of MATHUSLA's sub-modules with significantly higher spatial tracking resolution to look for LLPs with masses below $10-100 \mev$, or with the above-mentioned layer of material.

\section{Secondary Physics Goal: MATHUSLA as a Cosmic Ray Telescope}
\label{s.cosmicray}

The design of MATHUSLA is driven by the requirements of reconstructing upward-traveling displaced vertices and distinguishing them from downward-traveling cosmic rays. It therefore comes as no surprise that MATHUSLA has all the qualities needed to act as an excellent cosmic ray telescope. In fact, MATHUSLA's particular combination of robust tracking and large area allow it to make many unique measurements that could address important and long-standing questions in astroparticle physics. The study of cosmic rays is therefore an important secondary physics goal of MATHUSLA. These measurements, which do not interfere with the primary goal of LLP discovery,  represent a ``guaranteed physics return'' on the investment of the detector, as well as an opportunity to establish a cosmic ray physics program.

The cosmic ray physics program at MATHUSLA warrants in-depth examination, and some initial studies will be presented elsewhere \cite{MathuslaCRWP}. 
Here we only briefly comment on the  qualities that make MATHUSLA a uniquely interesting cosmic ray experiment, and outline some possible measurements that are of particular interest to the astroparticle physics community. 
To this end, we first review some basic facts about cosmic rays and how they are detected.

\subsection{Basic Principles of Cosmic Ray Detection}

\begin{figure}
\begin{center}
\includegraphics[width=0.7\textwidth]{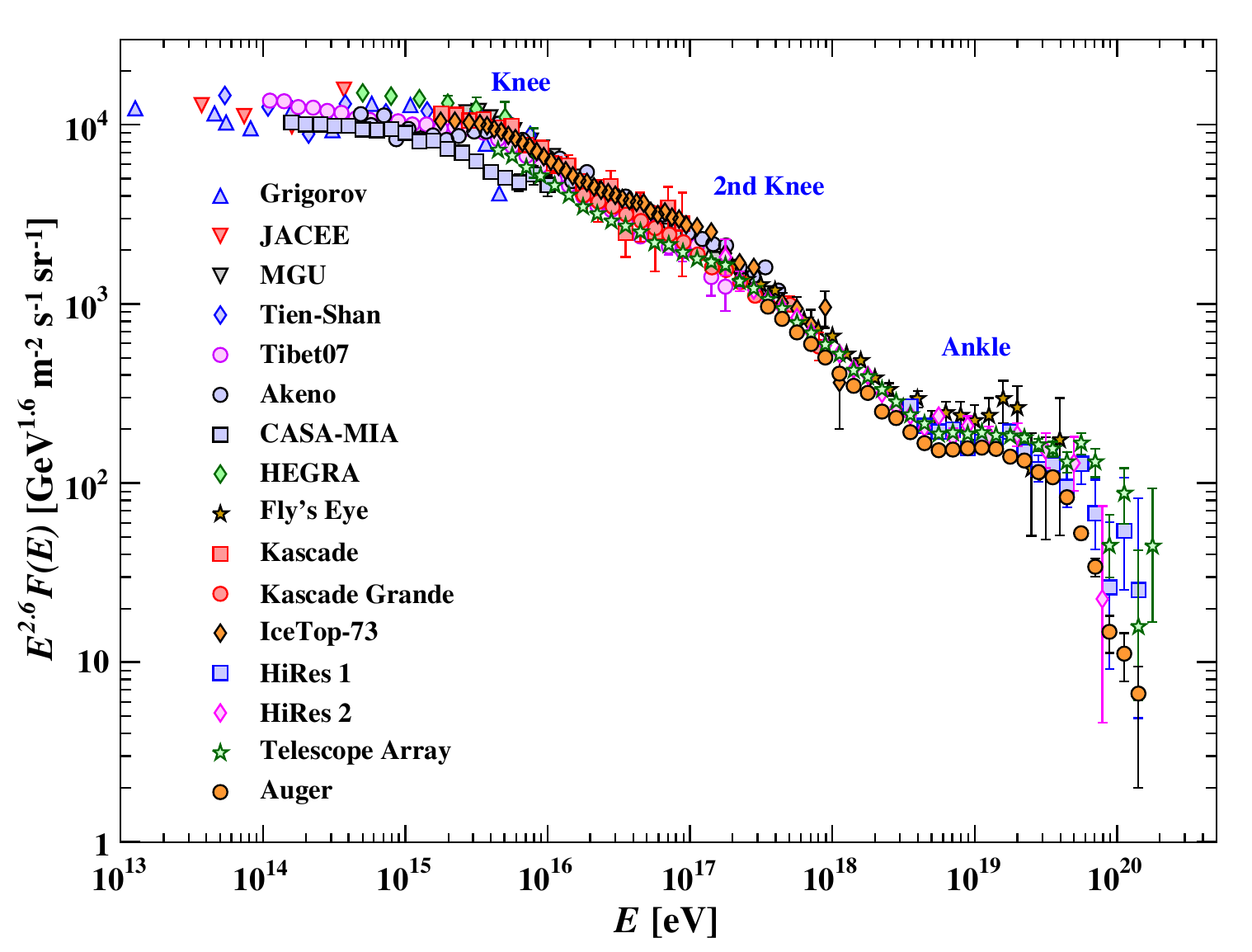}
\end{center}
\caption{
Global view of the all-particle cosmic ray energy spectrum (figure taken from \cite{Patrignani:2016xqp}). The total spectrum decreases quickly according to a power-law formula $E^{-\gamma}$, where $\gamma$ (the spectral index) varies from roughly $2.6$ to $3.3$ \cite{Roulet17}. Between $10^{15} \, \ev$ and $10^{19} \, \ev$, the spectrum exhibit three distinctive features created by a change in the value of $\gamma$: the knee, which is located at $\sim 4 \, \pev$ \cite{Patrignani:2016xqp, Kulikov1959}, the second knee, close to $100 \, \pev$ \cite{Apel2012, Aartsen2013, Prosin14} and the ankle, around $4 \, \eev$ \cite{Settimo2012, Abu2013}. There exists also a weaker structure called the low energy ankle \cite{Roulet17} at  $\sim 10 \, \pev$ \cite{Apel2012, Aartsen2013}. MATHUSLA is expected to be sensitive to hadronic EAS with primary energies around the knee, i.e. in the interval $E \sim 10^{14} - 10^{17} \, \ev$ according to its size $200 \times 200 \,\mathrm{m}^2$ and the atmospheric depth at which it will be located ($\sim 1000 \,\mathrm{g}/\mathrm{cm}^2$).
}
\label{f.CRspectrum}
\end{figure}

Cosmic rays, dominantly protons and heavier atomic nuclei, arrive at earth with an energy that spans some 12 orders of magnitude, from a few hundred MeV ($10^8$ eV) to 100 EeV ($10^{20}$ eV) \cite{Engel2009, Greiner15, Roulet17}, see for example, Fig.~\ref{f.CRspectrum}. 
They are produced in violent astrophysical scenarios  within our own galaxy (for energies $E \lesssim  10^{18} \, \mathrm{eV}$) and beyond (for $E > 10^{18} \, \mathrm{eV}$). At the highest energies, however, the origin of CRs is still mysterious, since propagation in galactic magnetic fields means CRs do not point back to far away sources~\cite{Roulet17, AugerScience2017}. 
In general, details of CR acceleration mechanisms,  composition, propagation through  space, and  features in their spectrum are not completely understood \cite{Roulet17, Giacinti15, Gaisser13, Deligny16}. The study of CRs offers a unique window on the most energetic natural phenomena of the cosmos \cite{Ackermann13, Abramowski16, Fang2018}, and their collisions with the atmosphere probe energies far in excess of the TeV scale \cite{Aab2015a,Aab2016a,Aab2017a}.

Primary CRs with relatively low energy can be directly characterized by balloon- or space-borne particle physics experiments equipped with trackers and calorimeters, like AMS-02 \cite{Kounine2012,MAguilarAMS} and CREAM \cite{Ahn2007,CREAM2010} among
others (see, for example, \cite{Seo2012, Mocchiut2014, Maestro2015}). The small size of these detectors restricts this approach to energies below $\sim 100 \tev - 1 \pev$, both because the CR flux drops dramatically above this threshold and because the detector's magnetic field and radiation depth limits the maximum energy which can be reconstructed. 

To study higher energy CRs above $\sim 10^{14} \ev$, the atmosphere is used as a calorimeter  \cite{Engel2009, Roulet17, Kampert12,Engel2011}. The detection technique consists in observing the Extensive Air Showers (EAS) of SM particles that CRs induce in collisions with the atmosphere. The EAS typically originates  from $15 \,\mathrm{km}$ to $35 \,\mathrm{km}$ above sea level \cite{Engel2011, Grupen2012} and spreads out as the particle front travels towards the ground. 
%
%
%
%
%
%
At sea level, an average EAS consists primarily of electrons/positrons and photons 
with $\sim 10\%$  muons and a smaller fraction of hadrons (for vertical incidence) \cite{Grieder2009}. Depending almost linearly on the energy in logarithmic scale, the total number of particles in the EAS is in the range $\sim 10^4 - 10^{10}$, for primary energies $E = 10^{14} \, \,\mathrm{eV} - 10^{20} \,\mathrm{eV}$, and it is mostly contained within a cone of $\mathcal{O}(0.1 - 1 \,\mathrm{km})$ in diameter at ground level~\cite{Kampert12, AugerScience2017}.
Measurements of the EAS allow the primary CR's direction, composition and energy to be reconstructed~\cite{Grieder2009}. 
Earth-bound CR experiments employ two classes of techniques to probe the EAS: (1) particle detectors on the surface, like particle counters, trackers, and calorimeters to sample the air shower front and (2) various telescopes to observe electromagnetic emissions from the shower or from the interaction of the EAS with the atmosphere (Cherenkov radiation, radio, fluorescence light). 
Many experiments use a combination of both techniques like the Pierre Auger Observatory \cite{Abraham2004}, 
TUNKA \cite{Prosin14} and TALE \cite{Abbasi2018}. 
EAS observatories monitor in general large areas to compensate the low CR flux at high energies. For example, at $E \sim 10^{15} \, \ev$, CRs are received at a rate of $\sim 1 \,\mathrm{particle}/\mathrm{m}^2\cdot\mathrm{year}$ \cite{Engel2009} and shower arrays with areas of order 
$\gtrsim \mathcal{O}(10^4 \,\mathrm{m}^2)$ are required, like the KASCADE air shower detector ($200 \times 200 \,\mathrm{m}^2$) \cite{Antoni2003}. On the other hand, at extremely high energies $E \sim 10^{20} \, \ev$, the CR flux is so low ($\sim 1 \,\mathrm{particle}/\mathrm{km}^2\cdot\mathrm{century}$ \cite{Engel2009}) that sufficient exposure requires installations with very large areas of the order of 
$\mathcal{O}(10^3 \,\mathrm{km}^2)$, like the $3000 \,\mathrm{km}^2$ Pierre Auger Observatory \cite{Abraham2004, Abraham2010}.

EAS telescopes/antennas are able to observe the longitudinal development of the air shower, while surface detectors
measure the lateral structure of the EAS at the atmospheric depth of the site \cite{Grieder2009}. In the latter, 
particle detectors are arranged in arrays and are spaced at regular intervals to optimize the measurements for 
the energy range of interest. Hence, in most cases, they sample only a small fraction of the shower at 
the observation level. For instance, in case of KASCADE, the main array of 252 e/$\gamma$ detectors covered only $1.22\%$
of the total surface, the muon array of 192 detectors, only $1.55\%$,
while the muon tracking detector and the hadron calorimeter covered just $0.64\%$ and
$0.76\%$, respectively \cite{Antoni2003}. Just in a few cases, full coverage was achieved as in the case of the
ARGO-YBJ detector, which consisted of a $74  \times 78 \,\mathrm{m}^2$ carpet of RPC's with an active area of almost $93\%$ \cite{Aielli2006}.
In general, most surface detectors are insensitive to the energy of a single charged particle (although a counter can be equipped with shielding or buried underground to implement a desired minimum energy threshold). Rather, the focus is on collective shower properties, like the spatial and temporal distribution of particles, as well as some basic particle ID to separate the e/$\gamma$,
muon and/or hadron components in the EAS. This data is used to characterize the primary CR, assuming a certain hadron interaction model which governs the evolution of the EAS in the atmosphere \cite{Engel2011, Knapp2003}.

Hadron interaction models are a crucial part of air shower Monte Carlo simulations. They are tuned to available high energy physics data but rely on extrapolation in certain regions of phase space, in particular, the forward region at very high energies. This introduces unavoidable uncertainties in the determination of the properties of the primary cosmic rays \cite{Ostapchenko2006, Pierog2006, Antoni2005}. Verifying and tuning these hadron interaction models is therefore of fundamental importance to CR physics \cite{Engel2011}. Tests of hadron interaction models
can be performed with the same data from EAS observatories. That requires, however, the simultaneous measurement of different observables of the air showers. 
KASCADE made important contributions to this topic \cite{Antoni1999,Antoni2001,Antoni2003a, Antoni2005a, Antoni2006, Apel2007, Apel2009}, because of both the quality of its measurements and its different detection
systems, like the full-coverage central tracker and calorimeter \cite{Antoni2003}. This central detector only had less than $1\%$ the 
area of the full experiment but was crucial in allowing for more detailed analysis of the shower \cite{Antoni1999,Antoni2001,Antoni2005a, Apel2007, Apel2009}.

\subsection{Cosmic Ray Physics with MATHUSLA}
\label{s.mathuslaCR}


We can now understand why MATHUSLA would make important contributions to cosmic ray physics. 
MATHUSLA200 is the size of KASCADE and will be also located at a comparable atmospheric depth ($\sim 1000 \,\mathrm{g}/\mathrm{cm}^2$). 
Therefore, it is expected to be sensitive to a similar CR energy range of $10^{14} - 10^{17}$ eV.
After a few years of exposure ($\sim 3 \, \mathrm{yr}$), roughly $10^6$ air showers with $E > 10^{15} \ev$ would be recorded with 
their cores traveling through the MATHUSLA detector. 
Unique for a CR experiment of this size, MATHUSLA has full-coverage robust tracking with excellent position and timing resolution. 
The scintillator planes which enclose the LLP decay volume would supply additional information, especially for highly inclined showers. 
Even without track-by-track $e/\mu$ discrimination or calorimetry, this would allow for very detailed EAS measurements, including highly granular analysis of the shower's temporal and spatial structure, which has never been undertaken at this size scale at PeV energies.
Furthermore, for roughly half of those CR events, the so-called ``golden events'',  part of the shower's high-energy muon component ($E^{th}_\mu \gtrsim 50 \, \gev - 70 \, \gev$ for vertical incidence) also passes through the underground ATLAS or CMS detector and could be simultaneously registered with special CR triggers during the LHC runtime.
Therefore, whether working in standalone mode, or in tandem with the main underground detector, MATHUSLA constitutes a powerful cosmic ray experiment with unique capabilities that will open an era of precision EAS measurements in the PeV energy region.

The particular CR measurements which MATHUSLA can perform will be studied in more detail in a future document \cite{MathuslaCRWP}. Some of the most compelling targets for CR measurements include:
\begin{itemize}
\item \emph{Primary CR spectra and composition:}
The spatial and temporal distribution of charged particles in the shower would probe the energy and composition of the primary CR, potentially addressing open issues like the exact position of the ``light knee'' (a change in spectral index  in the spectrum of protons and helium,  
which ARGO-YBJ found at $\sim 700 \, \tev$ \cite{Bartoli2015} and KASCADE, around $3 \, \pev - 4 \pev$  \cite{Antoni2005, Apel2013}) and the shape 
of the spectra of the heavy elemental components of primary CRs at PeV energies \cite{Antoni2005, Apel2013}.

\item \emph{Cosmic Ray Anisotropies:}
MATHUSLA's enhanced resolution, compared to previous CR experiments, could allow for improvements to the measurements on the dipole component of the anisotropy in the diffuse CR flux at PeV energies, which has been poorly investigated (see \cite{Roulet17, Deligny16} and references therein).
Another important aspect is the search for local point sources in the northern celestial hemisphere, which would provide vital clues about the presence of nearby galactic accelerators of very high energy CRs \cite{Antoni2004,Over2008,Kang2015}.

\item \emph{Highly inclined showers:}
The vertical scintillator planes, if installed, enclosing MATHUSLA's decay volume, together with its precise  full-coverage tracking, would allow for the study of highly inclined air showers at large zenith angles $\theta > 60^\circ$. These showers are interesting for a variety of reasons. If they originate from charged primary CRs, they are dominated by muons since their electromagnetic component is attenuated after traversing large distances in the atmosphere. Observations of such events could help to study the high-energy muon content of EAS and to test hadron interaction models by looking for 
anomalies in this sector at PeV energies, such as those which have been observed at higher energies by observatories like the KASCADE-Grande
detector \cite{Apel2011, Apel2017}, the Pierre Auger observatory \cite{Aab2015a, Aab2014} or the Yakutsk experiment \cite{Abu2000}. 
The former has observed, for example, that the actual attenuation length of shower muons at $10^{17} \, \ev$ is bigger than the predictions of hadron interaction models \cite{Apel2017}, while the latter ones have measured an excess of muons in ultra-high-energy EAS in comparison with the models, a problem which is known as the \emph{muon puzzle} \cite{Petrukhin2014}.

Following \cite{Aab2015b}, atmospheric and/or astrophysical neutrinos with energies above $10^{15} \ev$ could also be detected in this way if they scatter deep in the atmosphere or interact with the rock of the nearby Jura mountains.  Such neutrinos could induce very inclined young EAS,
which could be distinguished from regular old showers produced by CRs due to MATHUSLA's superior tracking resolution. Young EAS are characterized by a richer electromagnetic component, a broader time signal and a larger EAS front curvature. Thus measurements of the particle content and the spatial/temporal structure of inclined EAS in MATHUSLA could allow the search for neutrino signals. MATHUSLA might also offer
a tool to look for upward-going EAS from Earth skimming $\nu_\tau$'s \cite{Aab2015b, Feng2002} and upward-going muons from $\nu_\mu$'s interacting with the rock below the detector or inside the MATHUSLA's instrumented volume \cite{Gaisser1995, Ambrosio2003}.  If MATHUSLA is able to detect also upward-going muons resulting from muon-flavored neutrinos interacting in rock, MATHUSLA could also provide complementary measurements for $\nu-$oscillations (see, for example, \cite{Fogli1998, Jung2001, Giacomelli2013}).

\item \emph{Study of EAS and tests of hadronic interaction models:}
Detailed measurements of the spatial and temporal structure of EAS, as well as data on the charged particle attenuation length and the muon components of highly inclined showers, may provide a number of clues to understand  several outstanding ambiguities in hadron interaction models \cite{Apel2011, Apel2017, Aab2014, Abu2000, Aab2015b}. If there is a linkage to the LHC main detector, additional constraints would be supplied by correlating MATHUSLA's measurements with detection of the high-energy muon component in the LHC main detector.

\item \emph{High-multiplicity Muon Bundles:}
Muons with an energy greater than $\sim 50 \gev$ will penetrate down into the rock and reach the main detectors at CERN. ALEPH/DELPHI at LEP \cite{Grupen2003,Ridky2005} and ALICE at the LHC \cite{Adam2016} have studied this high-energy muon component of EAS's, observing events with more than 100 muons in the underground detector. During the LEP era, these high-multiplicity muon bundles could not originally be explained by hadron interaction models, and several BSM explanations were proposed \cite{Kankiewicz2017, Ridky2017}. Today, the data by ALICE points towards iron-rich CR primary composition at energies above $10^{16}$ eV, but the data has very low statistics and further measurements are needed to understand the origin of these muon bundles and their impact on primary CR studies. 

Muon bundles could be detected by the LHC main detector using the main detector's triggering system and correlated in time with data from MATHUSLA. This would give a much more complete picture of these special CR events and could allow for their origin to be unambiguously determined. These events are also valuable for constraining hadron interaction models. If the triggers of MATHUSLA and the LHC main detector were linked that would improve this measurement further. 
\end{itemize}
It is worth noting that these measurements could be significantly improved if $e/\mu$ discrimination capability were added to MATHUSLA. Apart from making measurements of CR primary composition and spectra less dependent on hadron interaction models, the separate muon data would allow for many additional detailed probes of hadron interaction models, with great benefit to all future CR measurements at other experiments, e.g. \cite{Abbasi2018, Joshi2017, Budnev2017, Aartsen2017}.
This non-exhaustive lists of physics targets demonstrates that MATHUSLA could supply data which will be of unique value to the astroparticle and CR physics communities.

\section{Background Studies}
\label{s.backgrounds}

The most important backgrounds to LLP searches at MATHUSLA come from Cosmic Rays, muons from the HL-LHC collisions point, and neutrino scatterings, see Fig.~\ref{f.bgsummary}. 
It is vital to the LLP reach of MATHUSLA that these and other backgrounds can be rejected to near-zero levels.
The various background rates and their rejection strategies were first estimated in~\cite{Chou:2016lxi}, establishing the plausibility of reaching this very low background regime.
Here we present the results of several much more detailed background studies, all of which confirm the original conclusion that background rejection at MATHUSLA is possible.
For new background studies which are still in progress, we review the original results or arguments presented in~\cite{Chou:2016lxi} and comment on the status of the ongoing efforts. 

At this stage of development, these background studies are very \emph{technology-independent}, either not requiring detector simulation or, as may be the case for cosmic rays, being studied in a generic simulated MATHUSLA detector that resembles the idealized design of Figs.~\ref{f.bgsummary} and \ref{f.mathuslageometry}. This is sufficient to establish their rates in much more detail than in~\cite{Chou:2016lxi}, and helps inform more realistic detector designs.

\subsection{Cosmic Rays}
\label{s.CRBG}

By far the largest background is charged particles from cosmic ray air showers, see Fig.~\ref{f.bgsummary} (e). They have a rate of $\mathcal{O}(10\  \mathrm{MHz})$ over the area of MATHUSLA200~\cite{Patrignani:2016xqp}, resulting in  $\sim 10^{15}$ charged particle trajectories integrated over the whole the HL-LHC run. 
A very high rejection efficiency of this background is possible using a combination of timing and trajectory reconstruction, due to the highly distinct signatures produced by cosmic ray charged particles compared to the decay of an LLP.

In the next subsection, we review why the basic design of MATHUSLA allows this background to be rejected~\cite{Chou:2016lxi}. We then outline several detailed ongoing studies using  full cosmic ray simulations using CORSIKA  \cite{CORSIKAref:1998}, fast parameterized simulations to study combinatoric aspects of this background, as well as aspects of the associated detector simulation.

It should be noted that these studies go hand-in-hand with exploring the physics potential of studying cosmic rays with MATHUSLA, see Section~\ref{s.cosmicray}.

\subsubsection{Basic Principles of Cosmic Ray Backgrounds at MATHUSLA}

The most important aspect of CR rejection is their direction of travel. The vast majority of CR charged particles travel down, not up like final states from the decay of an LLP originating at the $pp$ IP. 
%
The chance that a \emph{single} downward-traveling cosmic ray will be mis-identified as 
originating within the detector volume and traveling upward is less than $10^{-15}$.
%
This rate was estimated by assuming the tracking system consists of five RPC layers, separated by 1 m, with a Gaussian timing resolution of 1 ns. 
If the hermetic scintillator veto is present and has an order percent chance of not detecting the passing charged particle, this rejection is achieved with only three out of the four layers firing.
Without the scintillator veto, this rejection is surpassed by assuming four out of the five layers fire.
At least \emph{two} such fakes must occur to fake an upwards-traveling LLP decay. Furthermore, the LLP signal requirement would actually include that these two (or more) upwards-traveling final states fire all five tracking layers, not just three or four. 

The above clearly demonstrates that timing is an extremely powerful rejection strategy for cosmic rays, making the zero-background regime plausible. 
Furthermore, spatial information was not used at all in this estimate: we only required that each RPC hit occurs within two timing resolutions of a consistent time for an upwards traveling particle with constant relativistic speed. This timing requirement, which could possibly be loosened by taking spatial information into account, does not significantly reduce signal efficiency at our level of precision. Spatial information will be necessary to help reject cosmic rays that enter from large polar angles (close to parallel to the surface of the earth at MATHUSLA).

This small rate suggests that the only realistic way for cosmics to possibly fake the signal are correlated showers of many charged particles coincident on MATHUSLA. 
%
These showers lead to a large number of charged particles occupying MATHUSLA near-simultaneously. This is easily rejected with very little ``blind time'' for the LLP search.  

The required very high efficiency rejection of cosmic ray backgrounds for LLP searches during the HL-LHC run will be a major driver of the detailed MATHUSLA design. The precise required timing and spatial resolutions of the tracker and scintillators, as well as the exact number of tracking layers, is the key question to be resolved by our current ongoing studies, described below.
However, the unique geometrical nature of the LLP DV signal, and the multitude of coincidences that would be required for cosmic rays to fake the signal, makes us confident that this background can be controlled. Furthermore, this background can be measured in detail in the absence of HL-LHC collisions, and as part of the CR physics program outlined in Section~\ref{s.cosmicray}. Such non-collision data will aid in defining and/or validating rejection strategies.


\subsubsection{CORSIKA Cosmic Ray Simulations}

In order to study the response of the MATHUSLA detector to cosmic rays, and
to evaluate the corresponding background of penetrating particles (muons 
and neutrinos) on the searches for LLPs, MC simulations of EAS are being 
performed.

The development of the EAS will be simulated with the program   
CORSIKA 7.64 \cite{CORSIKAref:1998}, while the high-energy ($E \geq 200 \, \mbox{GeV}$) hadronic
 interactions of the primary nuclei and the secondary shower particles, with 
 the recent post-LHC hadronic interaction models, i.e., QGSJET-II-04 \cite{qgsjetii04}, EPOS-LHC \cite{eposlhc} and SIBYLL 2.3c \cite{Sibyll2.3c}. At lower energies, on the  other hand, the hadronic interaction models Fluka \cite{Fluka:2018} and GEISHA \cite{GHEISHAref:85} will be used.  We plan to include several  models in the studies to account for the uncertainties due to the hadronic  interaction models, as it is well known that they are one of the main source of systematic uncertainties in cosmic ray studies \cite{Ostap06}. 
 
 CORSIKA (COsmic Ray SImulations for KAscade) is a program to simulate extensive air showers (EAS) produced by cosmic ray primary particles. Interactions of the primary with the air nuclei or the decays are reproduced following several models depending on the energy of the reaction. It may be used up to and beyond the highest energies of 100 EeV \cite{CORSIKAref:1998}.
   
   To simulate high and low energy hadronic interactions in the EAS, CORSIKA
 makes use of phenomenological models, which are generally calibrated
 with collider data at low energies are extrapolated to higher energies and
 into the forward region~\cite{Ostap06}. Among the high-energy models used by CORSIKA we found EPOS-LHC,
 QGSJET-II-04 and SIBYLL 2.3c work best. For low energies, CORSIKA can be used
 with FLUKA and GEISHA. QGSJET-II-04 is based on the Gribov-Regge theory \cite{Gribref:2001}. EPOS-LHC  is based on the QGSJET \cite{QGSJETmodel:1997} and the VENUS\cite{VENUSref:1993} models and considers hard interactions and nuclear and high-density effects. It is also improved to reproduce in detail LHC data from various experiments \cite{EPOSref:2013}. SIBYLL 2.3c is a minijet model for efficient simulation of hadronic multiparticle production\cite{SIBYLLref:2017}. On the other hand, at lower energies GHEISHA describes the interaction of particles with matter considering results from atomic and nuclear scattering experiments \cite{GHEISHAref:85} while Fluka describes particle transport and interactions with matter \cite{Fluka:2018}.
    

\begin{figure}
\begin{center}
\begin{tabular}{rr}
\includegraphics[width=0.48\textwidth]{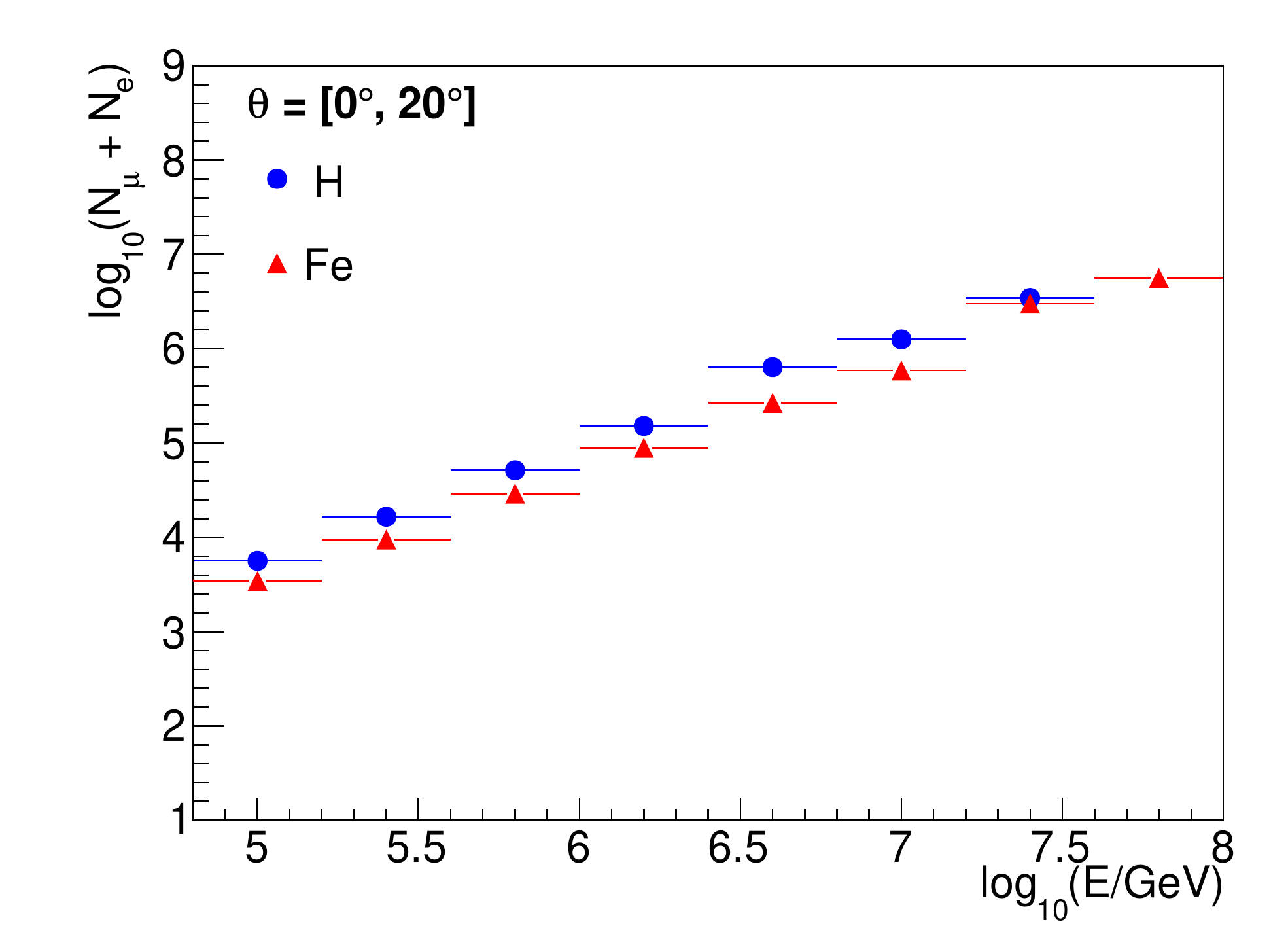}
&
\includegraphics[width=0.48\textwidth]{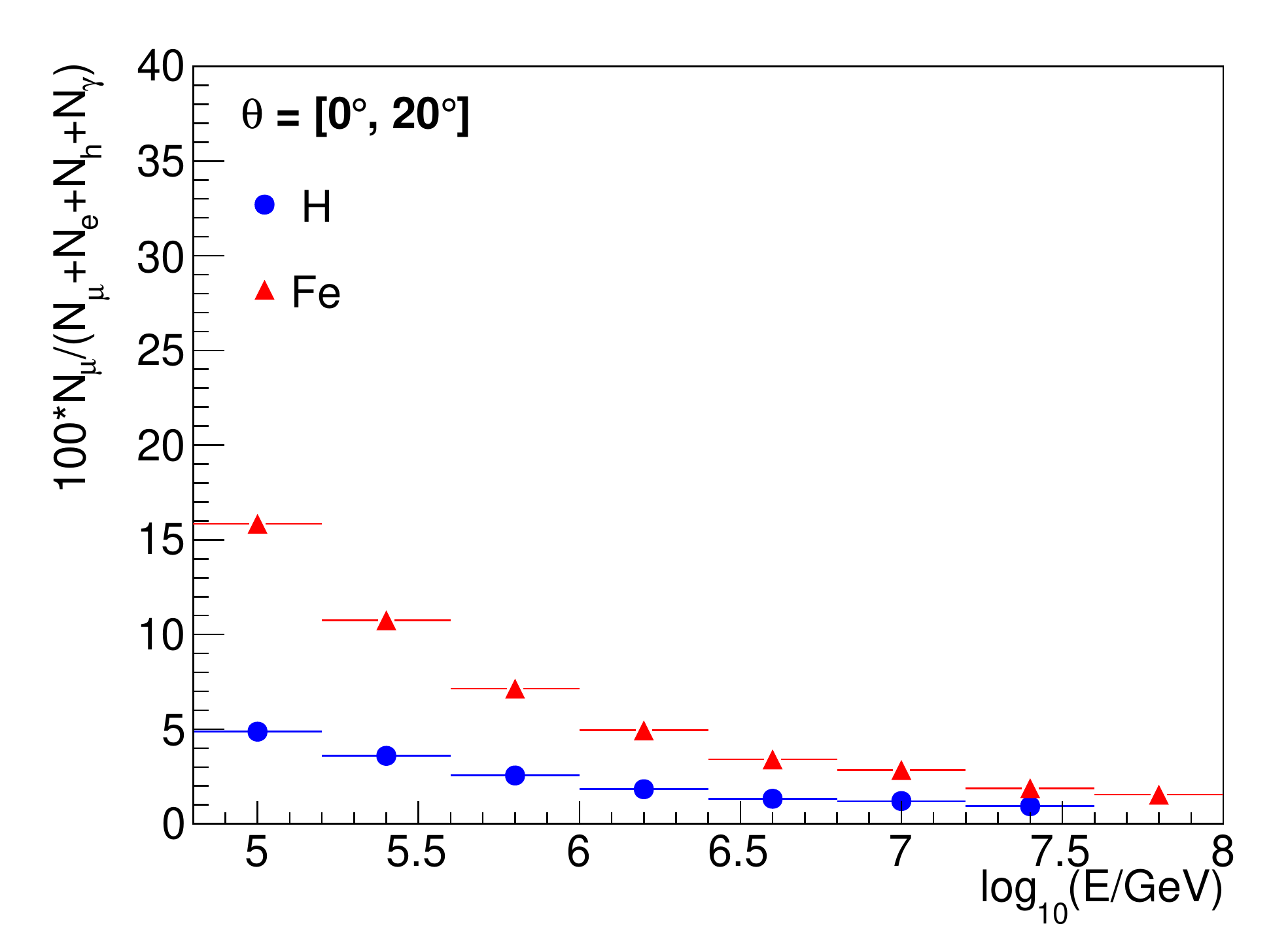}
\end{tabular}
\end{center}
\caption{
 CORSIKA simulations for air showers induced by cosmic rays 
 assuming protons (blue circles) and iron nuclei (red triangles) 
 as primary nuclei in the interval $E = 10^{4.8} - 10^{8} \, \mbox{GeV}$ and for
 arrival zenith angles ($\theta$) smaller than $20^{\circ}$.
 As high/low-energy hadronic interaction models QGSJET-II-04 
 \cite{qgsjetii04} and GEISHA \cite{GHEISHAref:85} were
 used, respectively. EAS are followed until ground level
 above the site of the ATLAS experiment.
 \textit{Left}: Mean logarithm of shower 
 size (total number of shower electrons plus muons) versus 
 logarithm of the primary energy. Here we have considered
 electrons with threshold $E_{th} > 3 \, \mbox{MeV}$ and muons 
 with $E_{th} > 100 \, \mbox{MeV}$. \textit{Right}: Fraction
 of muons (in percentage) in the air shower as a function
 of the logarithm of the primary energy.
}
\label{EASverticalcorsika}
\end{figure}

  To generate the simulations with CORSIKA, we will set at the moment
  the MAHUSLA location at ground level ($436 \, \mbox{m}$ a.s.l.) above the site
  of the ATLAS experiment  (latitude $46^\circ 14' 4.8''$ N,  longitude
  $6^\circ 03' 11.4''$ E). The geographical coordinates were taken
  from the Elevation Map Net\footnote{https://elevationmap.net/cern-european-organization-for-nuclear-research-ch} 
  and the details of the local geomagnetic
  field ($B_x = 22.2 \, \mu T$, $B_z = 41.9 \, \mu T$) were obtained from the
  NOAA web page\footnote{https://www.ngdc.noaa.gov/geomag-web/\#igrfwmm}. 
  We will select a curved atmosphere, instead of a flat
  one, to have more realistic simulations for inclined EAS
  ($\theta > 60^\circ$). We will also use the U.S. standard atmosphere model as 
  parameterized by Linsley, which is the default choice in CORSIKA 
  \cite{CORSIKAref:1998}. We have also set the following threshold energies
  for the secondary particles at the observation level:
  $3 \, \mbox{MeV}$ for $e^{\pm}$ and $\gamma$'s, and $100 \, \mbox{MeV}$
  for $\mu^{\pm}$ and hadrons. We will also use a thinning of $10^{-6}$, which
  will allow us to save computing time and save disk space without losing
  too much information from the structures of the EAS \cite{TanguyICRC17}. The 
  thinning is defined as the fraction of the primary energy below which the 
  secondary particles are no longer tracked in the code. They are replaced by 
  one particle that is followed with an appropriate weight to keep the average 
  properties of the EAS.

  Events in CORSIKA are also defined by the primary nuclei, its energy,
  and arrival direction. The last two parameters are randomly thrown
  from a given range and distribution previously specified. The corresponding 
  values that we will take for the above parameters are explained in the following   lines.
\begin{itemize}
\item Five cosmic ray primary nuclei are taking into account: Proton, Helium, Carbon, Silicon and Iron. The energy range for all particles is $10^{13}-10^{18} \mbox{eV}$ and will follow an $E^{-2}$
spectrum. On the other hand, their angular dependence (non-flat configuration) will cover zenith angles in the range [$0^{\circ}, 90^{\circ}$] and azimuth ones in the interval [$0^{\circ}, 360^{\circ}$].
In a further step, a weight will be added to simulate an appropriate physical distribution.
\item  To properly generate high angle particles and allow CORSIKA to evolve the showers to the edges of the detector, a non-flat geometry is used in CORSIKA  to take into account the detector walls and volume.    
\end{itemize}

\begin{figure}
\begin{center}
\begin{tabular}{rr}
\includegraphics[width=0.48\textwidth]{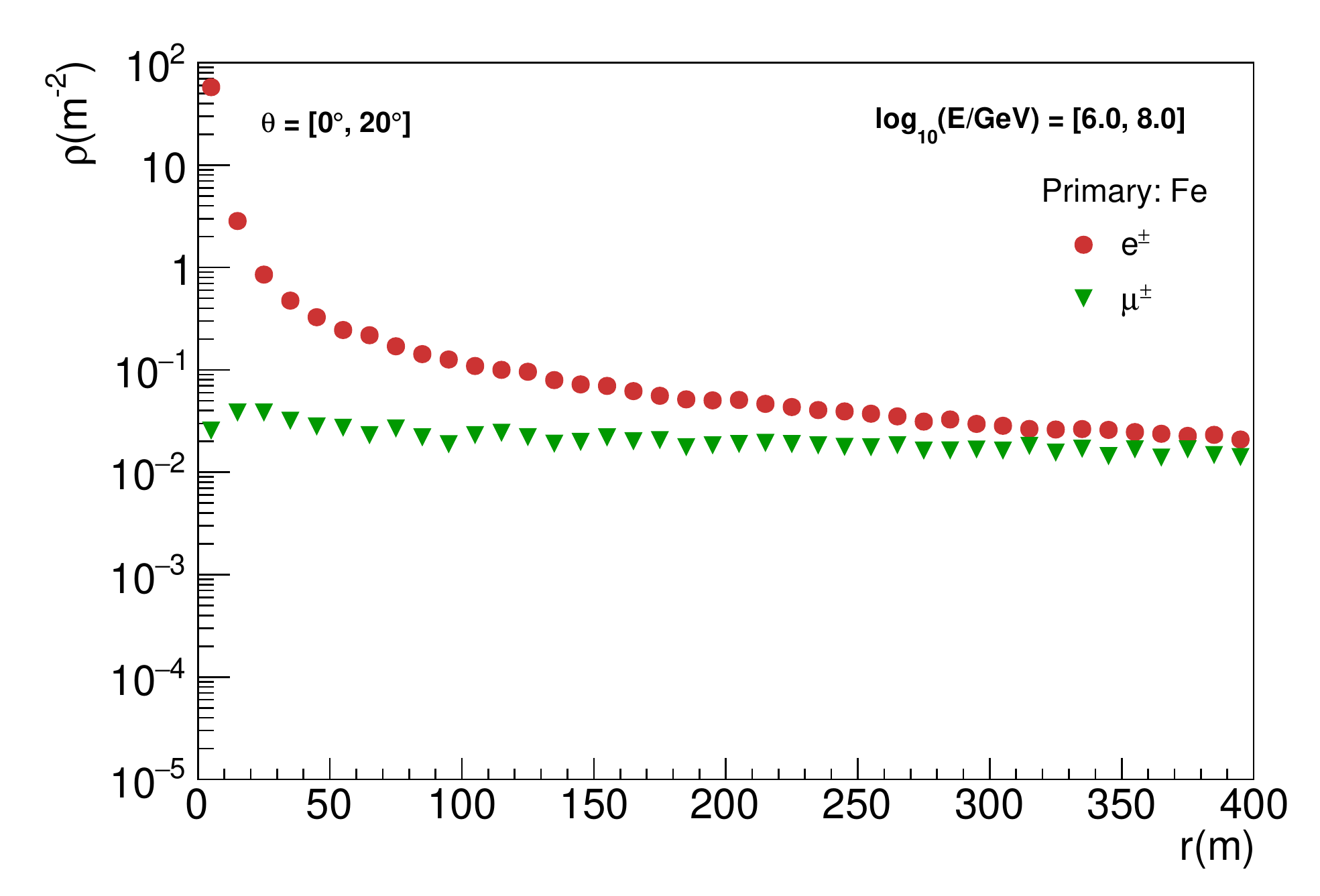}
&
\includegraphics[width=0.48\textwidth]{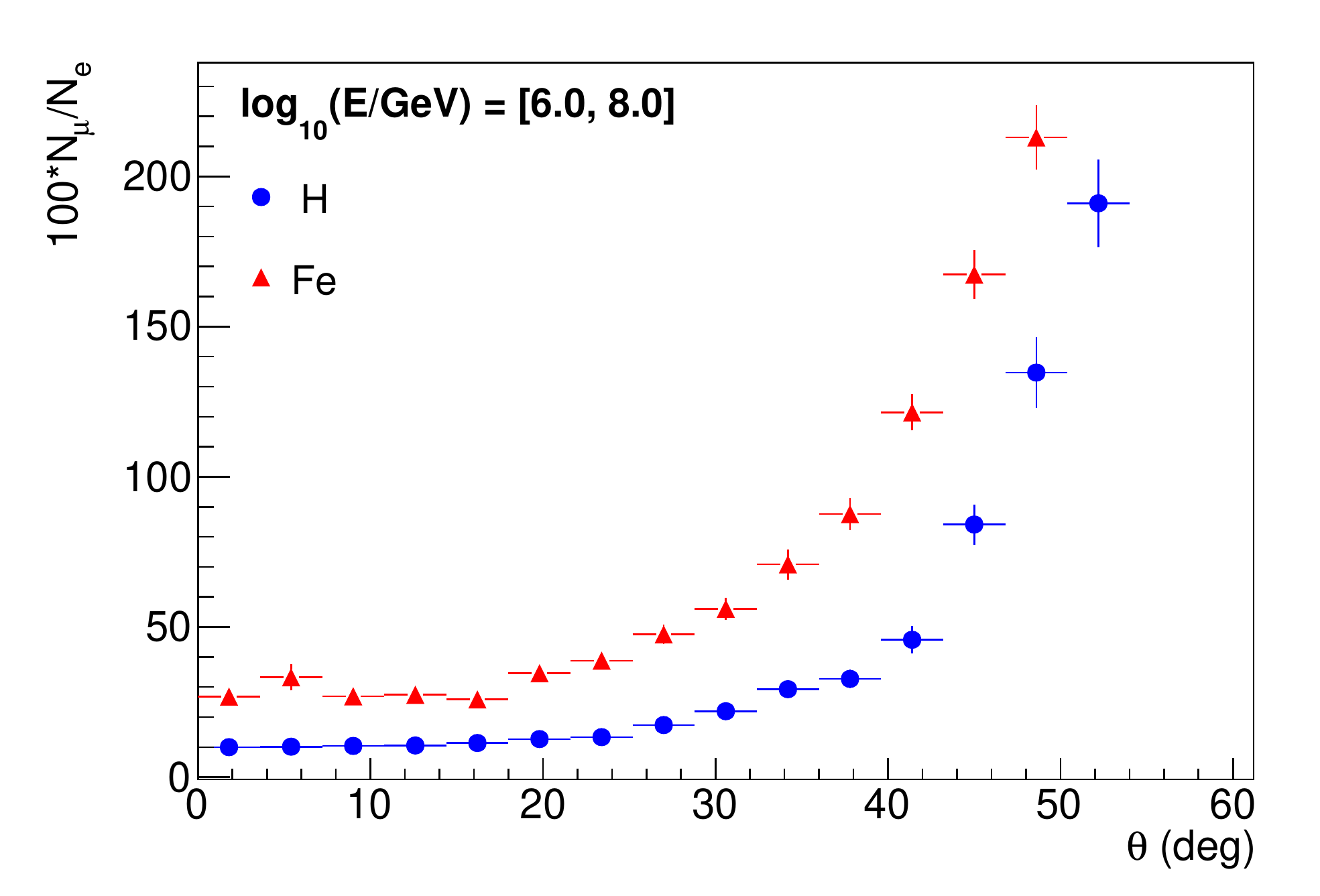}
\end{tabular}
\end{center}
\caption{
 \textit{Left}: Radial density distributions of shower
 electrons (red circles, $E_{th} > 3 \, \mbox{MeV}$) and muons 
 (green triangles, $E_{th} > 100 \, \mbox{MeV}$) at ground 
 level (MATHUSLA site) measured vs the distance to the core of the 
 EAS event for protons of energies $E = 10^{6}$ GeV $- 10^{8}$ GeV.
 \textit{Right}: Mean ratio of total shower muons, $N_\mu$, and
 total number of shower electrons, $N_e$, vs the arrival
 zenith angle for EAS induced by protons and iron nuclei with
 energy in the range  $E = 10^{6}$ eV $- 10^{8}$ eV.
 Simulations were obtained with CORSIKA using 
 QGSJET-II-04/GEISHA.
}
\label{EASverticalcorsika2}
\end{figure}

 The output of the CORSIKA program consists of several local and global
 variables associated with the properties of the EAS event at the observation level, such as arrival times, type, location and momenta of the secondary particles, the total number of muons ($\mu^{\pm}$), electrons ($e^{\pm}$), hadrons, neutrinos and $\gamma$'s above the specified energy threshold, among other useful shower information. It also keeps the input data of each event, e.g., the primary energy, the arrival direction (zenith and azimuthal angles), and the primary nuclei. CORSIKA simulation runs are at the moment
underway. Some results of the first runs are presented in Figs.~\ref{EASverticalcorsika} and \ref{EASverticalcorsika2} for primary cosmic ray energies 
relevant for the MATHUSLA detector using QGSJET-II-04 with GEISHA. 
In Fig.~\ref{EASverticalcorsika}, left we can see, for example, that 
the expected mean of $\log_{10}(N_e+N_\mu)$, where $N_e$($N_\mu$) is 
the total number of electrons (muons) in the shower, increases linearly 
with $\log_{10}(E/GeV)$. From this graph we observe that a cosmic ray
with primary energy of $10^{6}$ GeV produces an EAS with an average of
$10^{5}$ shower electrons and muons at the observation level of MATHUSLA.

On the other hand, the right plot of Fig.~\ref{EASverticalcorsika}
shows that the fraction of muons in vertical EAS is not bigger than 
$10 \%$ above primary energies of $1$ PeV. In general, the total number 
of muons is smaller than $N_e$ in vertical showers of high energy
($> 10^{16}$ eV) at the altitude of MATHUSLA. That can be appreciated
in Fig.~\ref{EASverticalcorsika2} left, where the radial density distributions (measured from 
the core of the EAS) for $e$'s and $\mu$'s in EAS induced
by \mbox{PeV} iron nuclei are shown. We see, however, that the muon/electron ratio depends on the distance to the core. In particular, 
this fraction increases for large radial distances. The same
happens at high zenith angles due to increment of the
atmospheric depth and the larger attenuation length of muons
in the atmosphere (see Fig.~\ref{EASverticalcorsika2}, right).

  For the present analyses, the idea is to model the cosmic
  ray intensity in the full energy range from $E = 1 \, \mbox{TeV}$
  up to $1 \, \mbox{EeV}$ using the above five nuclei primaries 
  ($H, He, C, Si, Fe$), which are representative of the distinct mass groups, 
  by using an appropriate cosmic ray composition model derived from fits to the 
  measured data (see, e.g., the H4a model \cite{Gaisser11} and the Global Spline Fit model \cite{Dembinsky})    
  to also take into account spectral features in the corresponding
  mass group spectra of cosmic rays. 
  
  This MC data sets will be used to extract the expected EAS, muon
  and neutrino backgrounds at MATHUSLA.

\subsubsection{Parameterized Simulation}
\label{s.parambackground}

The parameterized simulation will generate a library of {\em background} events for use in combination with other background studies and the detector simulation. These background events will play much the same role that pile-up events or overlay play in LHC experiments. The parameterized distributions will be extracted from CORSIKA as described by known models (see for example \cite{Agrawal96, Honda07, Biallas2009}). These background events are expected to be important in calculating an accurate fake rate for at upward going muons, for example.

\subsubsection{Detector Simulation}
\label{s.detectorsim}

A program for fast simulation and detailed simulation (including albedo and rudimentary material models) in Geant4 is under construction. Geant4~\cite{Agostinelli:2002hh} simulates detectors, including tracking particles through material and material interactions. Since this simulation will include full particle tracking it will be able to accurately track large high energy showers for the study of cosmic rays. By adding an appropriate concrete or dirt model for the floor of MATHUSLA it will also be able to predict MATHUSLA's ability to reject cosmic ray albedo (upward going particles from the downward-going cosmic rays). Further, the addition of support structures will help us estimate and backgrounds due to material scattering.

\subsection{Muons from the HL-LHC}
\label{s.muonBG}

In this section we present calculations of HL-LHC muon backgrounds in MATHUSLA using full Geant4 simulations of muon propagation in the rock between the IP and the surface. This represents a significant update on the initial estimates of~\cite{Chou:2016lxi}, which overestimated muon backgrounds using pessimistic analytical approximations of the muon penetration probability. 

The results  are summarized in Table.~\ref{t.muonBG}. 
Since they pertain only to rates of processes within the decay volume, they are independent of the details of the detector technology.
We find that fake DV rates from rare muon decays or inelastic scattering within the MATHUSLA air-filled decay volume are very small, contributing at most one expected event over the HL-LHC run for both MATHUSLA100 and MATHUSLA200. 
Depending on the detector design, inelastic scattering off the MATHUSLA support structure could yield $\mathcal{O}(10^2)$ fake DVs, but these can be rejected by applying a material veto from the position of the reconstructed DV within the support structure, with the main detector muon trigger providing additional handles.
Finally, passing muons can liberate relatively high-energy electrons from atoms (delta rays) in the air-filled decay volume, which can give rise to a DV. However, for electrons that are sufficiently energetic to leave a track, the opening angle is very small, and this background can be vetoed by requiring two-pronged DVs to have an opening angle greater than $2^\circ$.

None of the backgrounds we studied require a hermetic  veto for rejection. However, the cuts required to reject delta ray background may reduce MATHUSLA's sensitivity to very high-boost LLPs that decay to few final states. This is not MATHUSLA's primary physics target, but further investigation is needed to determine whether it justifies the inclusion of the hermetic veto.
Furthermore, while we expect their contributions to be small, it is possible that muon inelastic scatterings in the rock close to MATHUSLA, or combinatoric superposition with mis-measured cosmic ray tracks, may contribute to rare fake DVs, which could be countered with the hermetic veto or at least a layer of detector in the floor. We briefly discuss this in Section~\ref{s.otherBG}, but leave these processes for future study.

\subsubsection{Strategy}

For the MATHUSLA100 (MATHUSLA200) geometry, the distance a muon has to travel through the rock to reach the decay volume is between 142 and 229 (332) meters, depending on the direction. 
The minimum muon energy required to reach the decay volume with any significant probability is $\gtrsim 70 \gev$. 
The largest contributor to the production of such high-energy muons at the HL-LHC are DY processes $p p \to W \to \mu \nu_\mu$ and $p p \to Z \to \mu \mu$, which have total production cross section $\sim 10\mathrm{nb}$, dominated by muons below this energy threshold. 
Once a muon penetrates the surface and passes through MATHUSLA, its probability of either scattering or decaying is entirely determined by its kinetic energy (or boost) on the surface, and the length it traverses inside the decay volume. 
This suggests separating the muon background calculation into two parts: one to compute the muon probability of traversing various lengths of rock, and one to convolve this information with the kinematic distribution of produced muons at the IP, and the geometry of the decay volume. 
We outline each of these parts in more detail below.

\subsubsection{Simulating Muon Propagation to the Surface}

We assume 100 meters of rock in the vertical direction from the IP to the surface, ignoring the material of the main detector. We assume the rock is a composition of sandstone, marl (calcium and clay), and marl and quartz in fractions approximately consistent with survey data from ATLAS.

%
%
Passage of muons in various directions towards the surface, and with various initial kinetic energies $E_\mathrm{init}$ , is simulated using Geant4 10.4 \cite{Agostinelli:2002hh}. 
%
For a given direction of travel, the distance of rock $L_\mathrm{rock}$ traversed by the muon on the way to the surface, as well as the fractions of that path taken up by the various layers of different material is uniquely determined. 
The only relevant variables for muon propagation in the rock are therefore $E_\mathrm{init}$ and $L_\mathrm{rock}$. 

\begin{figure}
\begin{center}
\begin{tabular}{rr}
\includegraphics[height=6mm]{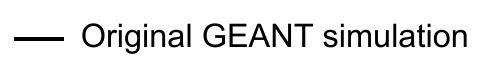}
&
\includegraphics[height=6mm]{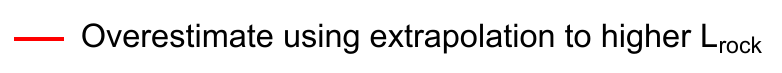}
\\
\includegraphics[width=0.45\textwidth]{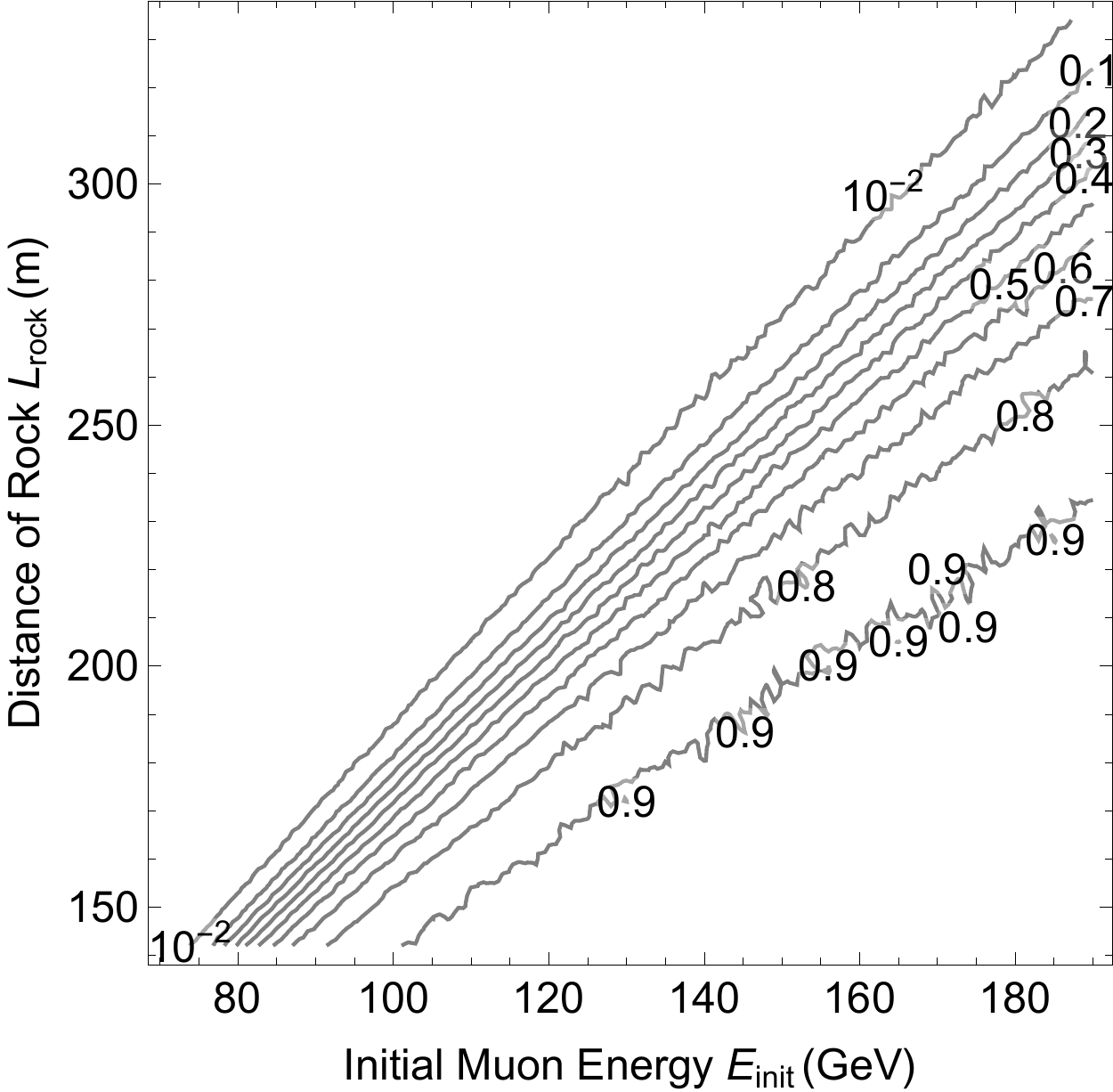}
&
\includegraphics[width=0.45\textwidth]{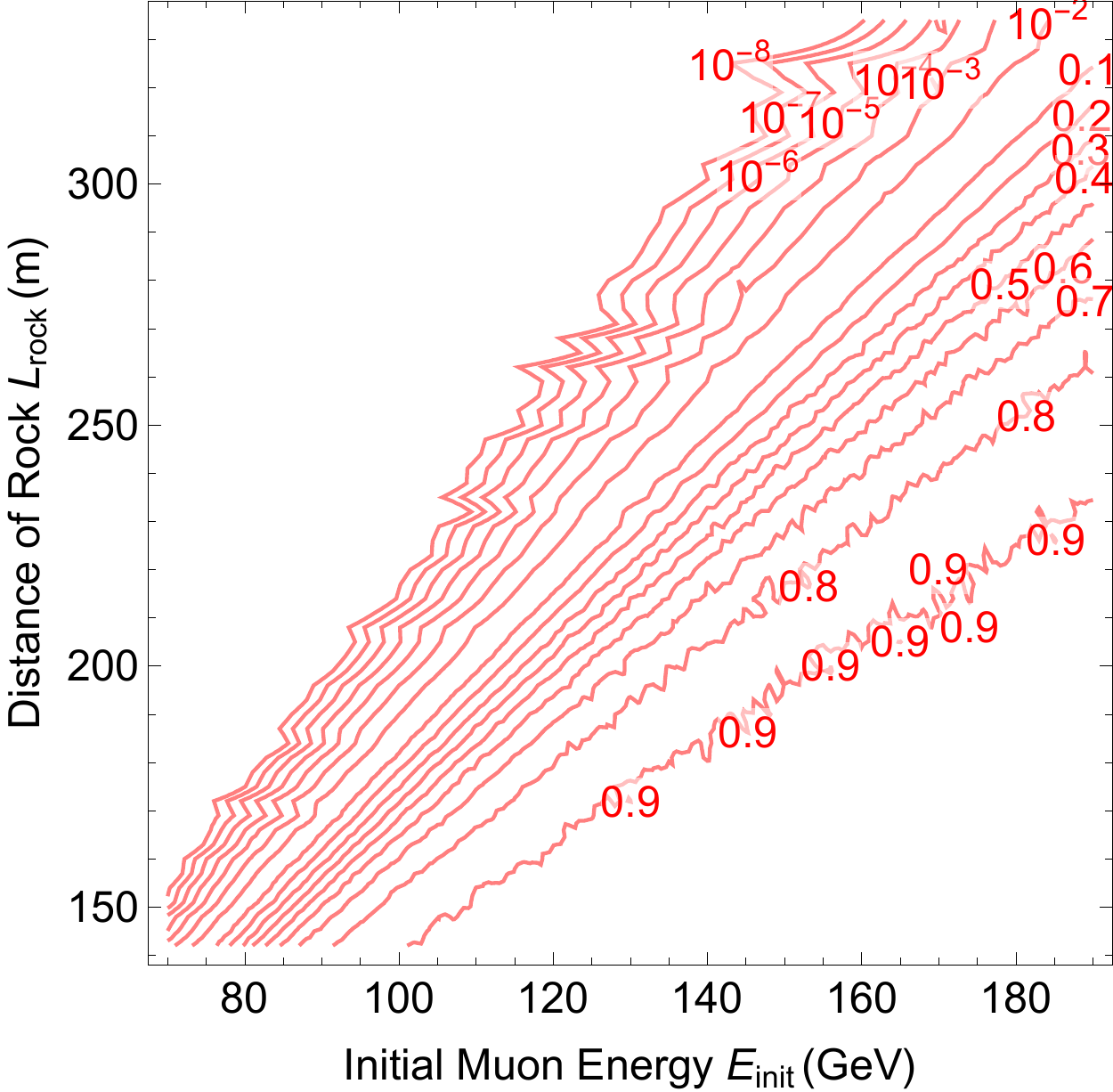}
\end{tabular}
\end{center}
\caption{
Probability $P(E_\mathrm{init}, L_\mathrm{rock})$ of a muon reaching MATHUSLA as a function of initial kinetic energy and distance traveled through rock. Left: original Geant4 simulation. Right: overestimate using extrapolation to higher $L_\mathrm{rock}$ than what is covered using Monte Carlo statistics of the Geant4 simulations.
}
\label{f.muonmap}
\end{figure}

\begin{figure}
\begin{center}
\hspace*{-14mm}
\begin{tabular}{ccc}
\includegraphics[width=0.35\textwidth]{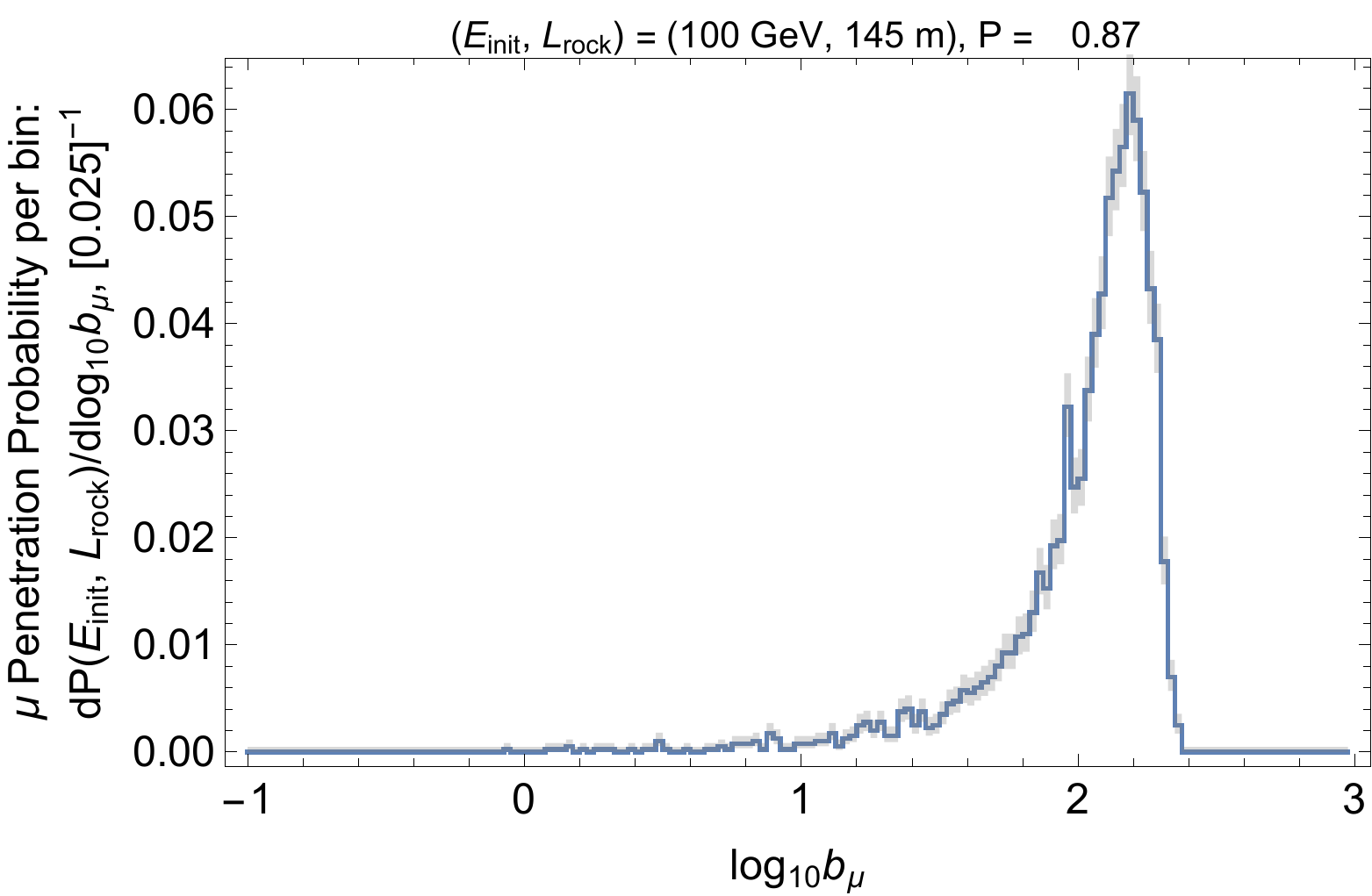}
&
\includegraphics[width=0.35\textwidth]{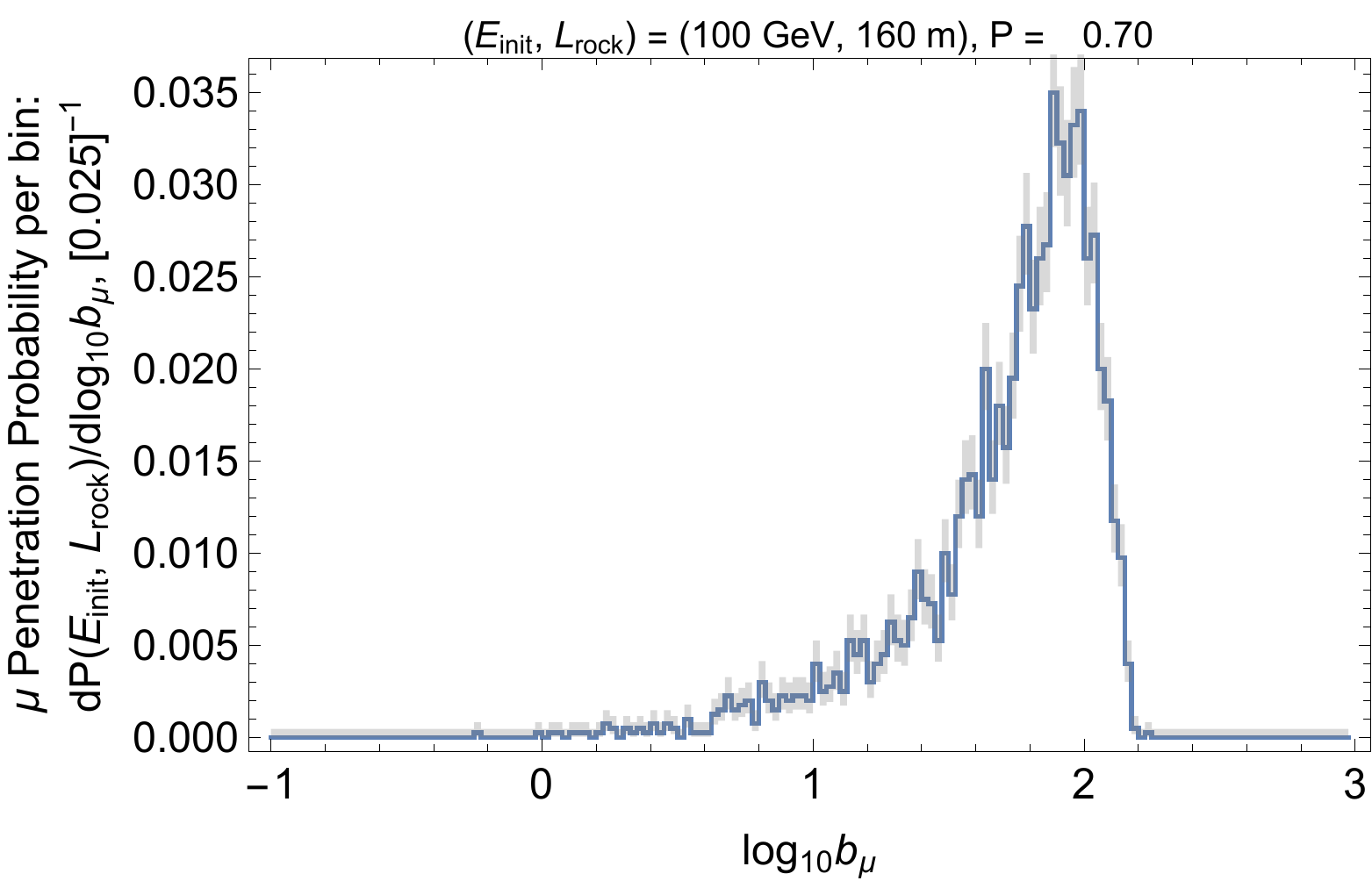}
&
\includegraphics[width=0.35\textwidth]{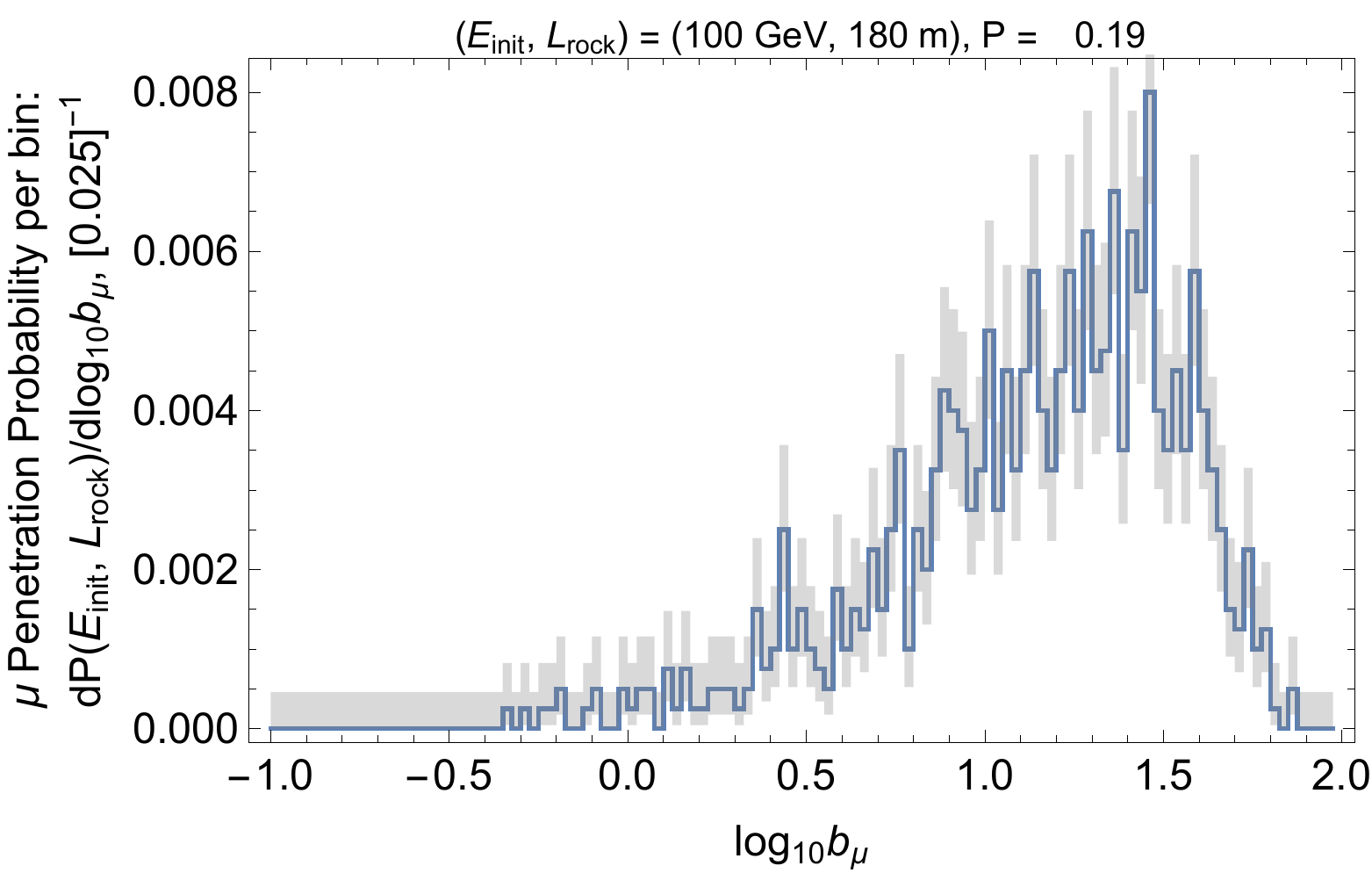}
\end{tabular}
\end{center}
\caption{
Spectrum of kinetic energies for muons that reach MATHUSLA, in units of log muon boost $\log_{10} b_\mu$, with $b_\mu = |\vec p_\mu|/m_\mu$, for initial energy $E_\mathrm{init} = 100 \gev$ and 145, 160 and 180 m of rock. Total penetration probability is indicated by $P$ at top of each plot, gray shading indicates statistical uncertainty of the Geant4 simulation. 
}
\label{f.surfacemuonboostspectra}
\end{figure}

We define the total probability for a muon to reach the MATHUSLA decay volume as $P(E_\mathrm{init},$ $L_\mathrm{rock})$, which satisfies
\begin{equation}
P(E_\mathrm{init}, L_\mathrm{rock}) = 
\int d E_\mathrm{final} \frac{d P(E_\mathrm{init}, L_\mathrm{rock}, E_\mathrm{final}) }{d E_\mathrm{final}} < 1 \ .
\end{equation}
The quantity inside the integral is simply the spectrum of possible muon kinetic energies ($E_\mathrm{final}$) on the surface, normalized to the total penetration probability, for a given initial kinetic energy $E_\mathrm{init}$ and traversed rock distance $L_\mathrm{rock}$. 
We compute $d P(E_\mathrm{init}, L_\mathrm{rock}, E_\mathrm{final})/d E_\mathrm{final}$ for a series of initial kinetic energies $E_\mathrm{init} \in (E_\mathrm{init}^\mathrm{min}, E_\mathrm{init}^\mathrm{max}) = (70, 190)$~GeV and rock distances $L_\mathrm{rock} \in (140, 330)$~m. 
The resulting penetration probability $P(E_\mathrm{init}, L_\mathrm{rock})$ is shown in Fig.~\ref{f.muonmap} (left).
Surface muon spectra, in units of muon boost $b_\mu = |\vec p_\mu|/m_\mu$ for initial energy $E_\mathrm{init} = 100 \gev$ are shown in Fig.~\ref{f.surfacemuonboostspectra}.

\subsubsection{Computing Muon Background Rates in MATHUSLA}

Computing the rates of muon decays and muon scatterings in the MATHUSLA decay volume is now straightforward.
Convolving the kinematic distribution of of muon production DY events, simulated in Madgraph5 \cite{Alwall:2011uj} at parton-level, with the geometry of the detector yields a differential $d\sigma/d E_\mathrm{init} d L_\mathrm{rock} d L_\mathrm{detector}$ distribution, where $L_\mathrm{detector}$ is the length of the muon trajectory in the decay volume. Convolving this distribution with $d P(E_\mathrm{init}, L_\mathrm{rock}, E_\mathrm{final})/d E_\mathrm{final}$ yields a differential distribution of just $d \sigma / dE_\mathrm{final} d L_\mathrm{detector}$ which can be used to directly compute all scattering and decay rates in the detector volume.


Before we present our results, we must discuss two important limitations of our Geant4 dataset that forms the basis of our estimate of $d P(E_\mathrm{init}, L_\mathrm{rock}, E_\mathrm{final})/d E_\mathrm{final}$, and how we can address them to obtain reasonable estimates of muon backgrounds.

First, the dataset only extends to a maximum initial kinetic energy $E_\mathrm{init}^\mathrm{max} = 190 \gev$. To estimate the effect of high energy muon tails on a given muon background $X$, we first treat all $E_\mathrm{init} > E_\mathrm{init}^\mathrm{max}$ as through they are equal to $E_\mathrm{init}^\mathrm{max}$ to capture some of their contributions and obtain an estimate of $N_X$ observed events. We then compute the same background by omitting all muons with $E_\mathrm{init} - 5 \gev$ to obtain a new estimate $N_{X, \mathrm{trunc}} < N_X$ with a truncated high energy muon tail. The difference between these two figures indicates the importance of the high energy muon tail, and we obtain 
\begin{equation}
N_{X, \mathrm{high-E max}} \equiv N_X + (N_X - N_{X, \mathrm{trunc}})
\end{equation}
as an estimated upper bound on the muon background from the estimated importance of high energy muons.

\begin{figure}
\begin{center}
\begin{tabular}{cc}
\includegraphics[width=0.35\textwidth]{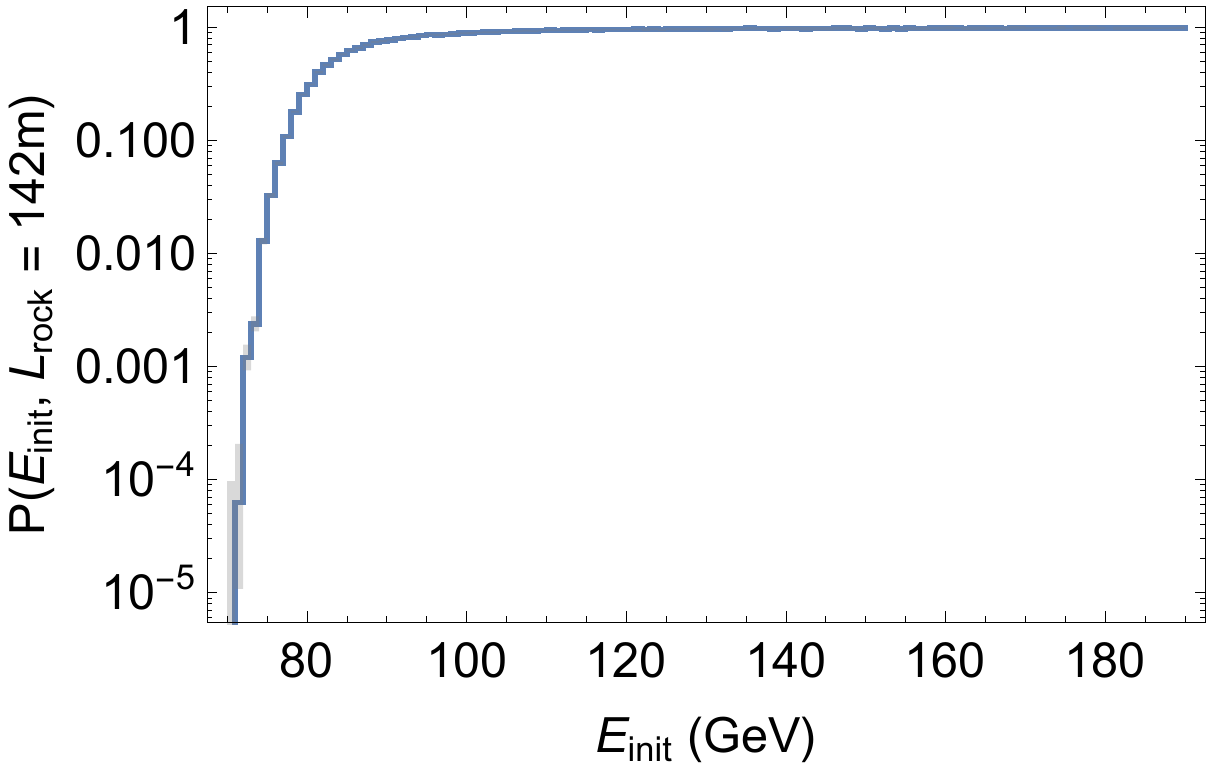}
&
\includegraphics[width=0.35\textwidth]{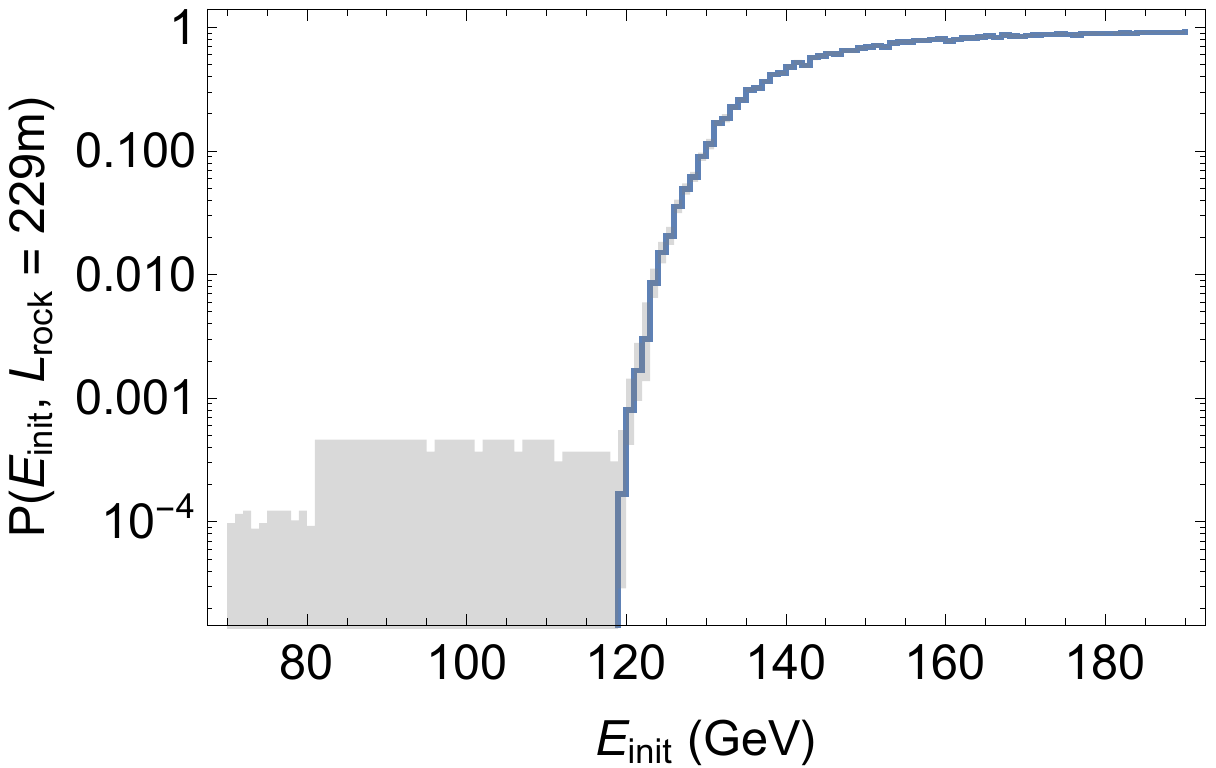}
\end{tabular}
\end{center}
\caption{
Total penetration probability of muons for fixed $L_\mathrm{rock} = 142$m (left) and 229 m (right) as a function of initial kinetic energy.
}
\label{f.penetrationprobfixedLrock}
\end{figure}

The first limitation will be addressed in future studies to extend the Geant4 dataset to higher energies, but the second limitation is more difficult to circumvent. 
Beyond some threshold of low $E_\mathrm{init}$ or high $L_\mathrm{rock}$, the muon penetration probability is extremely small, but it is not zero. This is very difficult to capture with Monte Carlo simulations. 
However, low energy muons dominate the total muon production at the HL-LHC, and since those muons which do make it to the surface are likely to have very low boost, they could contribute significantly to muon decay signatures. 
To estimate the contribution of these lower-energy muons, we extrapolate 
$P(E_\mathrm{init}, L_\mathrm{rock})$ obtained from the Geant4 dataset to lower energies / larger distances.
As shown in Fig.~\ref{f.penetrationprobfixedLrock}, 
for fixed $L_\mathrm{rock}$ and $P(E_\mathrm{init}, L_\mathrm{rock}) \lesssim 0.3$, the penetration probability drops very steeply. The shape of this low-energy tail can be captured accurately with a quadratic fit as a function of $\log_{10} E_\mathrm{init}$, where the fit region is varied for each value of $L_\mathrm{rock}$ to maximize the goodness-of-fit at low energies. We then vary all fit parameters within 1-sigma to obtain an \emph{overestimate} of the muon penetration function in the fit region. The resulting total penetration probability is shown in Fig.~\ref{f.surfacemuonboostspectra} (right).\footnote{This fit generally behaves sensibly but produces a few slightly spurious features, since in Fig.~\ref{f.surfacemuonboostspectra} (right) the penetration probability does not monotonically decrease with inreasing distance everywhere. However, the effect of these low-energy muons is so small that this will not affect our results.}

To probe the possible effect of low-energy muons, we also need some prediction for their kinetic energy distribution at the surface. For an initial estimate, we simply set the muon spectrum at the surface to $d P(E_\mathrm{init}, L_\mathrm{rock}, E_\mathrm{final}) = \delta(E_\mathrm{final} - E_\mathrm{final}^\mathrm{trial})$ when using the extrapolated fit of $ P(E_\mathrm{init}, L_\mathrm{rock})$. The surface kinetic trial energy $E_\mathrm{final}^\mathrm{trial}$ is varied between a boost of $10^{-3}$ (certain decay within MATHUSLA) and $10^2$ (maximizes scattering) to maximize the background $X$ that is being calculated.
This results in an upper bound on the low energy muon contribution to a given background $X$, resulting in estimates $N_{X, \mathrm{low-E}}$ and $N_{X, \mathrm{low-E}, \mathrm{high-Emax}}$, where the latter is computed as above to estimate the effect of the high energy muon tail.

These procedures allow us to set a range for the rate of a given background $X$, defined by the minimum/maximum of the values 
$N_X, N_{X, \mathrm{high-E max}}, N_{X, \mathrm{low-E}},  N_{X, \mathrm{low-E}, \mathrm{high-Emax}}$.



\begin{table}
\begin{center}
\hspace*{-11.5mm}
\begin{tabular}{|m{3.3cm}||m{3.2cm}|m{3.2cm}||m{6cm}|}
\hline
Process & MATHUSLA100 & MATHUSLA200 & Comments
\\
\hline \hline \hline
Upwards-traversing $\mu$ in decay volume
& 
$(2.0 - 2.5) \times 10^6$
&
$(3.1 - 4.0) \times 10^6$
&
Does not satisfy DV signal requirements.
Corresponds to rate of $\mathcal{O}(1)$ per minute.
\\ 
\hline
$\mu \to e \nu \nu$
&
$(3.0 - 3.2) \times 10^3$
&
$(5.5 - 6.8) \times 10^3$
&
Does not satisfy DV signal requirements.
Boost of decaying $\mu$ is 1-100 in decay volume, so $e$ may be detectably deflected off muon trajector.
\\
\hline \hline \hline
$\mu \to e e e \nu \nu$
& 
0.10 - 0.11
&
0.19 - 0.23
&
Genuine background to LLP DV search, but low expected rate can be vetoed, see caption.
\\
\hline
Inelastic Scattering (air in decay volume)
&
$0.3 - 0.6$
&
0.8 - 1.1
&
Ditto.
\\
\hline
Inelastic Scattering (support structure)
&
$(200-350)\times\left[\frac{\xi_\mathrm{iron}}{10\%}\right]$
&
$(490 - 680)\times\left[\frac{\xi_\mathrm{iron}}{10\%}\right]$
&
Can be partially vetoed, see caption. Can also be effectively rejected by applying material veto from position of reconstructed DV within support structure.
\\
\hline
Delta Rays \phantom{blablaaa} ($\mu$ liberating atomic $e$ with $E_e > 1 \gev$ in decay volume)
&120 - 160
&210 - 310
&
Can be partially vetoed, see caption.
Can be vetoed by requiring two-pronged DVs to have opening angle $\theta > 2^\circ$.
\\
\hline
\end{tabular}
\end{center}
\caption{
Summary of various signatures in MATHUSLA due to high-energy muons produced at the HL-LHC IP. 
Columns 2 and 3 show the number of expected events corresponding to 3000~$\mathrm{fb}^{-1}$ for the two different detector geometries.
Processes in rows 1 and 2 do not satisfy the LLP DV signal requirements, while rows 3-6 could naively fake a genuine LLP decay.
%
Additional rejection power would be possible with a hermetic veto around the decay volume, but all of these backgrounds can be rejected via other means. 
}
\label{t.muonBG}
\end{table}

Our results are summarized in Table.~\ref{t.muonBG}. Muon flux through MATHUSLA is much lower than first estimated in~\cite{Chou:2016lxi}, since the original estimate made several analytical approximations to set a conservative upper bound. The rate through the detector is only a few events per minute. Almost the entire uncertainty in this figure comes from the high energy muon tail and will be reduced in future estimates.

Most muons decay to a single electron and two neutrinos. The total number of those decays in the decay volume is a few thousand. The decaying muons have boost ranging from 1-100, meaning some decays will yield a single charged track significantly deflected from the initial muon trajectory, but since this does not reconstruct a DV, it does not as such constitute a background. 
Most of the uncertainty in the decay estimate comes from the low-energy muon tail that we estimate using the above extrapolation procedure, since slow muons have a larger chance of decaying in the detector. Even so, this uncertainty is modest, less than 10\%. 

The rare muon decay $\mu \to (3e) + (\nu_\mu \nu_e)$ has branching ratio $\approx 3.4 \times 10^{-5}$ \cite{Patrignani:2016xqp} and could in principle constitute a DV background if it was not vetoed. However, the rate of this background is very small, much less than one event over the HL-LHC run, and if it does occur it can be vetoed by correlating with the main detector muon trigger. 

Muons can scatter inelastically off air or support structure in the decay volume, giving rise to several charged tracks that might reconstruct a DV. Using the inelastic scattering cross sections in~\cite{Timashkov:2006kk}, we obtain less than one event in the air of the decay volume over the HL-LHC run. Furthermore, just as for rare muon decays, these scatterings can be vetoed with offline information from the main detector. 

Inelastic scattering off support structure can occur at significant rates. This depends on the final design and construction of the detector, but we can get a feeling for the size of this background by assuming that, on average, all muon trajectories spend fraction $\xi_\mathrm{iron}$ traveling through iron. Even if we make the assumption that $\xi_\mathrm{iron} = 10\%$, this only yields a few hundred events which can be rejected with a material veto due to the good position resolution of the reconstructed DV, with additional rejection possible by correlating with the main detector muon trigger. 
Most of the uncertainty in this estimate comes from the high energy muon tail, since scattering rate increases modestly with muon energy. As a result, refinements in the Geant4 dataset should make future estimates much more precise.

Finally, there is an appreciable rate of delta rays~\cite{Patrignani:2016xqp}. Electrons that are liberated from atoms in the air-filled decay volume by passing muons could give rise to a displaced vertex if the electron has enough kinetic energy $E_e$ to reconstruct a track. Assuming a threshold of $E_e > 1 \gev$, this gives rise to $\mathcal{O}(100)$ DVs in the air-filled decay volume. Again, most of the uncertainty comes from the high-energy muon tail. Fortunately, these scattering events are easily vetoed by requiring two-pronged DVs to have an opening angle larger than $2^\circ$. Reducing the electron energy threshold to 0.5 GeV roughly doubles the estimated delta ray background rate, but has very little effect on the opening angle cut required to veto this background.
Just as for the other scattering backgrounds, additional veto power comes correlating with the the main detector muon system off-line.
Delta rays can therefore be efficiently rejected.

In conclusion, all of the backgrounds from muons we studied can be efficiently rejected and do not form an obstacle to a zero-background LLP search in MATHUSLA. Furthermore, none of these backgrounds require a hermetic scintillator veto, though further study of other, possibly more exotic backgrounds is required to determine whether such a veto is ultimately needed.

\subsection{Neutrinos}
\label{s.neutrinoBG}

Below we reproduce the analytical estimates of neutrino scattering backgrounds in the MATHUSLA decay volume that were first presented in~\cite{Chou:2016lxi}. These estimates, as well as the investigation of rejection strategies, are conservative to overestimate the background rate at MATHUSLA200 (and can be rescaled for MATHUSLA100). A more detailed study using neutrino scattering events simulated with GENIE~\cite{Andreopoulos:2015wxa} is currently in progress to obtain more precise results on neutrino background rates and rejection efficiencies.

\subsubsection{Neutrinos from Cosmic Ray Showers}

The MATHUSLA detector is immersed in a diffuse neutrino flux generated by cosmic ray interactions in the atmosphere~\cite{Daum:1994bf, GonzalezGarcia:2006ay,  Halzen:2013dva}. This flux is isotropic below PeV energies. Atmospheric neutrinos scattering off nuclei in the air-filled decay volume are a potentially dangerous background, since they can result in a topology similar to an LLP decay, see Fig.~\ref{f.bgsummary}~(h). A careful examination of these scattering events indicates that they can be distinguished from LLP decays, by the tracking and time-of-flight information.

We restrict ourselves to neutrino energies greater than $\sim 400 \mev$, for reasons that will become clear below. In this region, the atmospheric muon neutrino flux as measured by Frejus~\cite{Daum:1994bf, Halzen:2013dva} is reasonably well approximated by
\begin{equation}
\frac{d \Phi}{d E_\nu} \sim 0.06 \left( \frac{\gev}{E_\nu}\right)^3 \ \gev^{-1} \mathrm{cm}^{-2} \mathrm{s}^{-1} \mathrm{sr}^{-1}
\end{equation}
(though we use the actual measured neutrino flux spectrum in our calculations). The electron neutrino flux is significantly lower and can be treated similarly, so we neglect it in this simple estimate.
For a given neutrino scattering process $i$ with cross section $\sigma_i$, the number of produced events \emph{per year} can be estimated as
\begin{equation}
N_i =  \int d E_\nu \  \frac{d \sigma_i}{d E_\nu} \ n_\mathrm{target}\  \left[ 4 \pi \frac{d \Phi}{d E_\nu}  \ V_\mathrm{target} \ T\right ]
\end{equation}
where $V_\mathrm{target}$ is the volume of the MATHUSLA detector as per Fig~\ref{f.mathuslageometry}, the exposure time $T$ corresponds to the fraction of the year the HL-LHC beams are running (assumed to be 50\%), and $n_\mathrm{target}$ is the number density of protons and neutrons in air. The various differential neutrino scattering cross sections $d\sigma_i/d E_\nu$ are theoretically and experimentally known at the better than $\sim 30\%$ level and are summarized in \cite{Formaggio:2013kya}. In practice we distinguish between protons and neutrinos, but we do not account for the fact that these are bound in nuclei. (As we note below, this underestimates our ability to reject these backgrounds.) We are interested in processes $i$ with at least two charged particles in the final state, including deep-inelastic scattering (DIS), so they can reconstruct a DV. We then divide those processes into two classes: those which are exclusively defined to contain a proton in the final state (PFS), like many quasi-elastic scattering (QES) processes, and those which are not (like DIS).

The number of scatters per year with protons in the final state is 
\begin{equation}
\sum_{\mathrm{PFS}} N_i \approx 
\left\{
\begin{array}{lll}
60  & \mathrm{at} & \mathrm{MATHUSLA200}\\
15 & \mathrm{at} & \mathrm{MATHUSLA100}
\end{array}
\right.  \ .
\end{equation}
which is  non-negligible. 
In most of these scatterings the final-state proton is measurably non-relativistic, and the final states are contained within a very narrow cone that generally does not point back to the LHC IP. 
A veto of DVs with a narrow final state cone that does not point back to the IP has minimal effect on signal acceptances: light and highly boosted LLPs give a cone which points back at the IP, while somewhat heavier LLPs give final states which are not confined to a single narrow cone. 
To make this more quantitative, assume these processes obey the differential cross section of QES.
%
We also assume, very conservatively, that a proton moving slower than $v_\mathrm{min} = 0.6c$ can be reliably distinguished from an ultra-relativistic muon by the MATHUSLA tracking system. (This corresponds to a 10 ns difference in travel time over 5 meters, which is a reasonable estimate for the depth of the tracker at the top.) Only neutrinos with energy above about 500 MeV can produce such protons (hence we only considered that part of the neutrino spectrum). For protons faster than $v_\mathrm{min}$, $2\to2$ kinematics dictates a maximum cone size for the final states, with a corresponding small probability that this cone's random orientation points back at the LHC IP. The number of scatters per year that passes these cuts and could actually fake a DV signal is
\begin{equation}
\label{e.Naftercuts}
\sum_{ \mathrm{PFS}} N_i^\mathrm{after\ cuts} \approx 
\left\{
\begin{array}{lll}
1  & \mathrm{at} & \mathrm{MATHUSLA200}\\
0.25 & \mathrm{at} & \mathrm{MATHUSLA100}
\end{array}
\right. \ .
\end{equation}
It is also important to point out that this low rate is a significant \emph{overestimate}, since (1)~our assumption that MATHUSLA could only distinguish $0.6c$  from $c$ is very conservative; (2) for non-QES processes the additional produced particle will result in slower final-state protons and/or narrower final-state cones; (3) we assumed that all scatters with protons faster than $v_\mathrm{min}$ had the largest possible final-state cone assuming $v_\mathrm{proton} = v_\mathrm{min}$; and (4) we neglected that the initial-state protons and neutrons are bound in nuclei, and the  slow-moving debris from the breakup of the nucleus could provide additional rejection handles.

We now turn to atmospheric neutrino scatters without a definite proton in the final state, for which the rate per year is
\begin{equation}
\sum_{\mathrm{not \ PFS}} N_i \approx 
\left\{
\begin{array}{lll}
10  & \mathrm{at} & \mathrm{MATHUSLA200}\\
2.5 & \mathrm{at} & \mathrm{MATHUSLA100}
\end{array}
\right.  \ .
\end{equation}
It is difficult to make detailed statements about these events without simulations, but they are dominated by DIS with neutrino energies in the several GeV range. We therefore expect that the final states will be confined in a relatively narrow cone, which only rarely points back to the LHC IP. 
Obviously, more detailed studies are required to make these predictions  robust, but from these estimates we expect that of all the different kinds of atmospheric neutrino scattering events in MATHUSLA can be rejected, with none or very few events surviving as irreducible backgrounds over the entire HL-LHC run. 
Furthermore, it is important to recall that this background, like charged cosmic rays, can be measured in detail when there are no colliding beams in the HL-LHC.

\subsubsection{Neutrinos from HL-LHC Collisions}

Cosmic rays are not the only source of neutrinos that are incident on MATHUSLA. They are also produced at the HL-LHC, see  Fig~\ref{f.bgsummary}~(i).
Direct production of neutrinos in $W$, $Z$, top and bottom quark decays can be estimated in MadGraph 5. The total number of these relatively high-energy neutrinos which scatter off air in MATHUSLA200, computed in a similar manner as for  muons and atmospheric neutrinos above, is less than $0.1$ events over the entire HL-LHC lifetime. 
In addition, there is a low-energy neutrino flux ($E_\nu \lesssim 1 \gev$) from decay of pions and kaons produced in primary and secondary hadron production in the LHC.  We estimate this rate by simulating minimum-bias collisions with Pythia 8~\cite{Sjostrand:2007gs} and propagating the particles through the approximately azimuthally symmetric material budget~\cite{Bayatian:2006zz} of the CMS experiment with Geant4.  
The vast majority of LHC neutrinos are produced by low-energy hadrons decaying in the LHC detectors.
This leads to a steeply falling neutrino spectrum for energies above a few hundred MeV.  These neutrinos have mainly quasi-elastic neutrino-nucleon scatterings in MATHUSLA. Imposing the same proton-time-of-flight cut used to derive \eref{Naftercuts} (but not imposing any geometrical cuts since the final state cone always points back to the IP), we arrive at an (over-) estimate of $\approx 1$ scattering event with protons in the final state passing cuts \emph{over the entire HL-LHC run at MATHUSLA200}. There are also $\approx 3$ scattering events (no cuts) from processes like DIS, which mostly do not have protons in the final state.
(At MATHUSLA100, these rates are reduced by a factor of a few.)
As discussed above, their rejection is harder to estimate, but given these very low rates, we are confident that rejection can be significantly improved with detailed study.

\subsection{Other possible backgrounds}
\label{s.otherBG}

Apart from pure cosmic ray backgrounds, and the relatively straightforward fake DVs from muons or neutrinos, more exotic backgrounds could also contribute to an LLP search in MATHUSLA. 

For example, the combination of a single upwards going muon from the LHC with a single mis-measured cosmic ray track may combine to form a fake DV that satisfies LLP signal requirements. This may seem unlikely, since only about $\sim 10^6$ muons from the HL-LHC travel through MATHUSLA compared to $\sim 10^{15}$ CR charged particles, but it has to be investigated in more detail using background events (see Section~\ref{s.parambackground}).

It is also possible that the large number of cosmic ray charged particles hitting the rock below MATHUSLA could eject the occasional energetic hadron upwards, which then either scatters or decays in the air of the decay volume. This is called albedo and will be studied using Geant4's material model of the earth and possible detector floor.

For example, this could occur due to lepton interactions giving rise to long-lived $K_L^0$ hadrons that can migrate into the detector and decay. 
The rate of these processes can be estimated using the cross sections in \cite{Formaggio:2013kya}, and is expected to be very low. Furthermore, a significant number of the $K_L^0$ will interact in the rock and not make it to the surface.  
However, if the rate of these events is significant, a layer of detector in the floor may be required to veto the charged particles that are produced in association with the neutral long-lived hadrons that are traveling upwards.

Further study is underway to estimate the rates of these processes at MATHUSLA. If their rates are problematic, it is likely that the basic detector design of Fig.~\ref{f.bgsummary} could be extended to improve their rejection, e.g. by adding additional veto layers on the floor of the detector.

\section{MATHUSLA Detector Design Concept}
\label{s.design}

\begin{figure}
\begin{center}
\begin{tabular}{cc}
\includegraphics[width=0.6\textwidth]{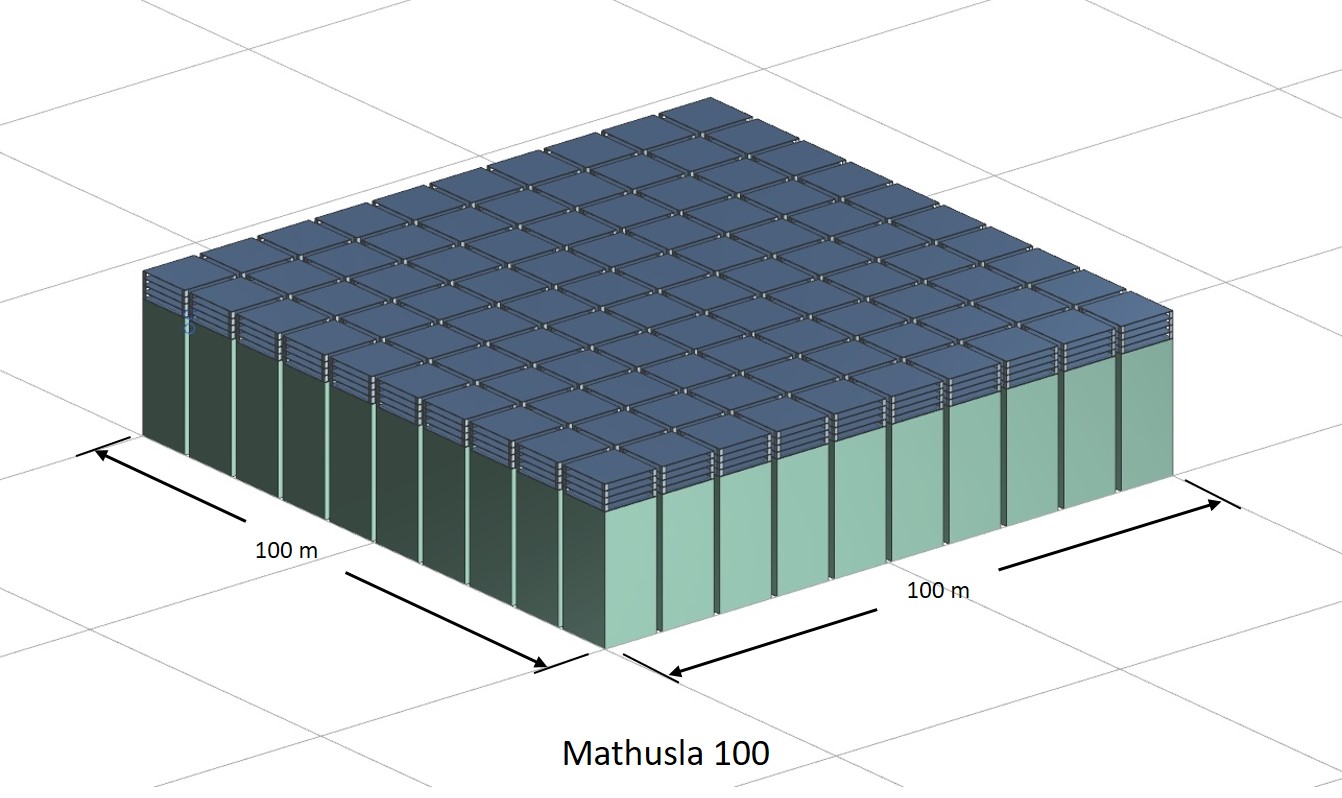}
&
\includegraphics[width=0.4\textwidth]{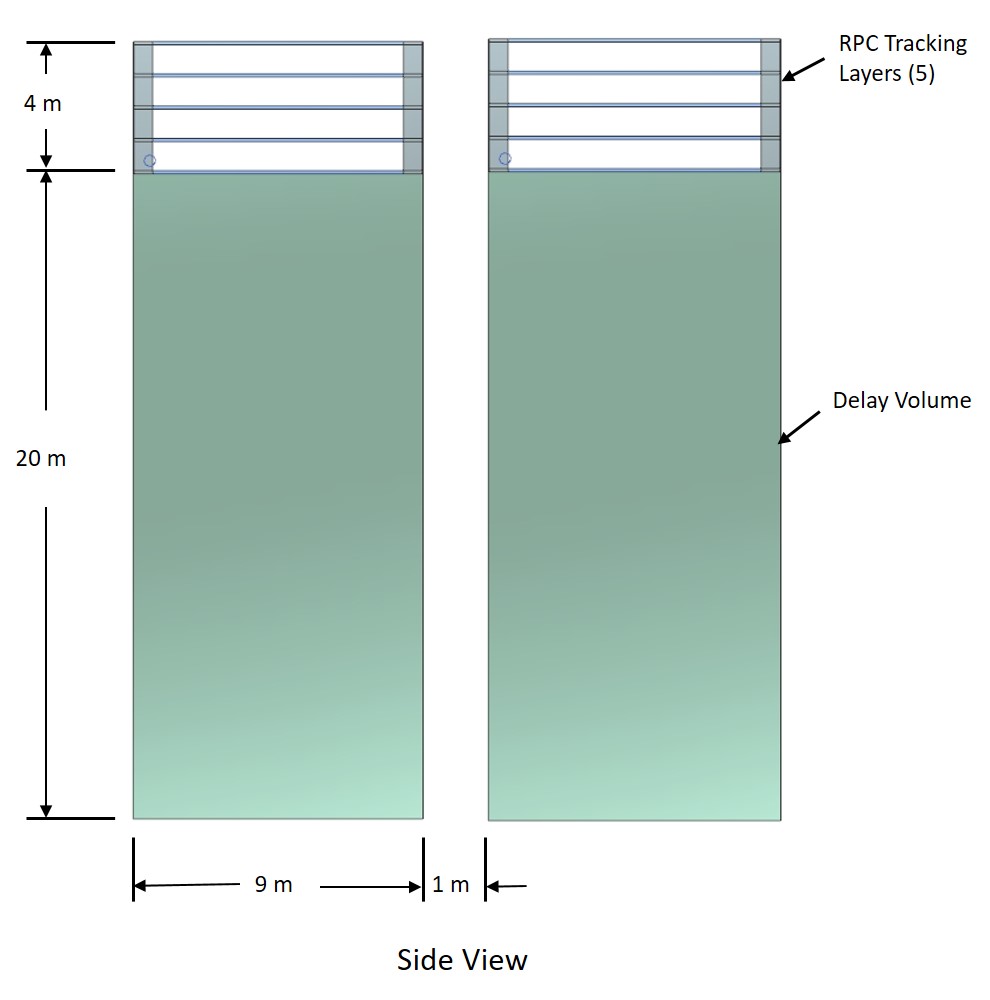}
\end{tabular}
\end{center}
\caption{
Layout of a modular implementation of the MATHUSLA100 geometry considered in this letter, with $10\times10 = 100$ modules, each with area $9 \times 9$ m, situated 1 m apart. The decay volume has a height of 20 m, topped by five layers of RPCs that are 1 m apart.
}
\label{f.mathuslamodular}
\end{figure}

Our current concept to implement the MATHUSLA100 geometry is to deploy an array of identical modules to cover the 100 $\times $100 $m^2$.  This facilitates construction and testing, allows easy adaptation of the layout to the eventual chosen experimental site, and is compatible with staging the full detector. Each module has a 20 m air-filled decay volume follow by 5 five planes of tracking detectors with 1-m separation between tracking layers as illustrated in Fig.~\ref{f.mathuslamodular}.  The total height of the module is 24 m.  The 20 m decay volume provides good sensitivity for LLP decays.  



%
%
The technical requirements of the tracker are modest: $\approx 1$ cm position resolution and $\approx 1$ ns time resolution, which is easily satisfied by many detector technologies. In this document, we focus on one particular option, the Resistive Plate Chamber (RPC), because it has been shown to be robust, economical and scalable. However, other options are not excluded. 

In the following subsections we discuss further details of the detector requirements, such as layout and Triggering/DAQ. We close by presenting some basic triggering and reconstruction efficiency studies for LLP decays, demonstrating the viability of this modular design.


\subsection{Detector requirements and technology}



The charged particle tracking unit located above the decay volume is envisaged to have five measurement planes that provide both coordinate and time information.  The tracking planes have 1-m separation between adjacent planes.  Five planes of tracking provides robust, efficient  tracking for background rejection and decay vertex reconstruction of LLP decays.  

The highest rate expected  background is from cosmic rays at $\mathcal{O}(2)$ MHz over the entire detector surface. They are downward going and can be separated from the upward going signal particles by timing within the main tracker. For example, for a given hit at the bottom of the tracker, the time difference at the top between an upward going signal particle and a downward going cosmic ray background particle is approximately $24$ ns over the 4 meters between the top and bottom tracking planes.  
A tracker time resolution of $\approx 1$ ns will provide efficient rejection of downward traveling cosmic rays.

The present benchmark design can be extended to include detectors at the bottom  and possibly also module walls that face the outside of the whole decay volume in order to veto cosmic rays entering from the sides. This veto could be implemented with scintillators or with additional RPC layers, and would provide additional rejection power for cosmic rays, as well as large-angle charged tracks entering the decay volume from the sides and muons from the LHC. The decision whether to include these extra detector layers will be informed by the outcome of ongoing detailed background studies.
%

A signal event is indicated by two or more upward going tracks that reconstruct to a 4-dimensional (space and time) decay vertex in the decay fiducial volume. 

The MATHUSLA100 detector baseline detector requirements include:
\begin{enumerate}
\item 100 m $\times$ 100 m $\times$ 20 m decay volume. 
\item A 4-m tall charged particle tracker with five detector planes located above the 20 m decay volume. 
\item Tracker hit spatial resolution of $\approx 1$ cm in each transverse direction and time resolution of $\approx 1$ ns.
\item Overall rate capability greater than approximately 2 MHz.
\item  For cosmic-ray shower detection , the tracker needs to be able to cope with hit densities that can exceed 10$^4$ m$^{-2}$ in the shower core.
\end{enumerate}
These technical requirements for the tracker are easily satisfied by many detector technologies. The Resistive Plate Chamber (RPC) has been chosen as our baseline  because of its good timing and position resolution, maturity and relatively low cost per  unit area.  RPCs are used as muon trigger chambers for both the ATLAS  and CMS detectors and they have performed reliably and with good efficiency, about 98\%. 

The construction procedure is well defined and has been industrialized by the Rome, Tor Vergata group resulting in a unit cost significantly less than  alternatives such as drift chambers.  Groups in China have also made RPC chambers using both glass and phenolic plates.  Both the Rome, Tor Vergata group and two Chinese groups with RPC expertise have signed this LoI. The area of RPC required for MATHUSLA is an order of magnitude larger than that of ATLAS and CMS RPC detectors combined. We recognize the challenge of scaling-up production, and the importance of having sufficient production sites involved.  We also view this as an opportunity to invest in design changes that will reduce the unit cost. 
The veto detectors on the sides and bottom of the decay volume, if needed, have looser performance requirements. While they have less area than the tracking planes combined, they still have to cover large areas and cost will be an important consideration. The current plan is to utilize the same RPC technology and design as the main tracker. 

\subsection{RPC Costs}

We do not have a final, complete design for the RPC units. The goal is to optimize the individual RPC performance-to-cost ratio. The Rome group is considering a design where each chamber has a single read-out panel with strips 3.2 m long and 25 mm pitch that gives the transverse coordinate.  They propose reading out the strips at both ends to obtain the longitudinal coordinate. The  gas volume is 3.20 $\times$ 0.80 m$^2$ with a 1-mm gap between two resistive electrode plates.  The groups in China are also investigating the design of large chambers.  

Because the design is evolving we do not have a complete cost estimate, but considering experience with construction of RPCs in Rome and  China we estimate a range of 150 to 260 Euro per square meter excluding electronics, gas, HV and LV supplies. 

\subsection{Layout}

As shown in Fig.~\ref{f.mathuslamodular}, the 100 m $\times$ 100 m area of the MATHUSLA100 detector is divided into a number of identical  modules.  Each modular is made weather-tight to avoid the need to build the enormous experimental hall. This approach means that each module can be fully tested and commissioned as it is assembled and that module installation and commissioning can be staged over several months or a couple of years if necessary.  The modular approach allows for beginning data taking with a partial detector.   


Each module will have the same geometry as MATHUSLA in the vertical direction, namely a 4 m thick tracker on top of a 20 m tall decay volume and a bottom veto, if needed. The transverse dimensions of a module are taken to be 9 m x 9 m here, with 1-m inter-module gaps. However, we emphasize that the module size, gap between modules, and possibly even gap between RPCs is subject to optimization as well as engineering constraints.

Individual modules will be completely self-contained and weather tight  except for power, RPC gases and other services. For example, each module will have its own HV power supply for the RPC, as well as temperature and pressure monitoring to dynamically adjust the HV setting to compensate for environmental changes. Each module will have its own DAQ subsystem servicing the local detectors. Information will be exchanged with nearby modules via optical links.  

\subsection{Triggering and DAQ}


Trigger information will be provided by the RPC chambers in the main tracker. A local trigger processor in each detector module aggregates the information from its own chambers as well as those in its nearest neighboring modules to look for hits within a time window that are spatially aligned. Each of these trigger primitives is classified as either upward going or downward going based on the RPC hit times.
	
Trigger primitives are then transmitted to a global trigger processor to form an overall MATHUSLA event trigger. All upward going triggers will be accepted while only certain downward going triggers will be kept, e.g. a high multiplicity indicating an air shower. Additional triggers are anticipated for calibration and alignment purposes. These steps are similar to what has been done in experiments such as ATLAS and CMS, and will benefit from their experience. Minor modifications will be needed since the LLP decay vertex, which plays the role of the LHC interaction point for ATLAS and CMS, is not known a priori. Given the size of the MATHUSLA detector, the distribution of a stable timing signal will require additional care. We will leverage the developments in accelerator facilities where tighter timing requirements and greater distances than MATHUSLA are already in use. 

The overall MATHUSLA event trigger will be distributed back to all the modules to initiate data acquisition.






\subsection{Efficiency Studies}
\label{s.efficiencystudies}

In order to demonstrate the viability of the modular detector design and the nearest-neighbor trigger strategy described in this section, we conducted a series of simple efficiency studies. 
Our results show that for hadronic LLP decays, the modular MATHUSLA100 implementation of  Fig.~\ref{f.mathuslamodular} should very closely approach the idealized, purely geometrical sensitivity estimates of Section~\ref{s.LLPphysics}.


We study triggering and reconstruction efficiency for a benchmark signal model where LLPs $X$ are produced in exotic Higgs decays $H \to XX$ and either decay hadronically $X \to b \bar b$ or leptonically $X \to \mu^+ \mu^-$. In the former, there are $\mathcal{O}(10)$ charged final states, which greatly aids in triggering and reconstruction compared to the two tracks of the former case.
Higgs production via gluon fusion and vector boson fusion is simulated in MadGraph and showered in Pythia.


We assume each of the 5 RPC layers has 100\% efficiency for detecting a single hit, and assume that any charged particle that crosses 5 layers can be perfectly reconstructed as a single track.
The track reconstruction efficiency is dependent on the precise RPC detector parameters, but will have to be very high to efficiently reject cosmic ray background. It should therefore not be the dominant factor in determining the overall LLP decay reconstruction efficiency. 
We also do not take the effects of finite spatial resolution of the RPC layers into account, but for resolutions in the cm range this is not a limiting factor for the LLP masses we study here.
We leave these issues for future detailed study.


In order for an LLP decay to pass the nearest-neighbor trigger defined above, at least one of its charged particle final states must cross 5 RPC layers within a $3\times3$ sub-grid of modules centered on the module where the particle crossed the lowest RPC layer. It is therefore possible for a charged track to travel between modules and only hit 4 or fewer layers, due to the finite gap between modules and between RPC layers.  In that case, that charged particle does not pass the trigger requirement. 
Once an LLP decay passes the trigger, we assume the whole detector is read out. In that case, we assume the LLP decay is fully reconstructed if it decays inside the modules' detector volume, and at least two of its charged particle final states pass through 5 RPC layers anywhere in the detector.

To study trigger and reconstruction efficiencies, we force LLPs that reach the surface to decay at a fixed height $y_\mathrm{decay}$ above the ground. The trigger efficiency $\epsilon_\mathrm{trigger}$ (reconstruction efficiency $\epsilon_\mathrm{DV}$) is defined as the number of LLP decays which pass the nearest-neighbor trigger requirement (and also pass the reconstruction requirement), normalized to the total number of LLPs that were forced to decay at height $y_\mathrm{decay}$ \emph{inside the detector volume}. 
The distribution of LLP decays within a given decay plane is determined by the rapidity distribution and hence the production mode of the LLP. The efficiencies we derive here are therefore applicable to LLPs produced in exotic Higgs decays, but they should be close to the efficiency for other processes as well. We will investigate additional benchmark models in future studies.




The efficiencies will depend on the gaps between modules and RPC layers. To investigate the importance of these physical parameters of the detector for signal reconstruction, we examine not only the geometry of Fig.~\ref{f.mathuslamodular}, but also modified geometries with smaller gap sizes or larger module sizes. These are summarized in Table.~\ref{t.modulegeometries}. 
(Note that in particular the RPC spacing will be restricted by the requirements of four-dimensional tracking, but depending on the time resolution, a gap size smaller than 1 m may be possible, so we investigate this degree of freedom	 here.)


\begin{table}
\begin{center}
\begin{tabular}{|l||l|l|l|}
\hline
Description & Module Size (m) & Module Gap (m) & RPC Gap (m) 
\\
\hline \hline
Default & 9 & 1 & 1 \\
\hline \hline
Smaller module gap & 9.5 & 0.5 & 1 \\
\hline 
Smaller RPC gap & 9 & 1 & 0.5 \\
\hline
Larger modules & 19 & 1 & 1 \\
\hline \hline
Monolithic detector & 100 & 0 & 1 \\
\hline
\end{tabular}
\end{center}
\caption{
Square module geometries of the MATHUSLA100 detector design for which we study triggering and reconstruction efficiencies of leptonic and hadronic LLP decays. The first geometry is the default benchmark of Fig.~\ref{f.mathuslamodular}. The four variations are considered to examine the impact of module size, module gap and gap between the five RPC layers. The lowest RPC layer is situated at a height of 20 m for all geometries.
}
\label{t.modulegeometries}
\end{table}



\begin{figure}[t]
\begin{center}
\hspace*{-12mm}
\begin{tabular}{m{0.4\textwidth}m{0.4\textwidth}m{0.25\textwidth}}
\includegraphics[width=0.4\textwidth]{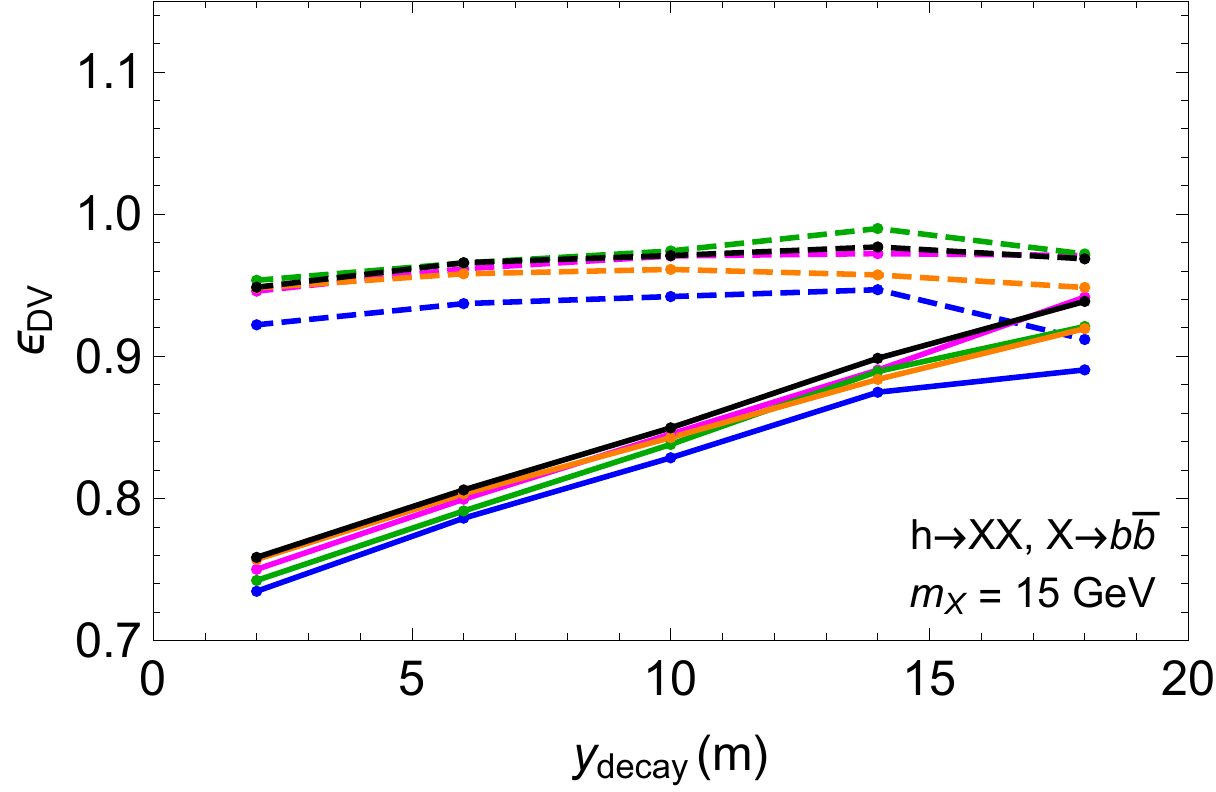}
&
\includegraphics[width=0.4\textwidth]{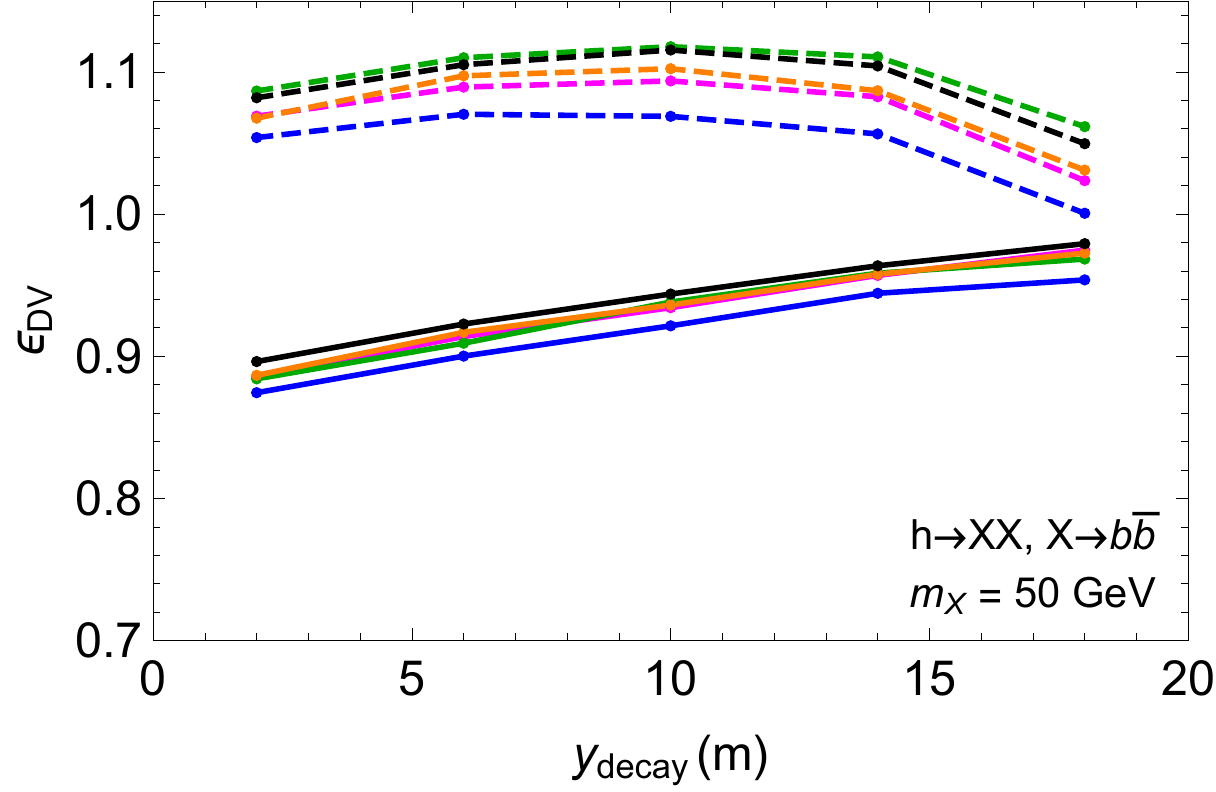}
&
\includegraphics[width=0.25\textwidth]{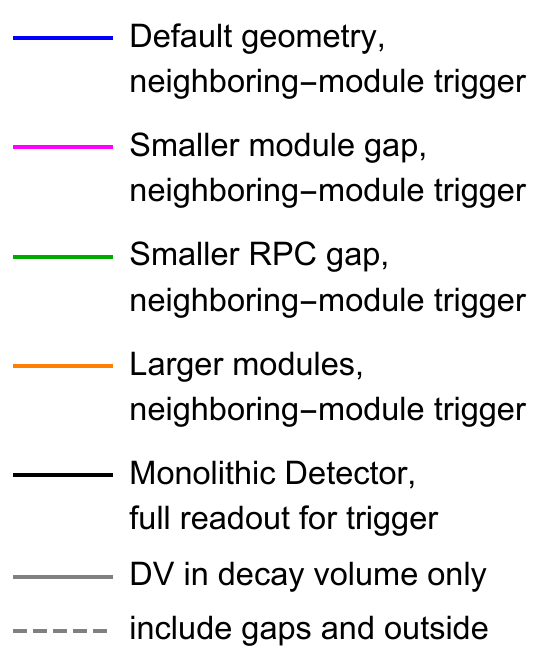}
\end{tabular}
\end{center}
\caption{
Reconstruction efficiencies for LLPs decaying hadronically a height $y_\mathrm{decay}$ above the ground in MATHUSLA100. The blue curve corresponds to the modular design of Fig.~\ref{f.mathuslamodular}. The magenta, green and orange curves show the effect of changing the gap size between modules or RPC layers, or the module size, see Table.~\ref{t.modulegeometries}. The black curve represents a non-modular monolithic detector that can trigger with full readout, representing the theoretical maximum efficiency for this detector geometry with 5 RPC layers separated by 1 m.}
\label{f.effbb}
\end{figure}

\begin{figure}[t]
\begin{center}
\hspace*{-12mm}
\begin{tabular}{m{0.4\textwidth}m{0.4\textwidth}m{0.25\textwidth}}
\includegraphics[width=0.4\textwidth]{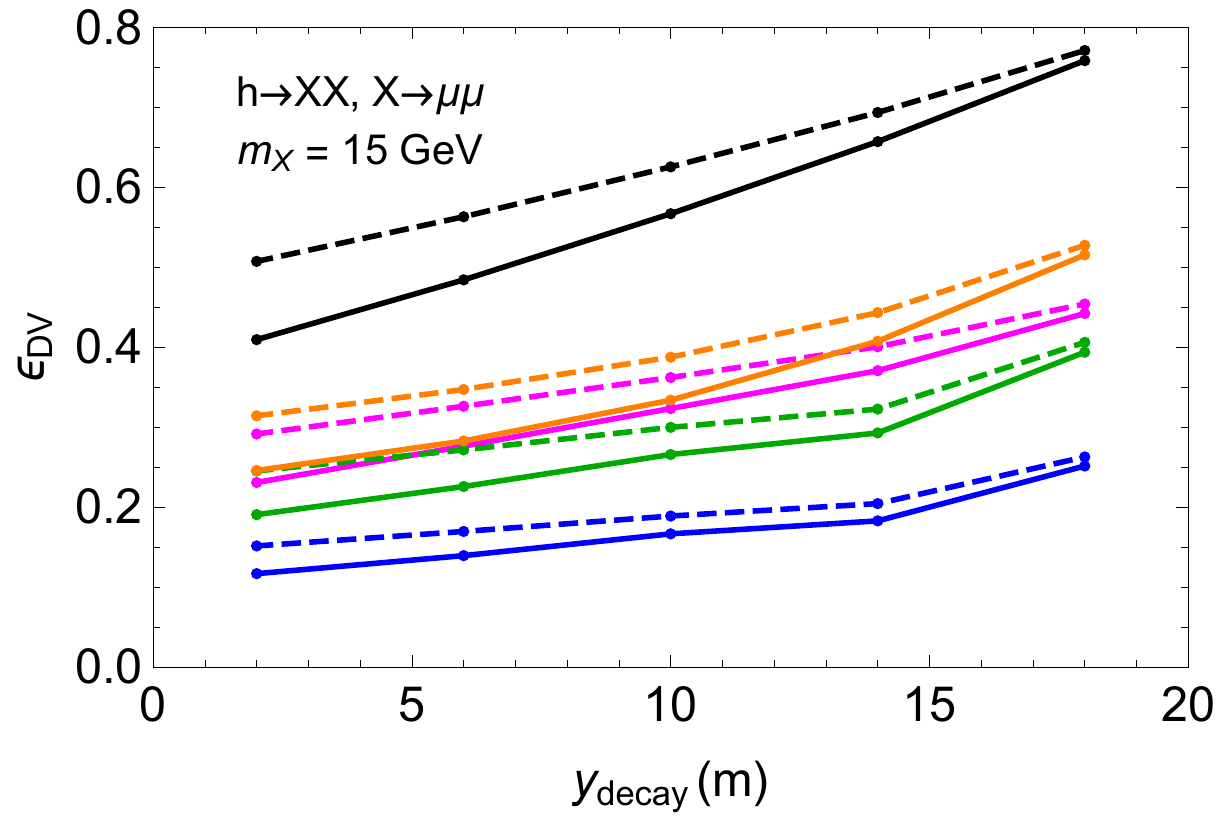}
&
\includegraphics[width=0.4\textwidth]{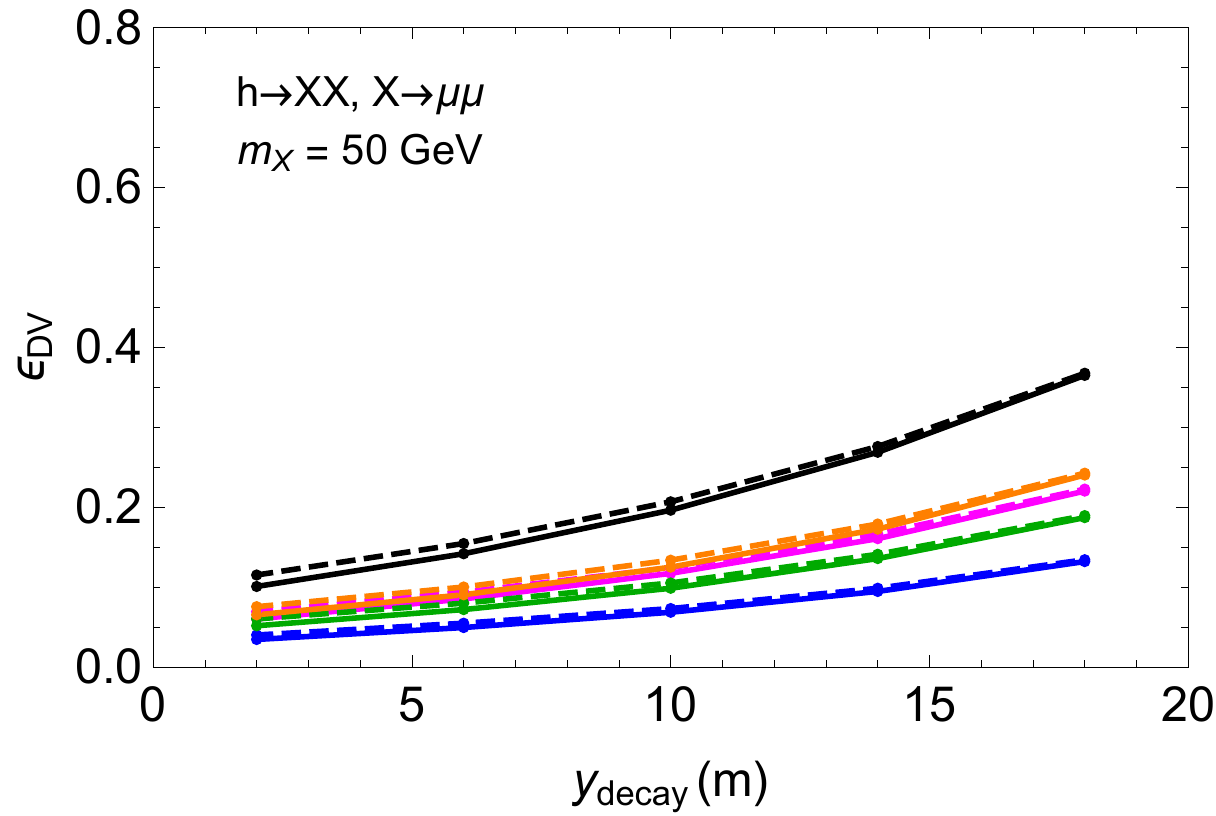}
&
\includegraphics[width=0.25\textwidth]{tightvslooseplot_allgap_legend2}
\end{tabular}
\end{center}
\caption{
Same as Fig.~\ref{f.effbb} but for leptonic LLP decays. 
}
\label{f.effmumu}
\end{figure}

We present our results for the reconstruction efficiency as a function of decay height $y_\mathrm{decay}$ in Figs.~\ref{f.effbb} for hadronic LLP decays. 
The blue curve represents the modular design of Fig.~\ref{f.mathuslamodular}, while the black curve represents the monolithic detector that triggers with full readout, representing the theoretical maximum for this detector geometry. The magneta, green and orange curves show the effect of reducing the module or RPC gap size, or increasing the module size. 
The benchmark modular design of Fig.~\ref{f.mathuslamodular} has very good efficiency  for reconstructing hadronic LLP decays, close to the theoretical maximum. The gaps between modules and RPC layers cause only very small reductions in the efficiency. This demonstrates that the idealized, purely geometrical signal estimates of Section~\ref{s.LLPphysics} are nonetheless a good indicator of MATHUSLA's sensitivity to hadronic LLP decays.


The standard ``tight'' LLP search requires the DV to be reconstructed inside the MATHUSLA100 module volumes. 
However, depending on the outcome of the more detailed background studies, this may be too conservative if MATHUSLA does not require a hermetic veto to reject cosmic ray and other backgrounds. 
We therefore also show efficiencies for a ``loose'' LLP search that allows for LLPs to decay in the gaps between modules and outside the detector, as long as its final states satisfy the same trigger and reconstruction criteria. These loose efficiencies are indicated as dashed lines in Fig.~\ref{f.mathuslamodular}. Both loose and tight efficiencies are normalized to the number of LLP decays inside the detector volume to clearly illustrate the achievable sensitivity gain, which is $\mathcal{O}(10\%$) for hadronic decays.

The reconstruction efficiencies for leptonic LLP decays are shown in Fig.~\ref{f.effmumu}. The monolithic detector efficiency (black curve) is $\mathcal{O}(50\%)$ for the 15 GeV LLP but much lower for heavier and hence slower LLPs. 
As expected, this efficiency is much worse than for hadronic decays with many charged final states. A large fraction of LLP decays have at least one muon exiting the decay volume through the sides or the floor.
The benchmark modular design (blue curve) has only about $1/3$ the efficiency of the monolithic detector. This can be understood due to the $\mathcal{O}(1)$ probability that a single angled track misses one RPC layer by passing between modules in the gaps if the gap size is $\mathcal{O}(10\%)$ the module size. 
We studied whether changing the trigger strategy to include next-to-nearest neighbors or even the full detector significantly increased sensitivity, but this was not the case: the nearest-neighbor strategy already saturated the possible sensitivity. The only way to increase reconstruction efficiency was to reduce the gaps in the detector (magenta, green, orange). However, it is clear that approaching the efficiency of the monolithic detector would require significantly smaller gaps than we are considering here.

\begin{table}
\begin{center}
\begin{tabular}{|l||l|l||l|l|}
\cline{2-5}
\multicolumn{1}{c||}{}
& 
\multicolumn{2}{c||}{$X \to b \bar b$}
& 
\multicolumn{2}{c|}{$X \to \mu^+\mu^-$} 
\\
\cline{2-5}
\multicolumn{1}{c||}{}
& $m_X = 15 \gev$ & $m_X = 50 \gev$ & $m_X = 15 \gev$ & $m_X = 50 \gev$ 
\\
\hline \hline
 {Default geometry} & 0.82, 0.93 & 0.92, 1.0 & 0.17, 0.20 & 0.076, 0.080 \\
 \hline
 {Smaller module gap} & 0.85, 0.96 & 0.93, 1.1 & 0.33, 0.37 & 0.13, 0.14 \\
 {Smaller RPC gap} & 0.84, 0.97 & 0.93, 1.1 & 0.27, 0.31 & 0.11, 0.11 \\
 {Larger modules} & 0.84, 0.95 & 0.93, 1.1 & 0.36, 0.40 & 0.14, 0.15 \\
 \hline
 {Monolithic Detector} & 0.85, 0.97 & 0.94, 1.1 & 0.57, 0.63 & 0.21, 0.22 \\
 \hline
\end{tabular}
\end{center}
\caption{
Average LLP reconstruction efficiencies $\epsilon_\mathrm{DV}$ for module configurations in Table.~\ref{t.modulegeometries}. The first (second) number in each entry corresponds to the tight (loose) search allowing only LLP decays in the decay volume (allowing decays outside), both normalized to the number of decays in the decay volume. Neighboring-module trigger except for Monolithic Detector. Note these efficiencies are normalized to the probability of decaying in decay volume under the RPCs.
}
\label{t.effsummary}
\end{table}


We summarize our results in Table.~\ref{t.effsummary}, where we show the reconstruction efficiencies for all studied decays, masses and module configurations by averaging $\epsilon_\mathrm{DV}$ over $y_\mathrm{decay}$, which corresponds to the average signal efficiency in the long-lifetime limit. 
Performance for hadronic decays is excellent, while relatively heavy leptonically decaying LLPs are more difficult. Fortunately, as discussed in Section~\ref{s.LLPphysics}, hadronically decaying and light leptonically decaying LLPs are the most important physics target, while heavy leptonically decaying LLPs will be well-covered by main detector LLP searches.


These studies will be refined further alongside the realistic detector design. For example, depending on backgrounds to two-pronged DV reconstruction, one might conduct a search at MATHUSLA for LLPs decaying to at least 3 or even more charged particles. This is likely to have very good acceptance for hadronically decaying LLPs with very strong additional background suppressions. 
To more effectively probe leptonically decaying LLPs, in particular for lower masses $\lesssim 10 \gev$, one might consider extending the design of Fig.~\ref{f.bgsummary} and \ref{f.mathuslamodular} to include a single vertical ``wall'' of RPC layers on the outer boundary of the decay volume facing away from the HL-LHC IP. This would only modestly increase the RPC sensor cost, but may significantly increase reconstruction efficiency for lower-multiplicity LLP final states. Quantitative studies of these scenarios are underway.




\section{The MATHUSLA Test Stand}
\label{s.teststand}

A test stand, with a detector layout similar to the one envisioned for the MATHUSLA detector, has been assembled at CERN. It was installed in the surface area above the ATLAS interaction point at LHC Point 1 in November 2017 and it has been taking data with different beam conditions. 
The main purpose of the test stand is to provide empirical information on potential backgrounds coming from the LHC as well as from cosmic rays.
The MATHUSLA detector has been conceived to be a background-free detector. This hypothesis can be studied making use of this test stand by taking data with different LHC beams conditions. On one hand, data taken with no beams running in the LHC will provide a good measurement of the background from cosmic rays as well as good information about their trigger and identification. On the other hand, data collected with beams running in the LHC can give an estimate of the background expected from LHC collisions. 
A precise measure of the charged particle flux in the test stand will provide the veto efficiency requirement for the main detector. The goal is to achieve a timing resolution sufficient to guarantee that no  cosmic particles can fake charged particle coming from LHC.
All the tests that will be performed until the end of LHC Run 2 will be fundamental to understand the cosmic ray rate and to extrapolate the LHC-related background rate from the test stand to MATHUSLA. 
All this information can be used to optimize the final design of the main detector.

As described in Section \ref{s.backgrounds}, several efforts are underway to develop simulations of the backgrounds expected in \mbox{MATHUSLA}. For muons and neutrinos traveling upwards, the idea is to create a ``gun'' that shoots particles into \mbox{MATHUSLA}, while for cosmic muons the standard cosmic ray simulations, CORSIKA, will be used. 
The test detector is crucial in the validation of these simulations. A comparison of real data with the simulations in the test stand will allow for any needed tuning in the simulations. It will also be a very useful cross-check to make sure that all sources of background have been properly taken into account.

\begin{figure}
\begin{center}
\subfigure[]{\label{f.mathuslateststand:a}\includegraphics[height=7cm]{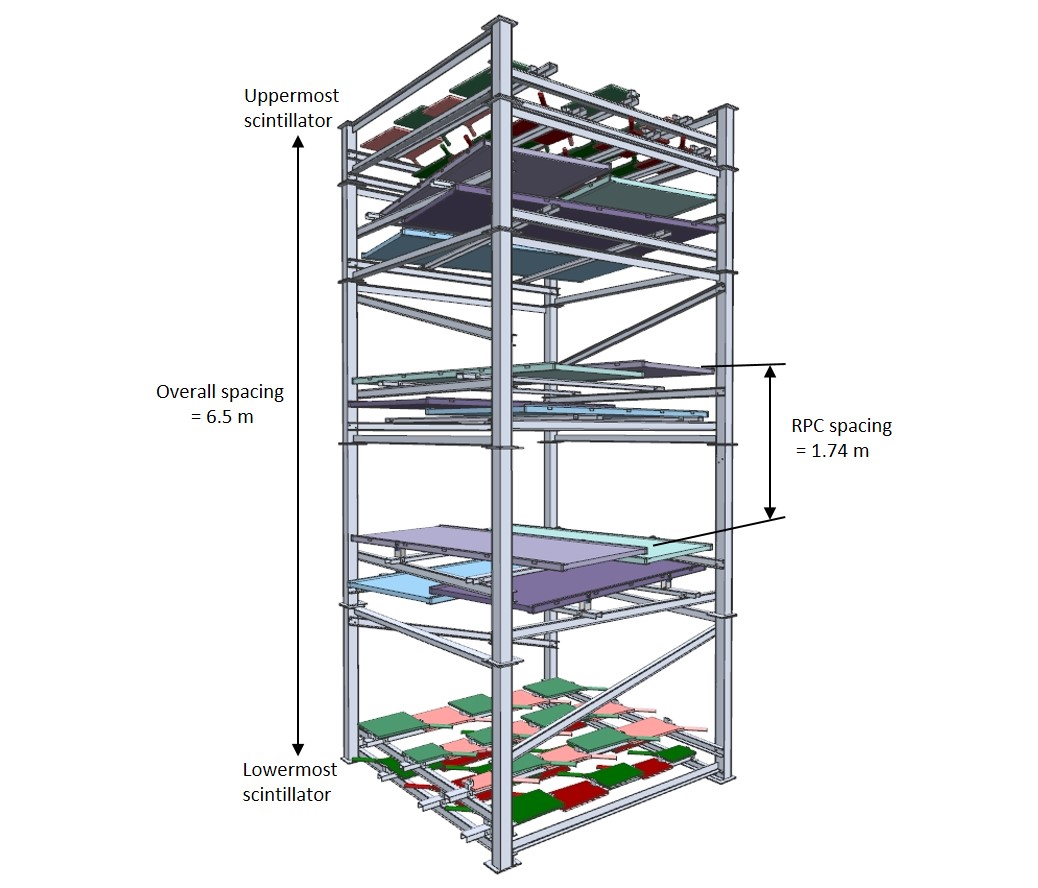}\label{f.mathuslateststand_a}}
\subfigure[]{\label{f.mathuslateststand:b}\includegraphics[height=7cm]{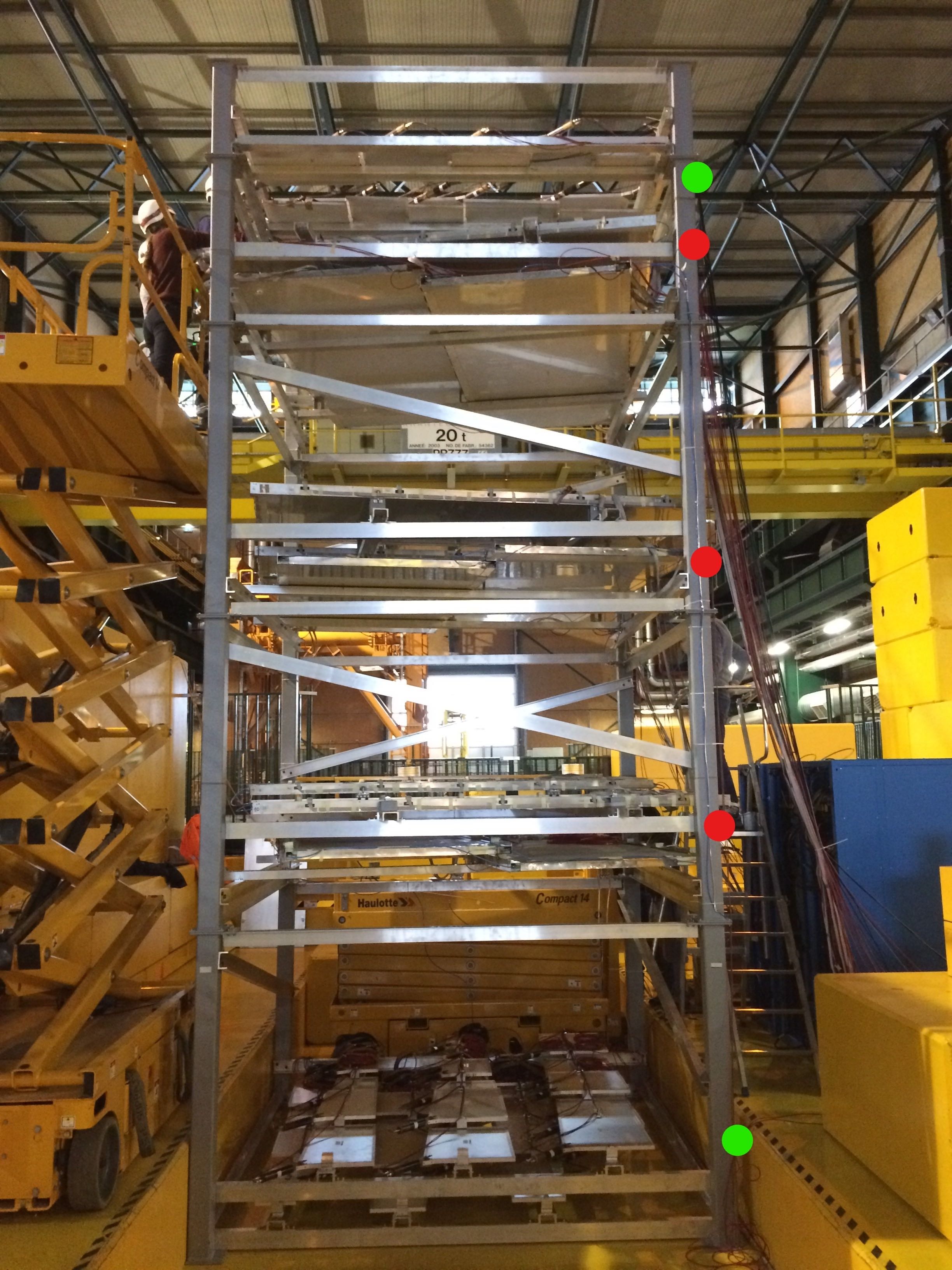}\label{f.mathuslateststand_b}}
\end{center}
\caption{
(a): schematic view of the MATHUSLA test stand. (b): picture of the final assembled structure in his test area in the ATLAS SX1 building at CERN. The green dots identify the two scintillator layers used for triggering, while the red dots the three RPC layers used for tracking.
}
\label{f.mathuslateststand}
\end{figure}

Following the concept of the main detector, the test stand is composed of two external layers of scintillators (one on the upper part of the detector, one on the lower part) with six layers of RPCs between them.
Figures~\ref{f.mathuslateststand_a} and \ref{f.mathuslateststand_b} show respectively the basic design of the test module and a picture of the final assembled structure in ATLAS SX1 building at CERN. The overall structure is $\sim 6.5$ m  tall, with an base of $\sim 2.5 \times 2.5$ m$^2$.

The test stand should not be considered a prototype of the main detector.
It has been built reusing pieces of other experiments, as detailed below, and hence the geometry of the test stand is not exactly the same as the one expected for MATHUSLA, where every subdetector will be custom-built. In addition, different detector technologies could be considered for the main experiment.
Even with these differences, the test stand will certainly provide very useful information for the design of the future MATHUSLA detector. 

The following subsections contain further information on the test stand design, trigger and data acquisition systems.

\subsection{Scintillators}

\begin{figure}[h!]
\begin{center}
\begin{minipage}[r]{0.35\textwidth}
\includegraphics[width=\textwidth]{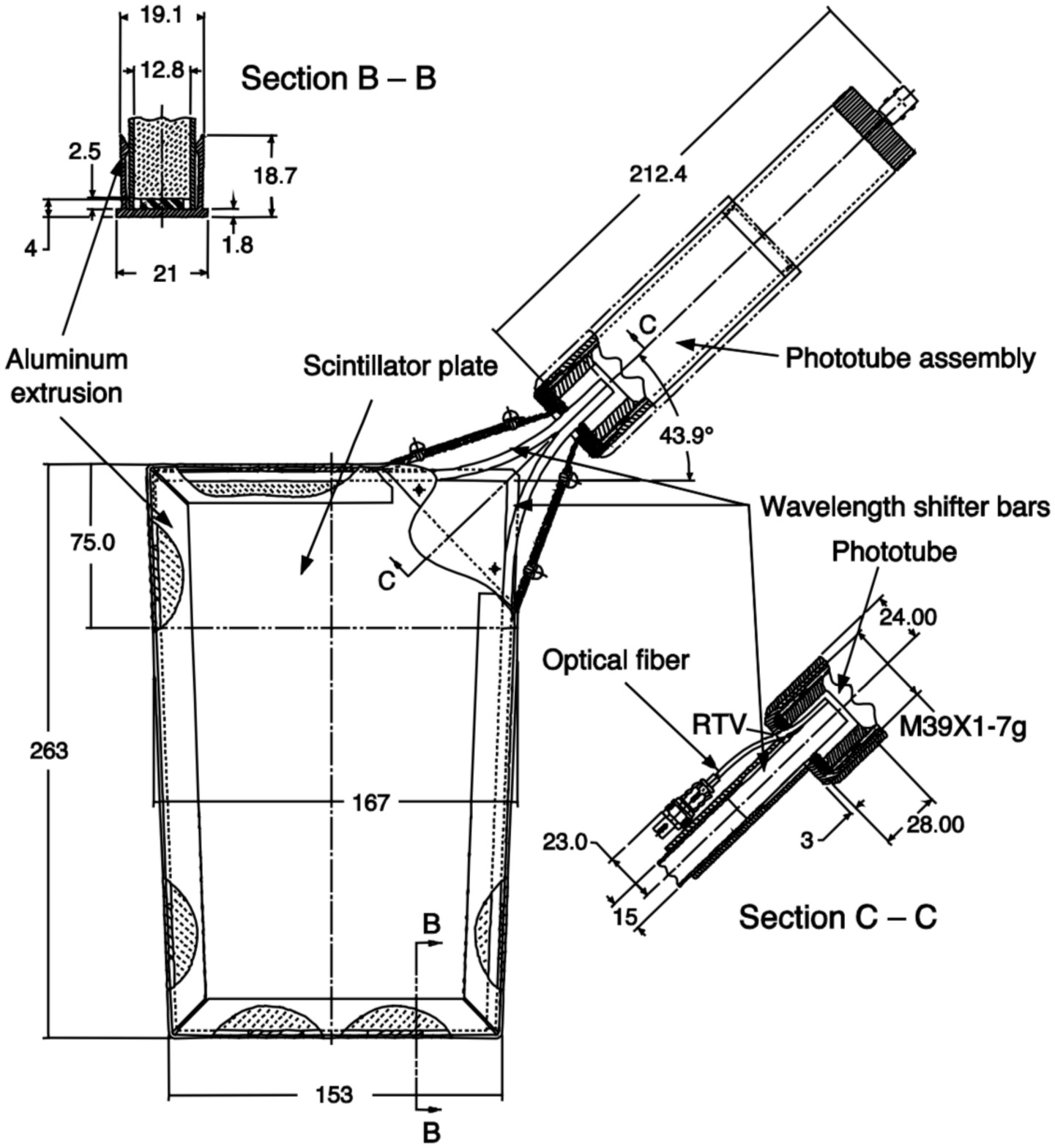}
\end{minipage}\hspace{0.5 cm}
\begin{minipage}[l]{0.30\textwidth}
\vspace{-0.8 cm}
\includegraphics[width=\textwidth]{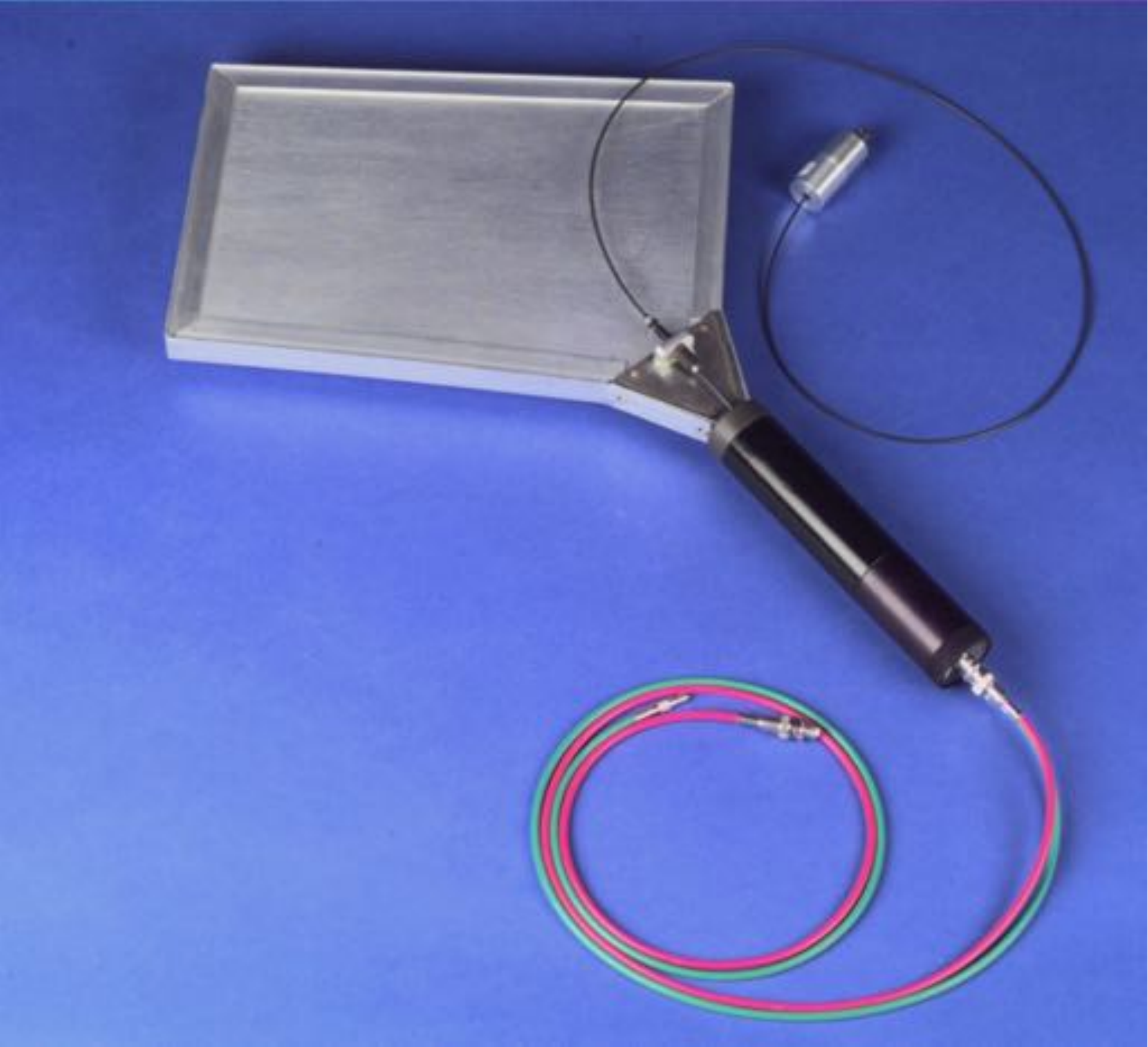}
\end{minipage}
\end{center}
\caption{
Details of the Tevatron D\O\,\, scintillators used for the MATHUSLA test stand.
}
\label{f.scintillator}
\end{figure}

Scintillator detectors (provided by Fermilab) are used in the test stand to trigger upwards and downwards going charged tracks. They are some of the 4214 trapezoidal-shaped scintillator counters used at the Tevatron D\O\,\, experiment for the forward muon trigger~\cite{D0:1997tda}. Figure \ref{f.scintillator} shows the details of one of the scintillators used for the test stand.

The size of the smallest counters is $9 \times 12$ cm$^2$ while the size of the largest counters is $60 \times 110$ cm$^2$. The counters use 12.8-mm-thick BICRON 404A scintillator plates of trapezoidal shape and wave length shifting (WLS) fibers for light collection. WLS bars material is SOFZ-105 and based on PMMA (polymethylmethacrylate) plastic containing wavelength shifting fluorescent dopant Kumarin 30.

\begin{figure}
\begin{center}
\subfigure[]{\label{f.mathuslateststand:a}\includegraphics[height=6cm]{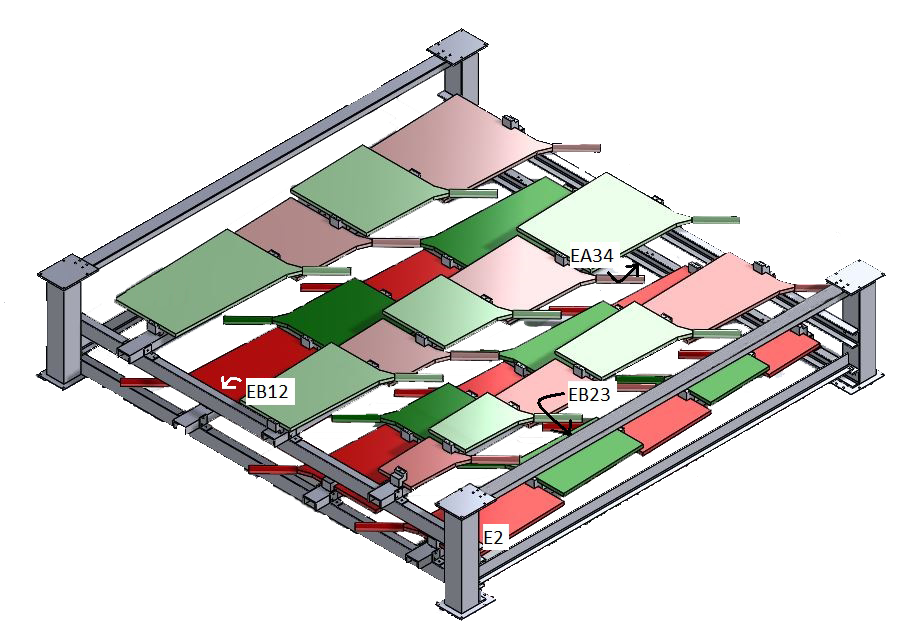}\label{f.scintillatorLayout_a}}
\subfigure[]{\label{f.mathuslateststand:b}\includegraphics[height=6cm]{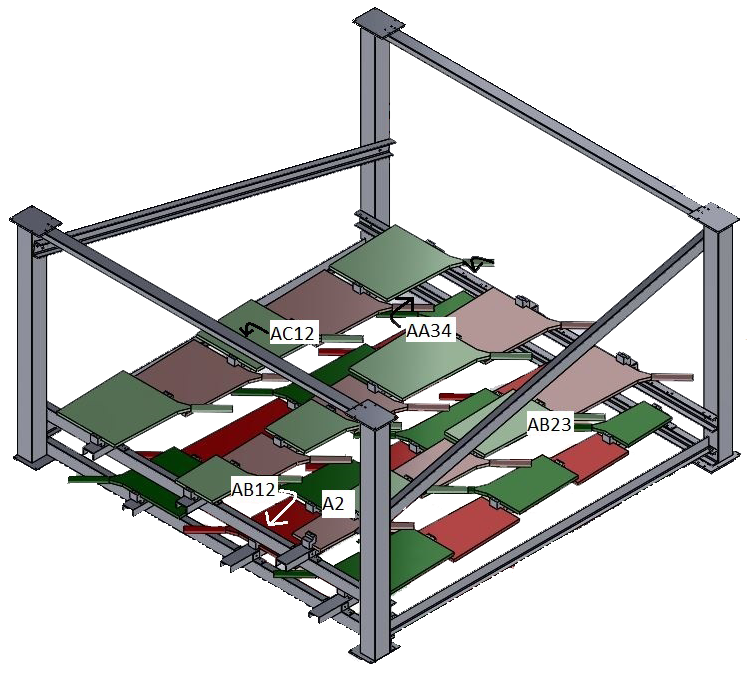}\label{f.scintillatorLayout_b}}
\end{center}
\caption{
Layout of the top (a) and bottom (b) scintillator planes used for the test stand.
}
\label{f.scintillatorLayout}
\end{figure}

The scintillators are mounted on two supporting structures as shown in Figure \ref{f.scintillatorLayout}, arranged in order to maximize the coverage. 

\subsection{RPC chambers}

The RPCs used for tracking (provided by University of Tor Vergata, Rome) are the same type of chambers used in the Argo-YBJ experiment~\cite{ref:Argo} at the YangBaJing Laboratory in Tibet (4300 m a.s.l.). Argo-YBJ chambers~\cite{ref:ArgoRPC} operated in streamer mode with a gas mixture of 15\% Argon, 10\% Isobutane and 75\% TetraFluoroEthane. The ATLAS RPC chambers operate in avalanche mode (to fulfill the harsh requirements of ATLAS trigger) using a gas mixture made of 95.2\% TetraFluoroEthane, 4.5\% of Isobutane, and 0.3\% of Sulfur Hexafluoride. Studies performed at University of Tor Vergata showed that Argo-YBJ chambers can be efficiently operated in streamer mode for test stand tracking purposes using the ATLAS RPC gas mixture with an addition of 15\% of Argon. This avoids the installation of an additional complex and expensive gas system in ATLAS Point 1.

\begin{figure}[h!]
\begin{center}
\begin{minipage}[r]{0.40\textwidth}
\includegraphics[width=\textwidth]{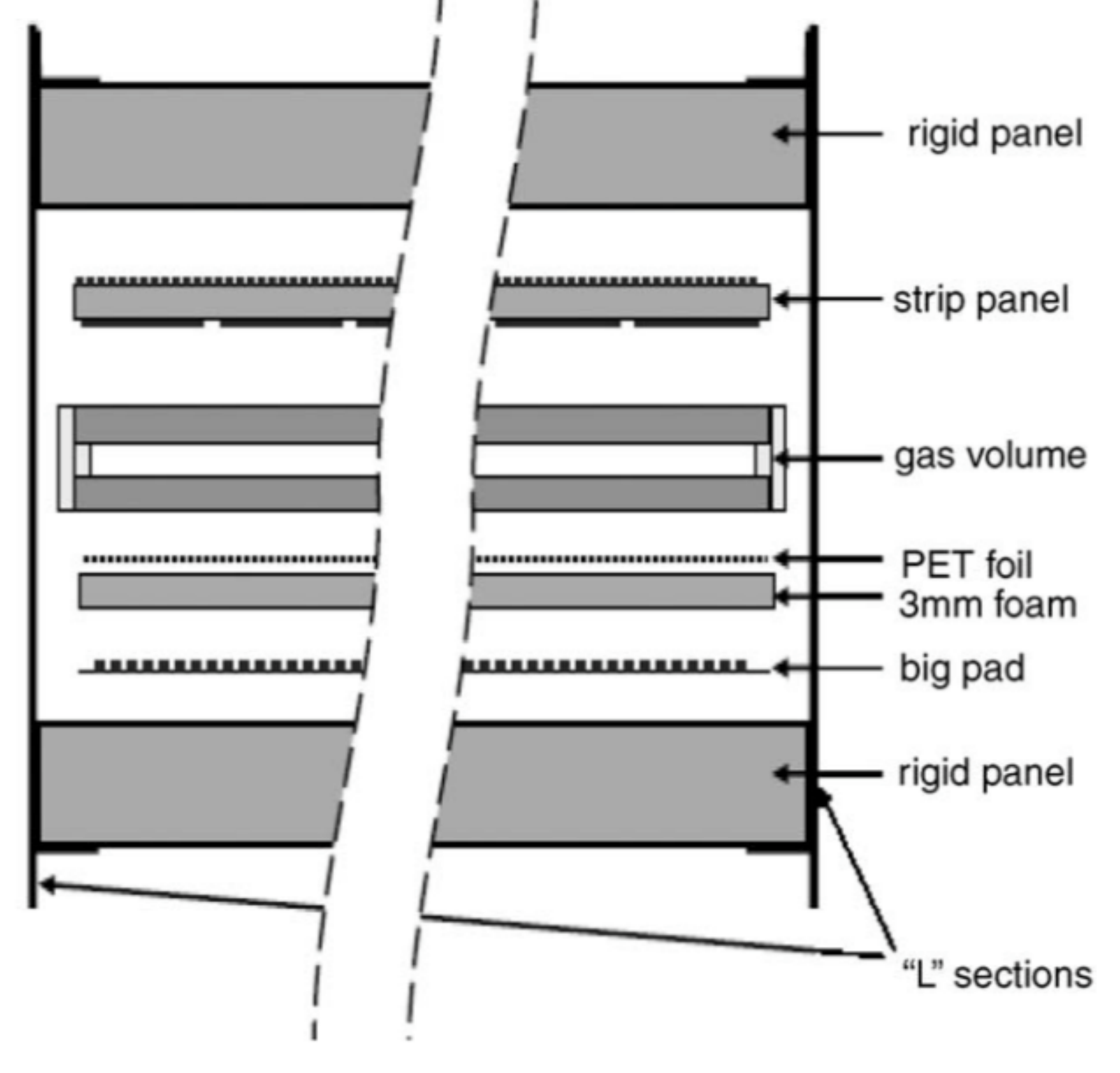}
\end{minipage}\hspace{0.5 cm}
\begin{minipage}[l]{0.49\textwidth}
\vspace{-0.8 cm}
\includegraphics[width=\textwidth]{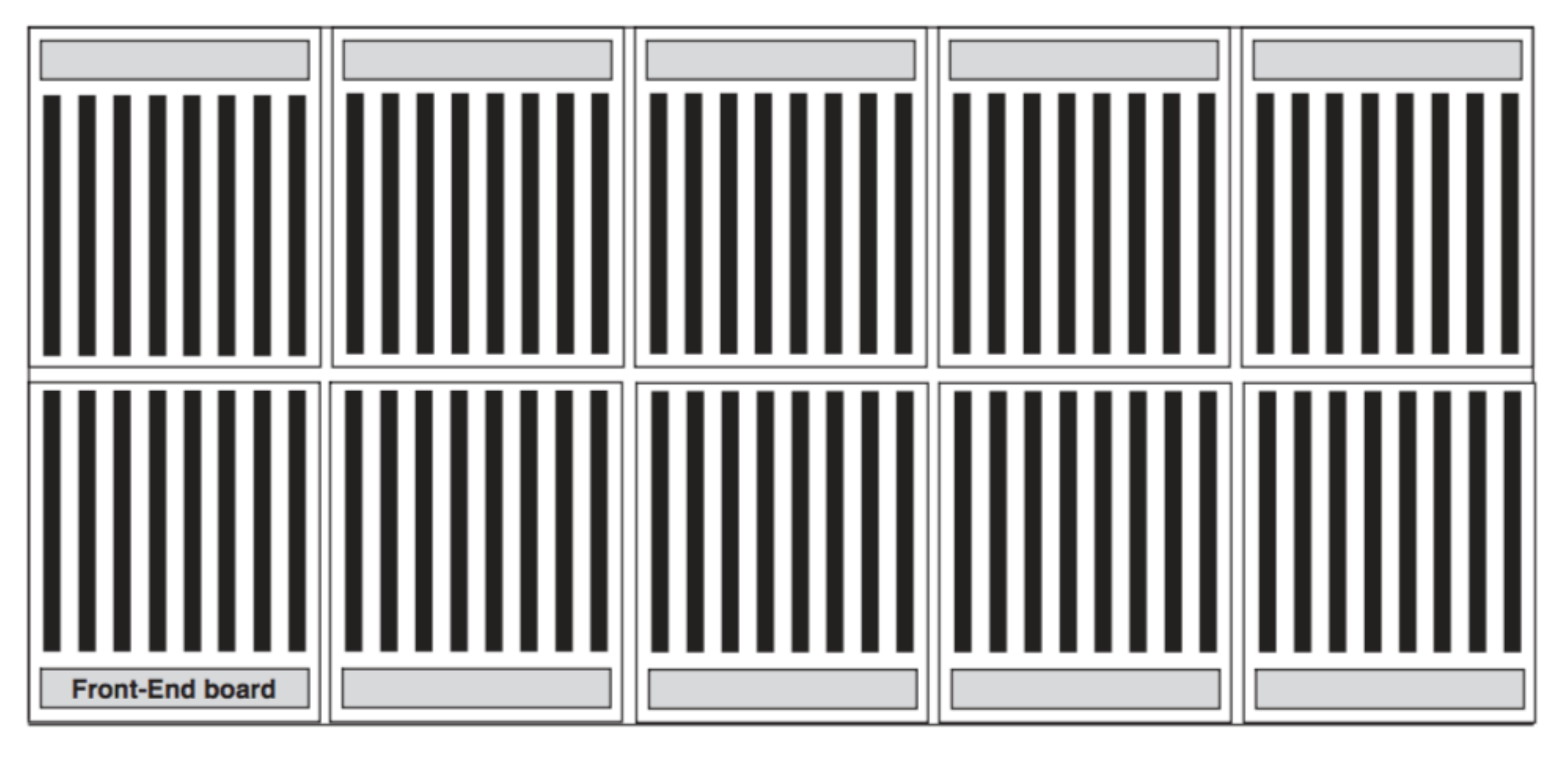}
\end{minipage}
\end{center}
\caption{
Cross-view of the Argo-YBJ chamber (left) and sketch of the strip panel used for read-out of the RPC (right).
}
\label{f.rpc}
\end{figure}

Figure \ref{f.rpc} shows a cross-view of the Argo-YBJ chamber and the sketch of the strip panel used for read-out of the RPC. Each counter is assembled in a box of $2850 \times 1225 \times 47$ mm$^3$. Two external shaped rigid panels consisting of an Aluminum (200 $\mu$m thick) foil, glued on a 1.5 cm foam layer, are used as a protection for the gas volume and to fix the chamber elements. Eight contiguous strips form a pad with a size $55.6 \times 61.8$ cm$^2$.

The gas volume consists of two Bakelite foils $2850 \times 1125 \times 2$ mm$^3$ with a resistivity $\sim 5 \times 10^{11}$ $\Omega$cm assembled to form a gas gap 2 mm wide. The strip panel consists of a sandwich of 17 $\mu$m copper foild glued on a 190 $\mu$m PET foil facing the gas volume and cut into 80 strips with a width of 6.75 cm and a length of 61.8 cm as shown in Figure \ref{f.rpc} (left), a 3 mm foam layer and a 17 $\mu$m copper foil glued on a 50 $\mu$m PET foil used as reference ground.

\subsection{Trigger and data acquisition}

\begin{figure}
\begin{center}
\subfigure[]{\label{f.mathuslateststand:a}\includegraphics[height=5cm]{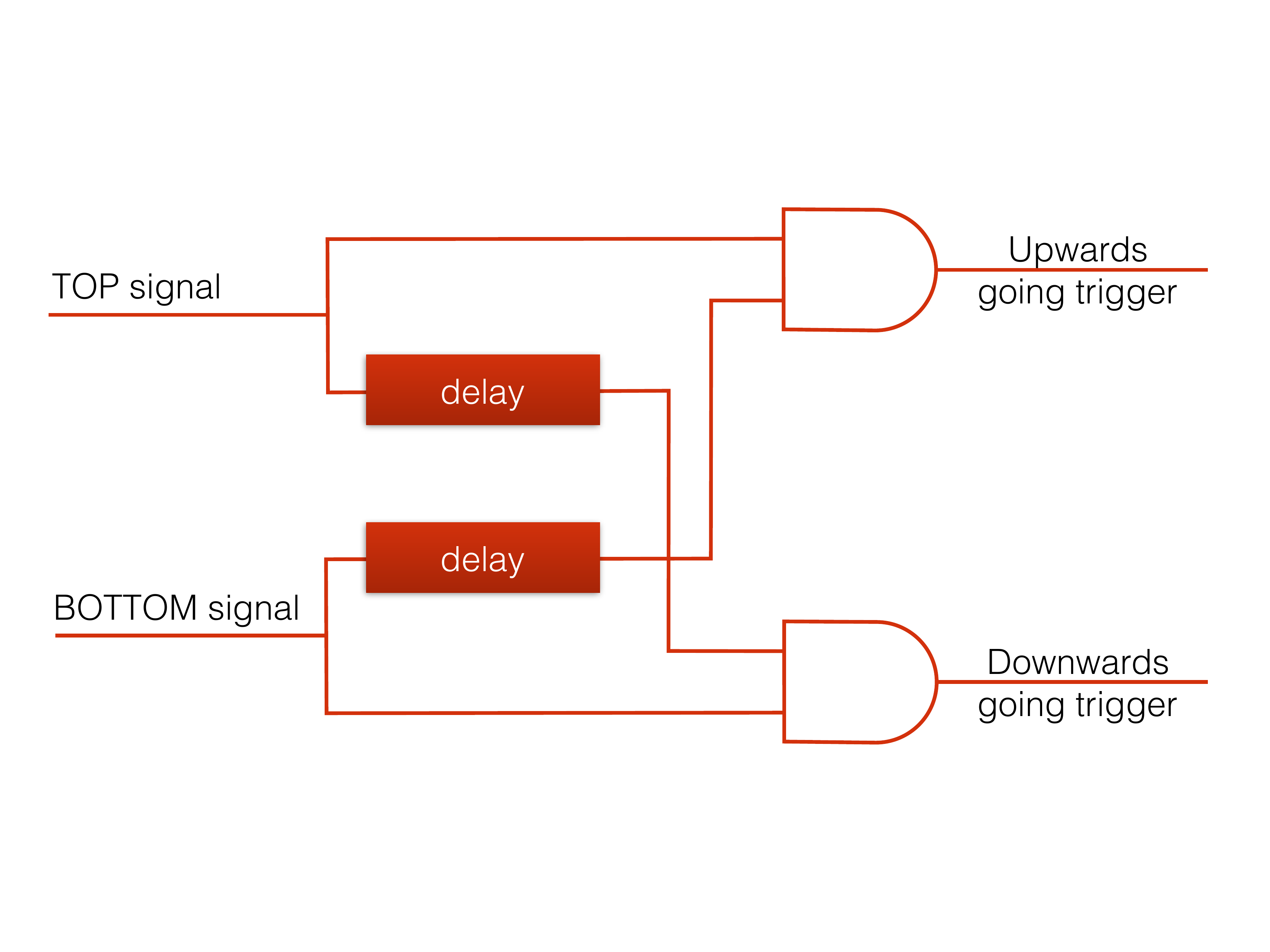}\label{f.TriggerDaq_a}}
\subfigure[]{\label{f.mathuslateststand:b}\includegraphics[height=4cm]{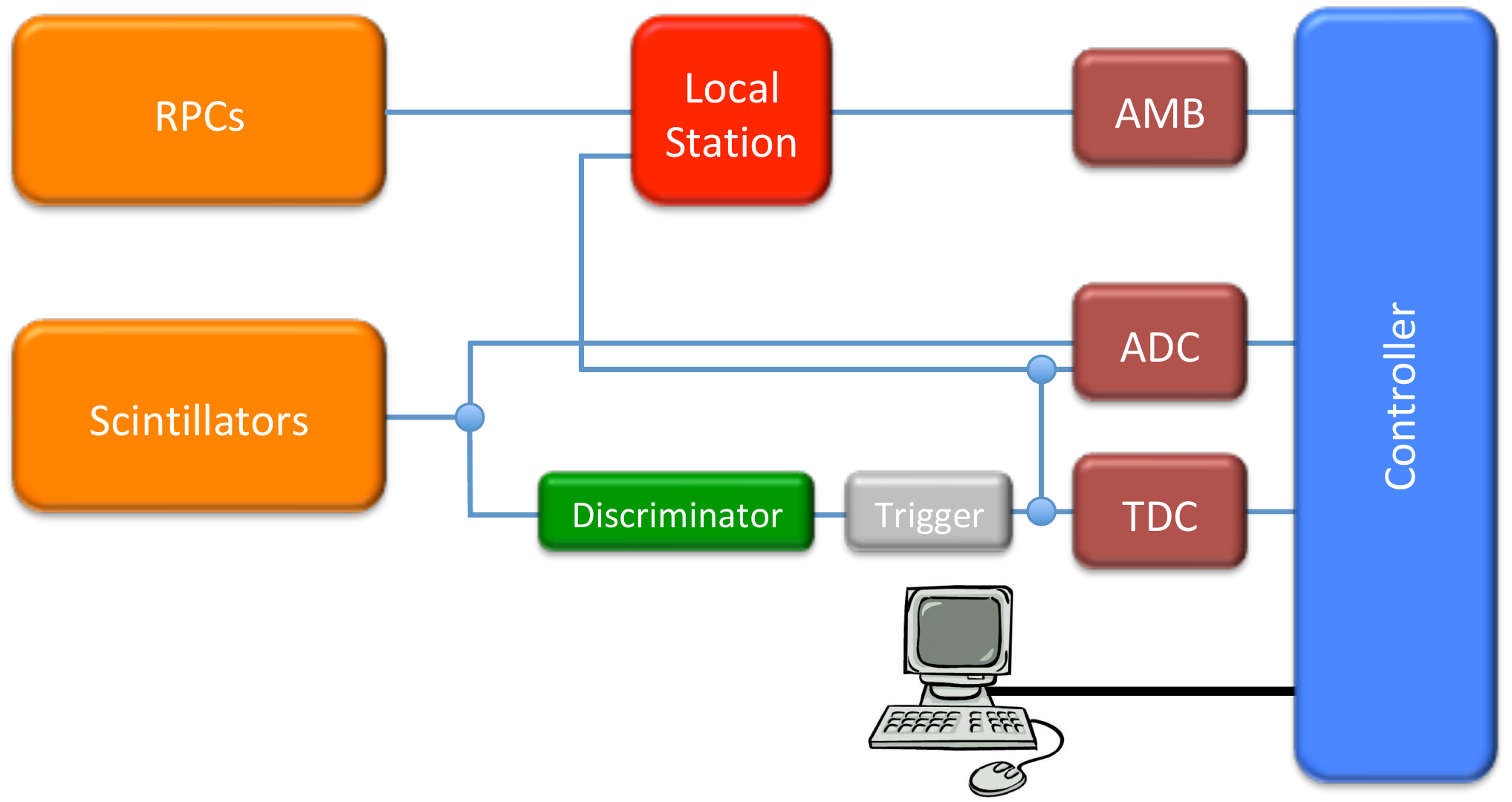}\label{f.TriggerDaq_b}}
\end{center}
\caption{
Schematic of the test stand trigger (a) and data acquisition system (b).
}
\label{f.TriggerDaq}
\end{figure}

Two primary triggers use the information from the scintillators to identify \emph{upwards going particles} and \emph{downwards going particles}.
Figure \ref{f.TriggerDaq} (left) shows the scheme of the trigger logics. The top (bottom) layer signal is defined as the logical OR of all the scintillators in the top (bottom) layer. The downward (upward) trigger is defined by the coincidence of the top (bottom) signal, with timing delayed by the time of flight of the crossing particle, and the bottom (top) signal. 

Figure \ref{f.TriggerDaq} (right) shows a schematic view of the MATHUSLA test stand data acquisition system. Scintillator data ADC/TDC are acquire using standard VME commercial electronics. Data from each RPC are acquired from a Receiver Card which reads out and digitizes the space and time information from 10 pick-up pads and gives out the pad multiplicity. The Receiver Cards are hosted in the Local Station used for Argo-YBJ experiment which can read up to 12 RPCs. When the Local Station receives a trigger, data are transferred to the AMB (Argo Memory Board).

\section{Next Steps}
\label{s.timeline}
We intend to write a full proposal that has mature background studies and cost estimates.  This still requires substantial work and some of the cost could be site dependent.  We are planning a work shop for the last week of August 2018 to be hosted by the Simons Center for Geometry and Physics at SUNY, Stony Brook, to which we have invited three experienced physicists to review our goals and current status.  We intend to submit by August several low-energy LLP model sensitivity estimates and a write up to the PBC working group to help them compare the MATHUSLA reach for low energy models to other proposals. We will continue to take test-stand data until the end of Run-2.







\bibliography{mathusla}
\bibliographystyle{JHEP}


\end{document}